\documentclass{egpubl}
\usepackage{eg2026}
\usepackage{amsmath, amssymb}
\usepackage[colorlinks=true,linkcolor={blue!70!black},citecolor={blue!70!black},urlcolor={blue!70!black}]{hyperref}
\usepackage{subcaption}
\usepackage{algorithm}
\usepackage{algpseudocode}
\usepackage{booktabs}
\usepackage{tabularx}
\usepackage{graphicx}
\usepackage{tikz}
\usetikzlibrary{
  positioning,
  arrows.meta,
  shapes,
  matrix,
  calc
}
\usepackage{caption}
\usepackage{xcolor}

\usepackage{amsmath,amsfonts,bm}

\def\eqref#1{equation~\ref{#1}}

\def\1{\bm{1}}

\def\rvb{{\mathbf{b}}}

\def\rvx{{\mathbf{x}}}
\def\rvy{{\mathbf{y}}}

\def\vzero{{\bm{0}}}
\def\vone{{\bm{1}}}

\def\vb{{\bm{\mathrm{b}}}}

\def\ve{{\bm{\mathrm{e}}}}

\def\vg{{\bm{\mathrm{g}}}}

\def\vm{{\bm{\mathrm{m}}}}

\def\vs{{\bm{\mathrm{s}}}}

\def\vu{{\bm{\mathrm{u}}}}

\def\vx{{\bm{\mathrm{x}}}}
\def\vy{{\bm{\mathrm{y}}}}

\def\mB{{\bm{\mathrm{B}}}}

\def\mD{{\bm{\mathrm{D}}}}
\def\mE{{\bm{\mathrm{E}}}}

\def\mG{{\bm{\mathrm{G}}}}

\def\mI{{\bm{\mathrm{I}}}}
\def\mJ{{\bm{\mathrm{J}}}}

\def\mM{{\bm{\mathrm{M}}}}

\def\mQ{{\bm{\mathrm{Q}}}}
\def\mR{{\bm{\mathrm{R}}}}

\def\mU{{\bm{\mathrm{U}}}}

\def\mY{{\bm{\mathrm{Y}}}}

\def\mLambda{{\bm{\Lambda}}}

\DeclareMathAlphabet{\mathsfit}{\encodingdefault}{\sfdefault}{m}{sl}
\SetMathAlphabet{\mathsfit}{bold}{\encodingdefault}{\sfdefault}{bx}{n}

\SpecialIssuePaper         

\CGFccby

\usepackage[T1]{fontenc}
\usepackage{dfadobe}

\biberVersion
\BibtexOrBiblatex
\usepackage[backend=biber,bibstyle=EG,citestyle=alphabetic,backref=true]{biblatex} 
\electronicVersion

\usepackage{egweblnk} 

\title[Differentiable Randers-Finsler Eikonal Solvers]{Differentiable Randers-Finsler Eikonal Solvers}

\author[B. Gahtan, J. Shpund \& A. M. Bronstein]{Barak Gahtan$^{1}$, Jacob Shpund$^{2}$, and Alex M. Bronstein$^{1,3}$\\$^1$ Department of Computer Science, Technion - Israel Institute of Technology, Haifa, Israel\\$^2$ The Hebrew University of Jerusalem, Jerusalem, Israel\\$^3$ ISTA Institute of Science and Technology, Vienna, Austria}

\pStartPage{1}

\begin{document}

\teaser{
\begin{tikzpicture}
  \node[anchor=south west, inner sep=0] (img) at (0,0)
    {\includegraphics[width=0.65\textwidth]{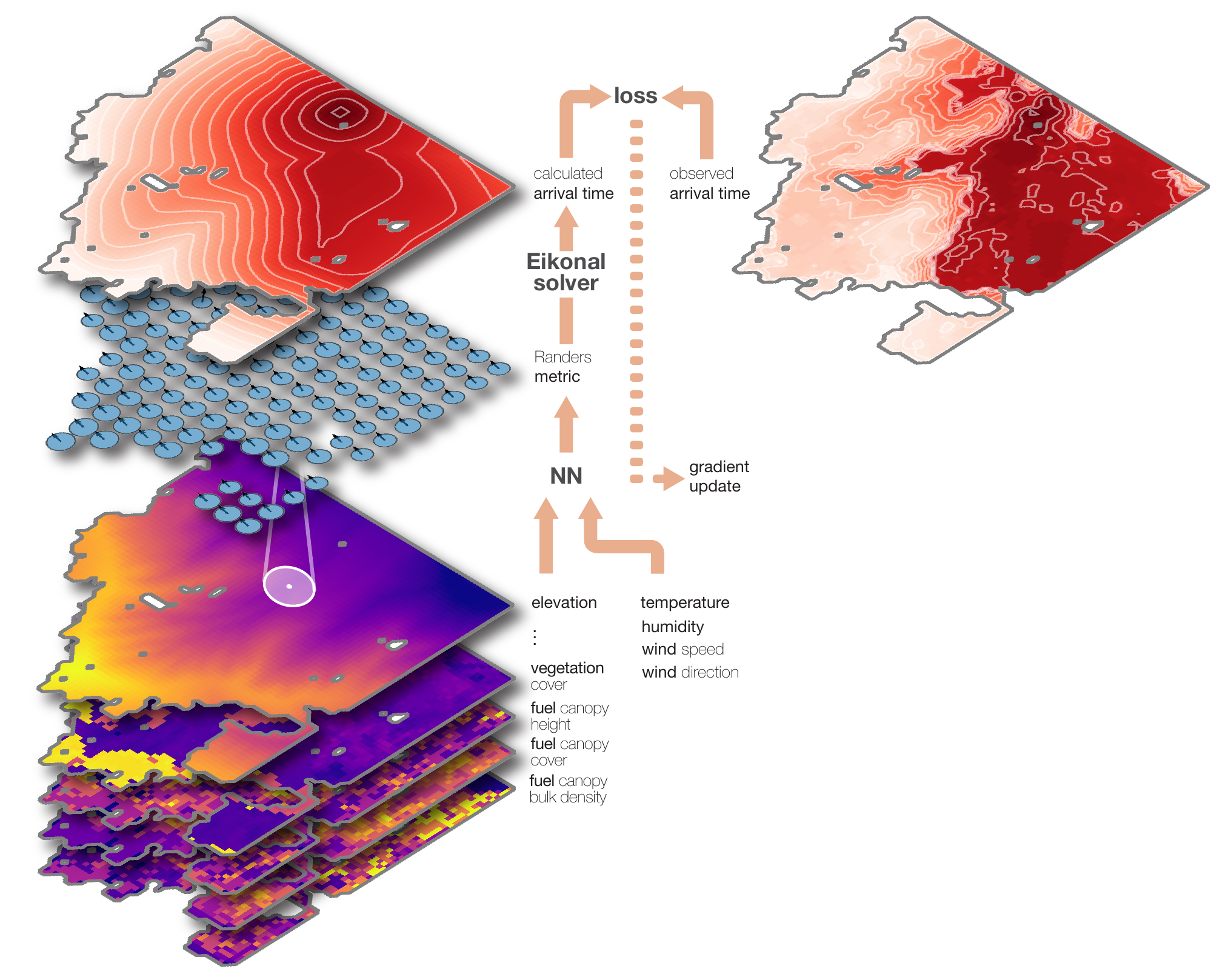}};
  \begin{scope}[x={(img.south east)}, y={(img.north west)}]
    \node[
      anchor=north west,
      fill=white,
      fill opacity=0.75,
      text opacity=1,
      rounded corners=2pt,
      inner sep=4pt,
      text width=0.5\linewidth,  
      align=justify                  
    ] at (0.65,0.55) {\textbf{Figure 1.} Overview of the differentiable eikonal framework for wildfire propagation. Environmental covariates including terrain properties (elevation, vegetation cover, fuel characteristics) and meteorological conditions (temperature, humidity, wind speed and direction) are processed through a neural network encoder that predicts spatially-varying Randers metric parameters at each grid location. The differentiable eikonal solver uses fast sweeping to compute predicted arrival times from these learned metrics. Training proceeds end-to-end by comparing predicted arrival times against observed fire progression data, with gradients flowing backward through implicit differentiation of the converged solution to update network parameters. This framework enables learning the complex relationships between environmental conditions and fire spread dynamics directly from observational data.};
  \end{scope}
\end{tikzpicture}

\captionsetup{labelformat=empty}
\caption*{} 
\captionsetup{labelformat=default}
\label{fig:framework_overview}
}

\maketitle
\begin{abstract}
Fast and differentiable solvers for anisotropic and asymmetric distance fields are a key primitive in geometry processing, enabling gradient-based optimization over metrics, drift fields, and downstream objectives that depend on geodesic distances and geodesics. We present a differentiable Eikonal solver for Randers–Finsler metrics on Cartesian grids that combines the efficiency of a GPU-friendly column/row fast sweeping with exact gradients obtained by implicit differentiation. Our forward pass uses local one- and two-point upwind updates selected by a causality-valid stencil; the backward pass exploits the induced arrival-time ordering to solve the adjoint system via a single reverse-time back-substitution, avoiding unrolling (i.e., recording and differentiating through every solver iteration) and substantially reducing memory and runtime. We derive closed-form derivatives of the discrete updates with respect to arrival times and Randers parameters, and we enforce metric feasibility with differentiable projections that guarantee positive definiteness and valid drift magnitude. Although stencil selection is piecewise-smooth, we show gradients are stable under small perturbations and match finite differences away from stencil boundaries. We demonstrate accurate forward solutions and enable inverse problems such as recovering spatially varying anisotropic metrics and drift fields from sparse arrival-time supervision. Finally, we apply the method to learning data-driven spread models on real wildfire perimeters, illustrating scalability and the practical utility of differentiable Randers distance fields.


\printccsdesc   
\end{abstract}  
\section{Introduction}
The eikonal equation $\|\nabla T\| = 1/F(\mathbf{x})$ governs wavefront propagation in isotropic media, with solutions representing arrival times or geodesic distances from source points. Efficient algorithms for solving this equation, notably the Fast Marching Method~\cite{sethian1996fast} and Fast Sweeping Method~\cite{zhao2005fast}, have become major tools in computational geometry, enabling applications from shape analysis and mesh processing to path planning and medical imaging. These methods solve the \emph{forward problem}: given a speed function $F(\mathbf{x})$, compute the arrival time field $T(\mathbf{x})$.

Many applications, however, require solving the inverse problem: given observed arrival times, recover the underlying propagation parameters. Seismic tomography estimates subsurface velocity structure from earthquake travel times. Diffusion tensor imaging infers tissue microstructure from anisotropic diffusion measurements. Robotics applications learn cost maps from demonstrated trajectories. In each case, the inverse problem requires differentiating through the eikonal solver to enable gradient-based parameter estimation.

Standard eikonal solvers are not readily differentiable. The Fast Marching Method maintains a heap-based priority queue with data-dependent ordering, while the Fast Sweeping Method involves discrete stencil selection where one of several update formulas is chosen based on the local solution structure. Both characteristics obstruct automatic differentiation: unrolling the iterations creates unwieldy computation graphs, and the discrete choices introduce points of non-differentiability.

We develop a differentiable framework for the Randers-Finsler eikonal equation, which generalizes the standard eikonal to include both anisotropic propagation (via a metric tensor $\mG$) and directional drift (via a vector field $\vb$). This formulation encompasses isotropic, Riemannian, and asymmetric propagation as special cases, making it suitable for applications where wavefronts exhibit direction-dependent speeds or systematic bias. Our framework consists of two components. The forward solver uses fast sweeping with eight triangular stencils to compute arrival times in $O(n)$ complexity per iteration for a grid with $n = N \times N$ points, converging in 2--3 iterations independent of grid size. The backward pass applies implicit differentiation to the converged solution, computing parameter gradients by solving a sparse triangular adjoint system whose structure mirrors the causality of the forward solve.

The key technical challenge is discrete stencil selection inherent to eikonal solvers. At each grid point, the solver chooses one of eight triangular stencils based on which neighboring values produce the smallest valid arrival time. This discrete choice means the solution operator is not everywhere differentiable. We analyze this issue and show that the problematic points---stencil boundaries where the optimal stencil changes---form one-dimensional curves in the two-dimensional domain, constituting a measure-zero set. At all other points, implicit differentiation produces exact gradients. Moreover, gradients remain stable under small parameter perturbations even when stencil boundaries are crossed, with variations below 1\% for 1\% parameter noise.

We validate the framework through extensive numerical experiments. The forward solver achieves sub-1\% relative error across isotropic, anisotropic, and combined Randers-Finsler configurations. Gradient verification confirms pointwise-exact agreement between implicit differentiation and finite differences at interior points, with the expected discrepancies only at stencil boundaries. For inverse problems, the framework recovers isotropic metrics to 5.6\% error and drift fields to under 3\% error from single-source observations, with multi-source configurations providing complementary geometric constraints that reduce errors by 45\%. Computationally, fast sweeping achieves up to 1400$\times$ speedup over Jacobi iteration, and implicit differentiation achieves 755$\times$ speedup over finite differences, making gradient-based optimization practical.

As a real-world application, we demonstrate the framework on wildfire propagation modeling using the Sim2Real-Fire dataset, where the Randers-Finsler formulation naturally captures terrain-induced anisotropy and wind-driven asymmetry. As illustrated in Figure~\ref{fig:framework_overview}, a neural network encoder maps environmental covariates (terrain, fuel, weather) to Randers metric parameters, the forward solver computes arrival times respecting wavefront physics, and gradients flow backward through implicit differentiation to train the system end-to-end. Experiments across 15 geographic scenes achieve mean test correlations of 0.824 for within-scene generalization, while a model trained on 11 scenes transfers to 4 unseen landscapes with correlation 0.766, demonstrating that learned covariate-to-metric mappings capture universal propagation physics rather than memorizing location-specific patterns.

Our contributions are: (1) a differentiable Randers-Finsler eikonal solver combining fast sweeping with implicit differentiation, (2) theoretical and empirical analysis of the stencil boundary non-differentiability showing it is measure-zero and gradient-stable, (3) comprehensive validation on synthetic inverse problems demonstrating accurate parameter recovery, and (4) application to wildfire modeling demonstrating both within-scene and cross-scene generalization, with capacity analysis establishing that learning generalizable covariate-to-metric mappings is the primary challenge rather than model expressiveness.

\section{Related Work}

\textbf{Eikonal Equations and Fast Solvers.} The two dominant approaches to solving the eikonal equation are the Fast Marching Method (FMM)~\cite{sethian1996fast} and the Fast Sweeping Method (FSM)~\cite{zhao2005fast}. FMM achieves $O(n \log n)$ complexity for $n$ grid points by propagating the solution outward using a heap-based priority queue, with the connection to Dijkstra's algorithm made explicit by Tsitsiklis~\cite{tsitsiklis1995efficient}. FSM provides an alternative achieving $O(n)$ complexity through alternating Gauss-Seidel iterations along characteristic directions; rather than maintaining a sorted front, fast sweeping performs a fixed number of directional passes through the grid, with convergence guaranteed by covering all possible characteristic orientations. Qian et al.~\cite{qian2007fast} extended fast sweeping to triangulated meshes, while parallel variants have been developed for GPU architectures~\cite{weber2008parallel, detrixhe2013parallel}. The choice between FMM and FSM involves tradeoffs: FMM provides exact ordering but requires heap operations, while FSM offers simpler implementation and better cache behavior but may require multiple iterations for complex geometries.

For surfaces and meshes, Kimmel and Sethian~\cite{kimmel1998computing} introduced geodesic computation via the eikonal equation on triangulated domains, enabling applications in shape analysis, surface parameterization, and mesh processing. The heat method of Crane et al.~\cite{crane2017heat} provides an elegant alternative using short-time heat diffusion, offering robustness to mesh quality at the cost of solving linear systems. These methods compute geodesic distances as a forward problem; our work also addresses the inverse direction: recovering the metric from observed distances.

\textbf{Anisotropic and Finsler Extensions.} Many applications require anisotropic propagation where speed depends on direction. Sethian and Vladimirsky~\cite{sethian2003ordered} developed ordered upwind methods for general Hamilton-Jacobi equations, handling anisotropy through enlarged stencils that respect the causality structure. Their approach extends FMM to settings where the characteristic direction is not aligned with the upwind gradient, requiring careful stencil selection to maintain convergence.

Finsler geometry provides the natural mathematical framework for direction-dependent metrics. In the general Finsler setting the metric need not arise from an inner product; in this work we restrict attention to the Randers case, where the metric combines a Riemannian (SPD) tensor with a drift vector, which suffices to model anisotropic and asymmetric propagation. The Randers metric, a specific Finsler structure combining a Riemannian metric with a drift vector, models situations where propagation is biased in one direction arising in optimal control under drift, robotics path planning with wind, and wavefront propagation in moving media.

Mirebeau~\cite{mirebeau2014anisotropic,braides2002riemannian} developed efficient algorithms for anisotropic eikonal equations using adaptive stencils derived from lattice geometry, achieving both accuracy and efficiency for strongly anisotropic metrics. His work on Randers metrics~\cite{mirebeau2014efficient} specifically addresses the asymmetric case relevant to our formulation. These solvers focus on the forward problem; we build on similar discretization principles but develop the backward pass for gradient computation through implicit differentiation.

\textbf{Forward Solver Comparison.} Mirebeau's FMASR~\cite{mirebeau2014anisotropic,mirebeau2014efficient} achieves logarithmic complexity in anisotropy ratio through adaptive stencils, while ordered upwind methods~\cite{sethian2003ordered} and FMM~\cite{sethian1996fast} use heap-based ordering. We employ eight-stencil fast sweeping, which offers simpler parallelization and, crucially, a transparent adjoint structure for implicit differentiation. The fixed sweep order avoids the data-dependent node orderings that complicate backward passes in heap-based methods. For moderate anisotropy ratios (up to 1:16), our discretization achieves sub-1\% accuracy while enabling efficient gradient computation.

\textbf{Differentiable Solvers and Inverse Problems.} Computing gradients through numerical solvers enables learning unknown parameters from observational data. Two main approaches exist: differentiating through the solver iterations via automatic differentiation, or applying implicit differentiation to the converged solution. The latter, rooted in PDE-constrained optimization~\cite{plessix2006review}, avoids storing intermediate states and handles iterative solvers with varying iteration counts.

Implicit differentiation treats the converged solution as satisfying $R(T, \theta) = 0$, where $\theta$ denotes the parameters of interest (in our setting, the metric tensor $\mathbf{G}$ and drift field $\mathbf{b}$), and applies the implicit function theorem to obtain $dT/d\theta$ without unrolling iterations. Amos and Kolter~\cite{amos2017optnet} popularized this approach for optimization layers in neural networks, while Chen et al.~\cite{chen2018neural} applied similar ideas to neural ODEs. For elliptic PDEs, the adjoint method yields sparse linear systems that can be solved efficiently~\cite{kochkov2021machine}.

Physics-Informed Neural Networks (PINNs)~\cite{raissi2019physics} offer an alternative by parameterizing solutions directly with neural networks and penalizing PDE residuals. While avoiding mesh-based discretization, PINNs can struggle with hyperbolic equations where characteristics create sharp gradients. Our approach combines classical numerical methods with implicit differentiation: we solve the eikonal equation using fast sweeping, then compute gradients by solving an adjoint system whose sparsity structure mirrors the causality of the forward solve.

The key technical challenge for eikonal equations is that the solver involves discrete stencil selection -- at each grid point, one of several candidate update formulas is chosen based on the local solution structure. This discrete choice creates points of non-differentiability. We analyze this issue and show that the problematic points form a measure-zero set (stencil boundaries), making implicit differentiation valid almost everywhere and stable under perturbation.

\textbf{Applications.} Inverse eikonal problems arise across diverse domains. In seismic tomography, subsurface velocity structure governs seismic wave travel times; the inverse problem recovers the slowness field from observed arrivals~\cite{rawlinson2003seismic}. Medical imaging applications include diffusion tensor imaging where anisotropic propagation reflects tissue microstructure~\cite{lenglet2009mathematical}. In robotics, the eikonal equation governs optimal paths under direction-dependent costs, with inverse problems arising when learning cost maps from demonstrated trajectories~\cite{ratliff2009chomp}.

We demonstrate our framework on wildfire propagation, where the Randers-Finsler formulation naturally captures terrain-induced anisotropy and wind-driven asymmetry. Fire arrival times derived from satellite observations provide supervision for learning the metric parameters as functions of environmental covariates. This application showcases the framework's ability to handle real-world data with complex spatial structure and partial observations.

\section{Methodology}\label{sec:methodology}
We develop a differentiable framework for learning spatially-varying propagation parameters from observed arrival time data. Our approach consists of three components: a Randers-Finsler eikonal formulation that captures anisotropic propagation with directional bias, a fast sweeping solver that computes arrival times in $O(n)$ operations for a grid with $n = N \times N$ points, and an implicit differentiation scheme that obtains parameter gradients by solving an adjoint system on the converged solution.

\subsection{The Randers-Finsler Eikonal Equation} \label{sec:eikonal}
Classical eikonal equations govern isotropic wavefront propagation where speed depends only on position. Many physical phenomena, however, exhibit fundamentally anisotropic behavior: seismic waves travel faster along geological strata, robotic path costs depend on terrain slope, and wavefronts in moving media experience directional bias. We capture these effects through a Randers-Finsler metric that generalizes the standard eikonal equation to include both anisotropy and directional drift.

The arrival time field $T(\mathbf{x})$ depends on two spatially-varying fields: a symmetric positive-definite metric tensor $\mathbf{G}(\mathbf{x}) \in \mathbb{R}^{2 \times 2}$ encoding anisotropic  propagation speed, and a drift vector $\mathbf{b}(\mathbf{x}) \in \mathbb{R}^2$ that biases the propagation direction. These fields enter through the static Randers-Finsler eikonal equation
\begin{equation}
(\nabla T - \mathbf{b})^\top \mathbf{G}^{-1} (\nabla T - \mathbf{b}) = 1,
\label{eq:randers_eikonal}
\end{equation} with boundary condition $T = 0$ at source locations. The metric tensor determines an elliptical speed profile at each location, with eigenvalues controlling the principal speeds and eigenvectors defining the preferred directions. The drift vector shifts this ellipse, enabling asymmetric propagation where the wavefront advances faster in one direction than its opposite.

This formulation includes several important special cases. When $\mG = \mI$ and $\vb = \vzero$, Equation~\ref{eq:randers_eikonal} reduces to the standard isotropic eikonal equation $\|\nabla T\| = 1$, producing circular wavefronts expanding at unit speed. Setting $\vb = \vzero$ with general $\mG$ yields a Riemannian eikonal equation with elliptical wavefronts whose shape varies spatially according to the metric. The full Randers-Finsler form with nonzero drift captures phenomena where propagation is inherently asymmetric, such as wavefront propagation in moving media or biased diffusion processes.

\subsection{Fast Sweeping Solver} \label{sec:fast_sweeping}

We discretize the Randers-Finsler eikonal equation on a Cartesian grid and solve it with a fast sweeping method. The key property is \emph{causality}: each node's arrival time depends only on neighboring nodes with smaller arrival times (upwind donors). The original FSM processes nodes in raster order within each sweep direction, creating sequential dependencies that limit parallelism. We adopt a column/row decomposition that processes entire columns (or rows) simultaneously, exposing $O(N)$ parallelism per step and making the algorithm GPU-friendly while preserving the same convergence guarantees. We update each grid node by taking the minimum over a small set of local, monotone candidates induced by adjacent neighbor pairs (``triangular stencils'') and, when necessary, one-edge fallbacks.

Concretely, for each node we evaluate (i) a two-donor update from a valid triangular stencil and (ii) one-donor updates along individual edges, and set $T$ to the smallest valid candidate. For the two-point update, let $\mathbf{M} = [\mathbf{m}_1 \mid \mathbf{m}_2]$ be the matrix of displacement vectors to the two donors with arrival times $T_1, T_2$. Define the stencil metric $\mathbf{E} = \mathbf{M}^\top \mathbf{G} \mathbf{M}$, its inverse $\mathbf{Q} = \mathbf{E}^{-1}$, and drift-adjusted times $S_i = T_i + \mathbf{m}_i \cdot \mathbf{b}$. The candidate arrival time $T_0$ is the larger root of the quadratic
\begin{equation}
    (\mathbf{S} - T_0 \mathbf{1})^\top \mathbf{Q} (\mathbf{S} - T_0 \mathbf{1}) = 1,
    \label{eq:two_point}
\end{equation}
where $\mathbf{S} = [S_1, S_2]^\top$ and $\mathbf{1} = [1,1]^\top$. This update is valid only when both donors satisfy $T_i < T_0$ (causality) and the update direction lies within the triangle (geometric validity); full derivation and validity conditions are provided in Appendix~\ref{app:solver_details}. When the two-point update fails validity, we fall back to one-donor updates along individual edges:
\begin{equation}
    T_0 = T_i + \vm_i \cdot \vb + \sqrt{\vm_i^\top \mG \vm_i}.
    \label{eq:one_point}
\end{equation}

We initialize $T=0$ at source nodes and $T=\infty$ elsewhere, then perform alternating directional sweeps (Figure~\ref{fig:sweep_directions})~\cite{weber2008parallel}. Each iteration consists of four sweeps corresponding to the four quadrant orderings, ensuring that regardless of wavefront geometry, at least one sweep propagates information consistently with the characteristics. In practice we observe convergence in 2--3 iterations independent of grid size, yielding $O(n)$ total work for $n = N \times N$ grid points. Pseudocode is provided in Appendix~\ref{app:solver_details}.
\setcounter{figure}{1} 
\begin{figure}[t]
    \centering
    \includegraphics[width=1\columnwidth]{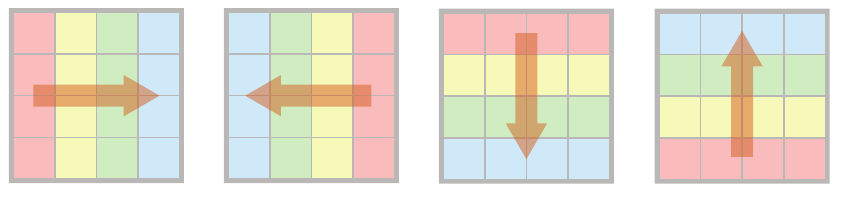}
    \caption{Four sweep directions in fast sweeping. Arrows indicate column or row traversal order; like-colored grid elements are updated simultaneously. Each sweep processes the grid corner-to-corner, ensuring at least one sweep direction contain characteristic direction.}
    \label{fig:sweep_directions}
\end{figure}

\subsection{Differentiable Backward Pass} \label{sec:backward}
We learn $\theta=(\mG,\vb)$ from observed arrival times by differentiating a loss $\mathcal{L}(T)$ through the converged eikonal solution. Backpropagating through the iterative sweeping procedure (unrolled differentiation) is undesirable due to (i) variable iteration counts, (ii) discrete stencil selection, (iii) long sequential dependencies, and (iv) prohibitive memory requirements, as storing intermediate states for grids of practical size ($300 \times 300$ and larger) exceeds available GPU memory. Instead, we differentiate the \emph{fixed point} defined by local residuals $R(T,\theta)=0$ using implicit differentiation.

Applying the implicit function theorem yields
\begin{equation}
    \frac{dT}{d\theta} = -\left(\frac{\partial R}{\partial T}\right)^{-1} \frac{\partial R}{\partial \theta},
    \label{eq:implicit}
\end{equation}
and therefore
\begin{equation}
    \frac{d\mathcal{L}}{d\theta}
    = \frac{\partial \mathcal{L}}{\partial T}\frac{dT}{d\theta}
    = -\boldsymbol{\lambda}^\top \frac{\partial R}{\partial \theta},
    \label{eq:param_grad}
\end{equation}
where $\boldsymbol{\lambda}$ solves the adjoint system
\begin{equation}
    \left(\frac{\partial R}{\partial T}\right)^\top \boldsymbol{\lambda}
    = \left(\frac{\partial \mathcal{L}}{\partial T}\right)^\top.
    \label{eq:adjoint}
\end{equation}

The key insight enabling efficient gradient computation is that the forward sweep's arrival-time ordering directly provides the structure needed for the backward pass. Crucially, when nodes are ordered by increasing arrival time, the Jacobian $\mathbf{J}=\partial R/\partial T$ is sparse and lower-triangular because each residual $R_j$ depends only on $T_j$ and its donors; hence $T_i$ affects $R_j$ if and only if $i \in \text{donors}(j)$. This structure enables solving the adjoint system $\mathbf{J}^\top \boldsymbol{\lambda} = \mathbf{g}$ efficiently by a single reverse-time back-substitution pass, processing nodes in decreasing arrival time order. We then compute per-node parameter gradients via $-\boldsymbol{\lambda}^\top(\partial R/\partial \theta)$. Full Jacobian expressions, stencil-boundary discussion, and backward-pass pseudocode are deferred to Appendix~\ref{app:solver_details}.

\textbf{Stencil Boundaries and Non-Smoothness.} The solution operator $T(\theta)$ is piecewise smooth: within regions where the active stencil configuration remains fixed, implicit differentiation yields classical derivatives. At boundaries where the minimizing stencil switches, $T$ remains continuous but non-differentiable. At such points, our method differentiates through the active (winning) stencil, yielding one element of the Clarke subdifferential~\cite{clarke1990optimization}---the convex hull of all limiting gradients. This suffices for convergence under stochastic subgradient descent~\cite{davis2020stochastic}. We note that the measure-zero property alone does not guarantee gradient stability; we therefore validate this empirically in Section~\ref{sec:gradient_validation}, where perturbation tests confirm that a 1\% metric perturbation produces only 0.48\% mean gradient variation, demonstrating that the subdifferential elements are well-behaved in practice. Smooth alternatives such as soft-min relaxations or gradient averaging under parameter jitter could eliminate these boundary discontinuities entirely; see Section~\ref{sec:gradient_validation}.

\section{Empirical Evaluation}\label{sec:validation}

We validate the forward solve, implicit gradients, and the inverse problem. Across configurations, the forward solver attains sub-1\% error, implicit differentiation matches finite differences away from measure-zero stencil boundaries, and the inverse problem recovers isotropic metrics to $\approx$6\% error and drift fields to $<3\%$ error.

\subsection{Forward Solver Accuracy}
\label{sec:forward_validation}
We verify numerical accuracy through a grid refinement study on a unit domain with isotropic metric $\mG = \mI$ and point source at $\vx_0$. Table~\ref{tab:convergence} reports errors against the analytical solution $T(\vx) = \|\vx - \vx_0\|$ for grids ranging from $25 \times 25$ to $400 \times 400$.

\begin{table}[t]
\centering
\small
\caption{Convergence study for the isotropic eikonal equation. Relative errors remain below 1\% across all resolutions.}
\label{tab:convergence}
\small
\begin{tabular}{ccccc}
\toprule
Grid Size & $h$ & $L_2$ Error & $L_\infty$ Error & Relative $L_2$ \\
\midrule
$25 \times 25$ & 0.040 & $4.77 \times 10^{-3}$ & $7.73 \times 10^{-3}$ & 0.84\% \\
$50 \times 50$ & 0.020 & $3.13 \times 10^{-3}$ & $5.18 \times 10^{-3}$ & 0.48\% \\
$100 \times 100$ & 0.010 & $1.96 \times 10^{-3}$ & $3.22 \times 10^{-3}$ & 0.29\% \\
$200 \times 200$ & 0.005 & $1.19 \times 10^{-3}$ & $1.93 \times 10^{-3}$ & 0.17\% \\
$400 \times 400$ & 0.0025 & $6.99 \times 10^{-4}$ & $1.15 \times 10^{-3}$ & 0.10\% \\
\bottomrule
\end{tabular}
\end{table}
A log--log fit gives an empirical rate $\alpha=0.69$, consistent with the known source singularity and first-order upwind behavior on Cartesian grids. Beyond the isotropic case, we obtain sub-1\% errors for diagonal anisotropy (0.18\%), rotated SPD metrics (0.14\%), constant drift (exact asymmetry match for $\|\vb\|=0.3$), and the combined Randers--Finsler setting (0.45\% via Richardson extrapolation); see Figures~\ref{fig:A2_anisotropic}--\ref{fig:A5_combined}.

The fast sweeping algorithm converges in 2--3 iterations independent of grid size, confirming $O(1)$ iteration complexity. Timing results (showing the expected $O(n) = O(N^2)$ total work) are reported in Table~\ref{tab:complexity} in Appendix~\ref{app:forward_solver}.



\subsection{Gradient Verification and Stencil Analysis}
\label{sec:gradient_validation}

Discrete stencil selection in fast sweeping raises questions about gradient validity. We verify correctness by comparing implicit differentiation against central finite differences ($\epsilon = 10^{-5}$) at 20 randomly sampled interior points per configuration. All test points achieve relative errors below $10^{-6}$ across isotropic, anisotropic, and drift configurations (Table~\ref{tab:gradient_accuracy}), confirming numerically exact gradients away from stencil boundaries.

\begin{figure}[t]
\centering
\includegraphics[width=\columnwidth]{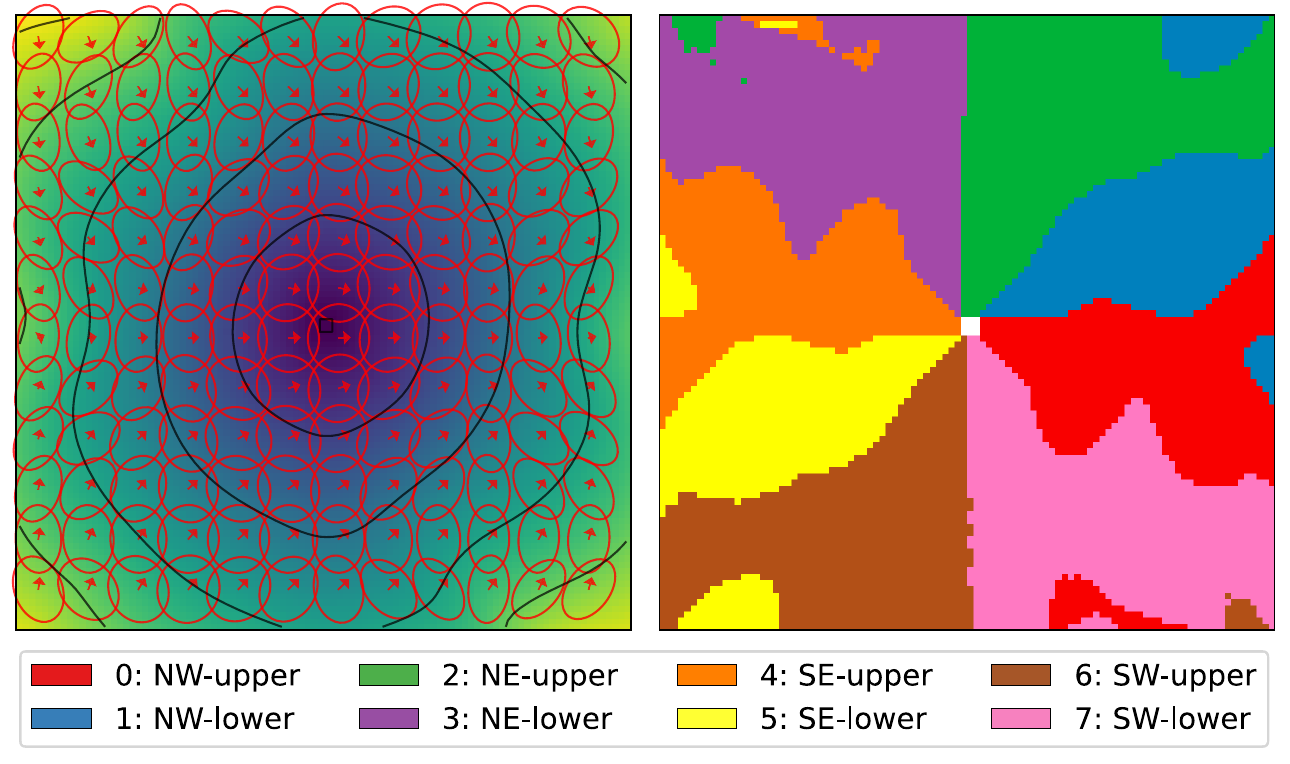}
\caption{Stencil structure showing arrival times (left) and active stencil map (right). Boundaries form measure-zero curves tracing characteristics from $\vx_0$.}
\label{fig:stencil_boundaries}
\end{figure}

\begin{table}[t]
\centering
\caption{Pointwise gradient verification against finite differences.}
\label{tab:gradient_accuracy}
\small
\begin{tabular}{lccc}
\toprule
Configuration & Components & Points Tested & Max Rel.\ Error \\
\midrule
Isotropic & $\mG$, $\vb$ & 20 & $< 10^{-6}$ \\
Anisotropic & $g_{11}$, $g_{12}$, $g_{22}$ & 20 & $< 10^{-6}$ \\
Drift only & $b_1$, $b_2$ & 20 & $< 10^{-6}$ \\
\bottomrule
\end{tabular}%
\end{table}
Stencil boundaries where the optimal stencil changes form measure-zero curves in the domain (Figure~\ref{fig:stencil_boundaries}). Testing 100 random perturbation directions confirms that global directional derivatives show reduced accuracy (only 20\% achieve $<$10\% error) because perturbations cross boundaries somewhere in the domain. However, gradients remain stable: 1\% metric noise produces only 0.48\% mean gradient variation, confirming bounded, localized changes rather than catastrophic discontinuities. Additional gradient verification experiments appear in Appendix~\ref{app:gradient_verification}.

\subsection{Computational Efficiency}
\label{sec:efficiency}
Efficient computation is critical for gradient-based optimization with hundreds of iterations. We benchmark both forward and backward pass implementations to establish practical performance characteristics. Comparisons of algorithmic variants (fast sweeping vs Jacobi iteration, implicit differentiation vs finite differences) appear in Appendix~\ref{app:comparisons}.

\begin{figure}[t]
    \centering
    \includegraphics[width=1\columnwidth]{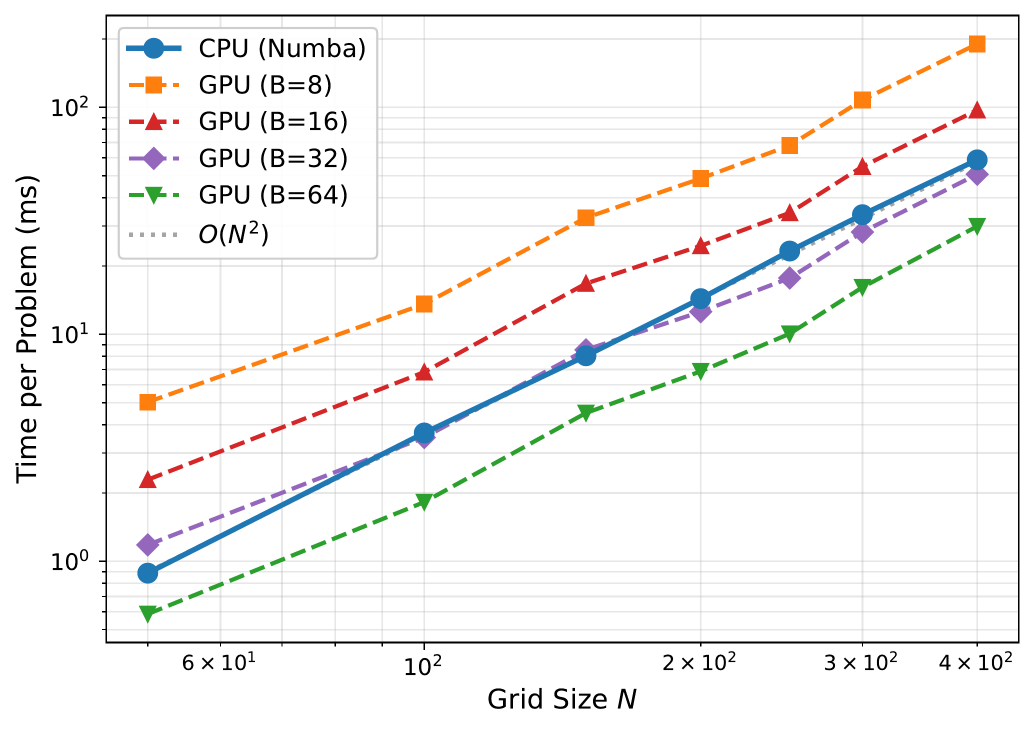}
    \caption{Forward solver runtime per problem instance on an $N \times N$ grid ($n = N^2$ total points). Both CPU and GPU implementations scale as $O(n)$, with GPU efficiency improving dramatically as batch size increases from B=8 to B=64. At $N=250$ with batch size 64, the GPU achieves 2.31$\times$ speedup over CPU (10.1ms vs 23.3ms per problem).}
    \label{fig:forward_benchmark}
\end{figure}

\textbf{Forward Solver Performance.} We benchmark fast sweeping on Randers-Finsler eikonal problems with realistic wildfire configurations: anisotropic metric tensors $\mG$ (rotation angles 0 to $\pi/2$, anisotropy ratios 1.5--2.0), small drift vectors $\vb$ (magnitude $\sim$0.02), and 3$\times$3 ignition regions at grid center. The solver runs for up to 30 sweeping iterations, though convergence typically occurs within 2--3. Figure~\ref{fig:forward_benchmark} compares CPU (Numba-optimized, sequential) and GPU (batched CUDA, parallel across problem instances) implementations across grid sizes.

The CPU implementation achieves single-problem runtimes from 0.89ms (N=50) to 58.7ms (N=400), confirming $O(N^2)$ scaling. The GPU implementation solves batches in parallel, with per-problem cost decreasing substantially as batch size increases. At N=250 with B=64, the GPU reduces per-problem time from 23.3ms to 10.1ms (2.31$\times$ speedup), while at N=400 the speedup is 1.96$\times$. This batched approach proves essential during training, where each gradient computation requires solving forward problems for multiple fire instances with different environmental conditions simultaneously.

\begin{figure*}[t]
    \centering
    \includegraphics[width=1\textwidth]{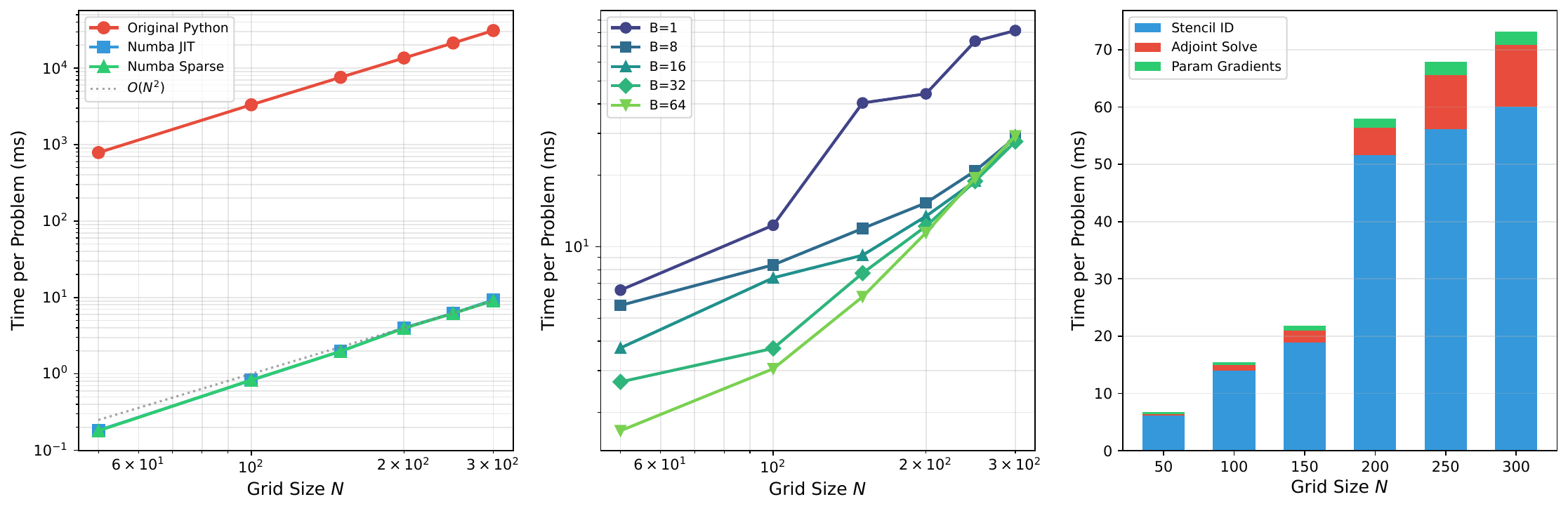}
    \caption{Backward pass performance on an $N \times N$ grid ($n = N^2$ points). Left: Adjoint solve time for three implementations showing Numba achieves 3700$\times$ average speedup over naive Python with maintained $O(n)$ scaling. Center: Batched backward pass demonstrating per-fire cost reduction with increased batch size, achieving 3.8$\times$ speedup at $N=250$ (B=64 vs B=1). Right: Component breakdown showing stencil identification dominates at 83\% of total time, while adjoint solve contributes 14\% and parameter gradients 3\%.}
    \label{fig:backward_benchmark}
\end{figure*}

\textbf{Backward Pass Performance.} Computing gradients through implicit differentiation requires solving the adjoint system and evaluating parameter gradients. We benchmark three implementations: naive Python (nested loops), Numba JIT-compiled, and Numba sparse (processing only burned pixels where $T < 10^5$). The test problem uses an anisotropic metric ($30^{\circ}$ rotation, 1.8 anisotropy ratio), small drift field, 3$\times$3 ignition region, and MSE-style loss gradients at 30\% of valid pixels. Figure~\ref{fig:backward_benchmark} compares implementation strategies and analyzes computational breakdown.

The naive Python implementation serves as baseline, requiring 786ms at N=50 to 31.1 seconds at N=300. Numba JIT compilation provides dramatic acceleration through loop fusion and type specialization, reducing times to 0.18ms at N=50 and 9.2ms at N=300, representing approximately 3700$\times$ average speedup. The sparse variant achieves nearly identical performance by skipping inactive pixels, offering no additional benefit when most of the domain is burned.

The backward pass decomposes into three components at N=250: stencil identification (56.1ms, 83\%), adjoint system solution (9.5ms, 14\%), and parameter gradient evaluation (2.3ms, 3\%). Stencil identification dominates due to neighbor queries and conditional logic determining the active stencil at each node. The adjoint solve, despite processing all interior nodes in reverse order, remains efficient through vectorized operations. Batched backward passes exploit parallelism similarly to the forward solver: at N=250, per-fire cost reduces from 73.5ms (K=1) to 19.5ms (K=64), achieving 3.8$\times$ speedup.

The backward pass requires approximately 3$\times$ the forward pass time (67.9ms backward vs 23.3ms forward at N=250), confirming that gradient computation remains practical. Combined with batched forward and backward solving, this enables training on thousands of fire instances with hundreds of optimization iterations within hours rather than days.

\section{Inverse Metric Problem} \label{sec:inverse_validation}
We validate the framework's ability to recover metric parameters from observed arrival times through synthetic experiments with known ground truth. The inverse problem seeks metric parameters $\theta = (\mG, \vb)$ such that the solution $T(\vx; \theta)$ of the eikonal equation matches observed arrival times $\hat{T}$ at a set of observation locations $\Omega_{\text{obs}}$.

\textbf{Optimization Formulation.} We formulate parameter recovery as the regularized least-squares problem
\begin{equation}
    \min_{\theta} \; \mathcal{L}(\theta) = \frac{1}{2}\sum_{i \in \Omega_{\text{obs}}} \bigl(T_i(\theta) - \hat{T}_i\bigr)^2 + \lambda_{\mG}\,\mathcal{R}(\mG) + \lambda_{\vb}\,\mathcal{R}(\vb),
    \label{eq:inverse_problem}
\end{equation}
where $\mathcal{R}(\cdot)$ denotes total variation regularization to promote piecewise-smooth solutions. We solve~\eqref{eq:inverse_problem} via projected gradient descent: at each iteration we compute the data gradient via implicit differentiation (Section~\ref{sec:backward}), add the TV subgradient, project $\mG$ to the positive-definite cone and $\vb$ to satisfy the Randers feasibility constraint, and update with step sizes $10^{-2}$ for $\mG$ and $5 \times 10^{-3}$ for $\vb$.

\textbf{Isotropic Metric Recovery.} The simplest inverse problem recovers an isotropic metric $\mG = g(\vx)\mI$ from arrival time observations. We test recovery of a piecewise constant metric with $g=1$ in the left half and $g=2$ in the right half of an $80 \times 80$ domain, placing a point source at $\vx_0$. With full observations (100\% of pixels), the optimizer achieves 5.6\% relative error after 300 iterations, accurately capturing both the values in each region and the interface location (Figure~\ref{fig:isotropic_recovery}). Reducing observations to 7\% of pixels increases error to 21.2\%, but the general spatial structure is still recovered.

\textbf{Drift Recovery.} Recovering the drift field $\vb$ with known metric tests sensitivity to directional propagation effects. For constant true drift $\vb = (0.15, 0.08)$, the optimizer achieves errors of 2.8\% for $b_1$ and 2.0\% for $b_2$. This superior performance compared to metric recovery confirms that drift creates a stronger observational signature through the characteristic asymmetry between upwind and downwind propagation.

\textbf{Regularization and Observation Density.} Total variation regularization is critical for well-posed recovery. Without regularization, the optimizer overfits to noise (22.4\% error); optimal regularization ($\lambda = 0.001$) reduces error to 8.1\%, a 2.8$\times$ improvement. Varying observation density reveals a phase transition near 50\% coverage: below this threshold, errors plateau around 21\%, while above it, performance saturates near 6\%. This transition separates underdetermined and well-determined regimes for the inverse problem.

\textbf{Multi-Source Recovery.} Multiple ignition sources create intersecting wavefronts that probe the domain from different directions. Table~\ref{tab:multifire} shows that combining data from multiple sources with different ignition locations substantially improves recovery accuracy. Going from one source (286 observations at 7\% density) to five sources (1430 observations) reduces error from 18.6\% to 10.1\%---a 45.6\% improvement that exceeds what would be expected from simply increasing observation count. The additional benefit arises because different source locations provide complementary geometric constraints on the metric tensor.
\begin{table}[t]
\centering
\caption{Multi-source metric recovery. Joint optimization across sources with different ignition locations improves accuracy beyond the effect of increased observations.}
\label{tab:multifire}
\small
\begin{tabular}{ccc}
\toprule
Number of Sources & Total Observations & Relative Error \\
\midrule
1 & 286 & 18.6\% \\
2 & 572 & 17.5\% \\
3 & 858 & 12.4\% \\
5 & 1430 & 10.1\% \\
\bottomrule
\end{tabular}%
\end{table}
These synthetic experiments establish that the differentiable eikonal framework enables accurate parameter recovery: isotropic metrics to 5.6\% error, drift fields to under 3\% error, with multi-source learning providing substantial gains through complementary geometric constraints.

\section{Wildfire Propagation Simulation}
\label{sec:wildfire}

\begin{figure}[t]
    \centering
    \includegraphics[width=1\columnwidth]{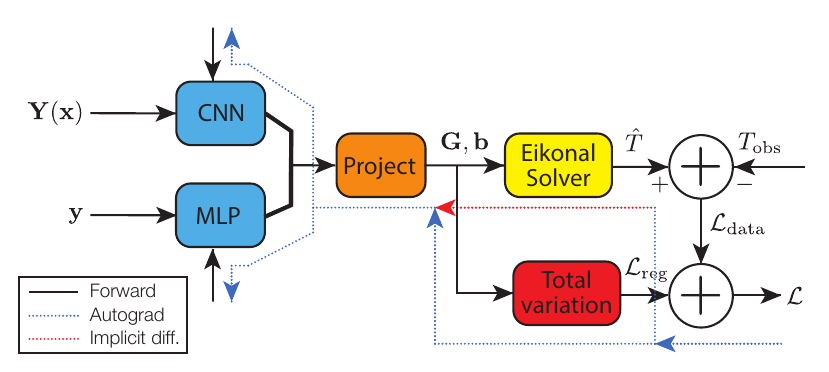}
    \caption{ Learning framework for the differentiable eikonal solver. Spatial covariates $Y(\rvx)$ are processed through a CNN while global features $\rvy$ pass through an MLP, with both pathways combined before projection to produce metric parameters $(\mG, \vb)$ that satisfy feasibility constraints ($\mG \succ 0$ and $\|\rvb\|_\mG < 1$). The eikonal solver computes predicted arrival times $\hat{T}$ without recording gradients. Two gradient pathways update the network: standard automatic differentiation for the regularization loss $\mathcal{L}_{\mathrm{reg}}$ (blue dotted), and implicit differentiation through the adjoint system for the data loss $\mathcal{L}_{\mathrm{data}}$ (red dashed).}
    \label{fig:framework}
\end{figure}

Having validated correctness on synthetic problems, we now evaluate practical utility on real-world wildfire data. We approximate the fire propagation physics with the Randers-Finsler eikonal equation and use a translation-invariant inductive neural model to decode the latent local metric parameters from the directly observable space-dependent covariates such as terrain elevation and vegetation density and global covariates such as wind direction and temperature. The learning setting is illustrated in Figure~\ref{fig:framework}.

%

\textbf{Encoder Architecture.} The neural encoder maps covariates to metric fields through parallel pathways: a global MLP processes domain-wide features $\vy$ to produce baseline parameters $(\mG_0, \vb_0)$, while a fully-convolutional network processes spatial covariates $\mY(\vx)$ to produce spatially-varying residuals combined as $\mG = \mG_0 + \alpha \mG_{\text{local}}$ and $\vb = \vb_0 + \alpha \vb_{\text{local}}$.

The network outputs five scalar fields per location: metric components $(g_{11}, g_{12}, g_{22})$ and drift $(b_1, b_2)$, projected to satisfy feasibility constraints ($\mG \succ 0$ and $\|\vb\|_{\mG^{-1}} < 1$) through differentiable eigenvalue clamping and norm rescaling. Full architecture details appear in Appendix~\ref{app:learning_details}.

\textbf{Loss Function.} The total loss combines data fidelity and spatial regularization,
$
    \mathcal{L} = \mathcal{L}_{\text{data}} + \lambda_G \mathcal{R}(\mG) + \lambda_B \mathcal{R}(\vb),
    \label{eq:loss}
$
where the data term
$\mathcal{L}_{\text{data}} = \frac{1}{2} \| \hat{T} - T_\mathrm{obs} \|_2^2, \label{eq:data_loss}$is evaluated over a subset of the domain on which $T_\mathrm{obs}$ is valid, and the regularization term $\mathcal{R}$ promotes piecewise-smooth parameter fields (Appendix~\ref{app:learning_details}).

\textbf{Gradient Computation.} The two loss components require different differentiation strategies (Figure~\ref{fig:framework}). The regularization terms depend only on the network outputs $\mG$ and $\vb$, so their gradients propagate through standard automatic differentiation. The data term, however, depends on the solution $T$ of the eikonal equation, which is computed by the iterative fast sweeping solver. We obtain gradients for this term using the implicit differentiation scheme described in Section~\ref{sec:backward}: given $\partial \mathcal{L}_{\text{data}} / \partial T$ from the loss, we solve the adjoint system to obtain $\partial \mathcal{L}_{\text{data}} / \partial \mG$ and $\partial \mathcal{L}_{\text{data}} / \partial \mB$, then backpropagate these through the network via a second forward pass. This hybrid approach combines the efficiency of automatic differentiation for the regularization path with the mathematical correctness of implicit differentiation for the solver path.

\textbf{Dataset Description} 
We use the Sim2Real-Fire dataset~\cite{li2024sim2real}, which provides dense temporal ground truth for simulated fires across diverse terrain and weather conditions. The dataset comprises over one million wildfire scenarios generated using FARSITE~\cite{finney1998farsite}, WFDS~\cite{meerpoel2023modeling}, and WRF-SFIRE~\cite{mandel2011coupled}, with five aligned covariate modalities (topography, vegetation, fuel, weather, and satellite-derived fire masks) at 30-meter spatial and hourly temporal resolution. We use 15 scenes (26,752 fires total) for per-scene evaluation and a disjoint 11-train/4-test split for cross-scene transfer; full dataset details appear in Appendix~\ref{app:dataset_details}.

\textbf{Experimental Questions.} Our wildfire experiments address two questions of increasing generalization difficulty. First, can the framework learn metric parameters that generalize to new ignition points within a single landscape, where terrain and fuel properties remain constant but weather and fire timing vary (per-scene evaluation, Section~\ref{sec:per_scene})? Second, can a single model trained on multiple landscapes transfer to completely new geographic regions with different terrain characteristics (cross-scene evaluation, Section~\ref{sec:cross_scene})?

\textbf{Evaluation Metrics.} We report three complementary metrics to characterize prediction quality. The Pearson correlation coefficient between predicted and ground truth arrival times measures how well the model captures the relative ordering and timing relationships across the burned region. Relative RMSE, computed as the root mean square error normalized by the maximum ground truth arrival time, provides a scale-invariant measure of absolute timing accuracy that is comparable across fires of different durations. The Intersection over Union at 50\% of maximum time (IoU@50) evaluates spatial accuracy by measuring the overlap between predicted and actual burned areas at the midpoint of the fire's progression.

\subsection{Per-Scene Evaluation: Within-Landscape Generalization}
\label{sec:per_scene}

\textbf{Experimental Protocol.} We conduct per-scene experiments in which a separate model is trained and evaluated for each of the 15 scenes in the dataset. Within each scene, we partition the available fire events using a 70/15/15 split for training, validation, and test sets respectively. This protocol evaluates the model's ability to generalize to unseen fires occurring in the same landscape, which represents a practically relevant scenario where historical fire data from a region inform predictions for future events in that region.

For each scene, we train our covariate encoder network to predict the Randers metric tensor field $\mG(x)$ and drift vector field $\vb(x)$ from the available topography, vegetation, fuel, and weather inputs. The forward eikonal solver computes predicted arrival times via fast sweeping, and gradients are backpropagated through implicit differentiation to update the network parameters. We employ early stopping based on validation correlation with a patience of 20 epochs and select the model checkpoint achieving the highest validation performance for final test evaluation.

\begin{figure*}[t]
\centering
\includegraphics[width=0.9\textwidth]{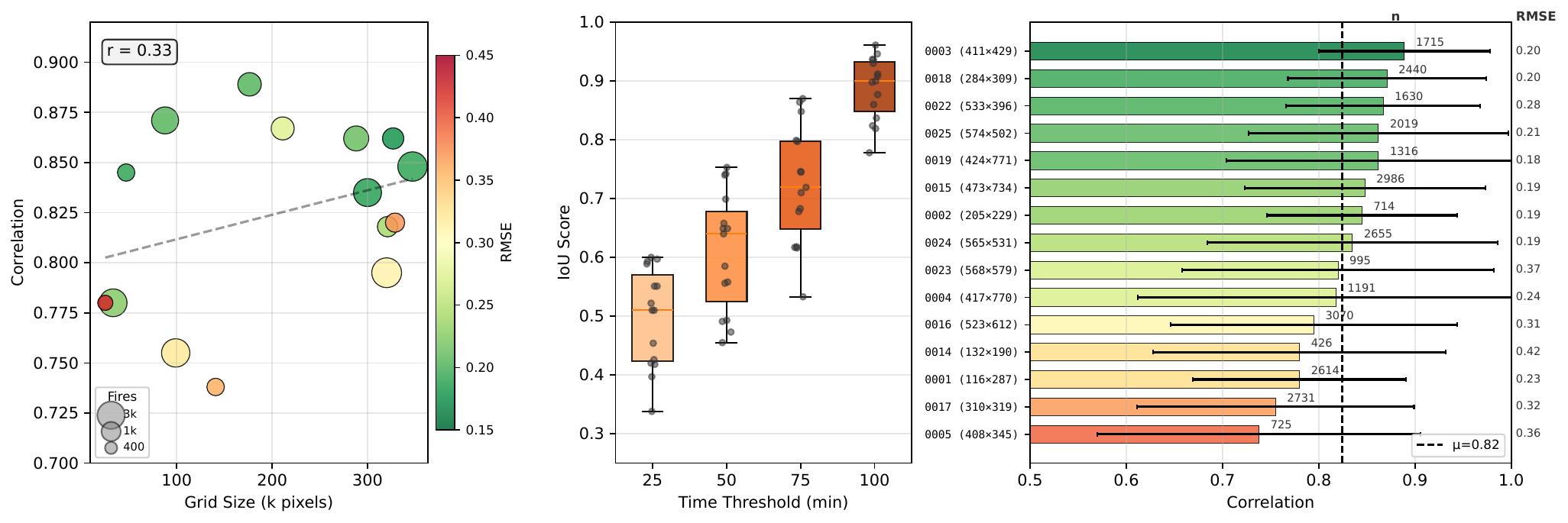}
\caption{Within-scene generalization summary on Sim2Real-Fire across 15 scenes. (a) Correlation versus grid size (bubble size indicates number of fires; color indicates relative RMSE). (b) IoU at increasing arrival-time thresholds, showing improved overlap as temporal tolerance increases. (c) Per-scene correlation, with the overall mean shown as a dashed vertical line (annotations indicate per-scene fire count and relative RMSE).}
\label{fig:per_scene_summary}
\end{figure*}

\textbf{Results.}
Figure~\ref{fig:per_scene_summary} summarizes within-scene test performance across the 15 scenes of Sim2Real-Fire. Detailed per-scene metrics are reported in Appendix~\ref{app:sim2real_additional} (Table~\ref{tab:per_scene_results}).

Figure~\ref{fig:per_scene_summary} shows consistently strong within-scene generalization across the benchmark, with a mean test correlation of $0.824 \pm 0.044$ aggregated over scenes. Panel (a) suggests only a weak relationship between grid size and correlation, indicating that spatial resolution alone does not explain performance; rather, scene-specific dynamics and covariate informativeness dominate. Panel (b) shows a monotonic increase in IoU as the arrival-time tolerance is relaxed, consistent with early-front prediction being more sensitive to local heterogeneity and ignition geometry than later-stage overlap. Panel (c) highlights the per-scene ranking and variability, while Appendix~\ref{app:sim2real_additional} reports the full per-scene metrics (Table~\ref{tab:per_scene_results}).

\begin{figure*}[t]
\centering
\includegraphics[width=0.8\textwidth]{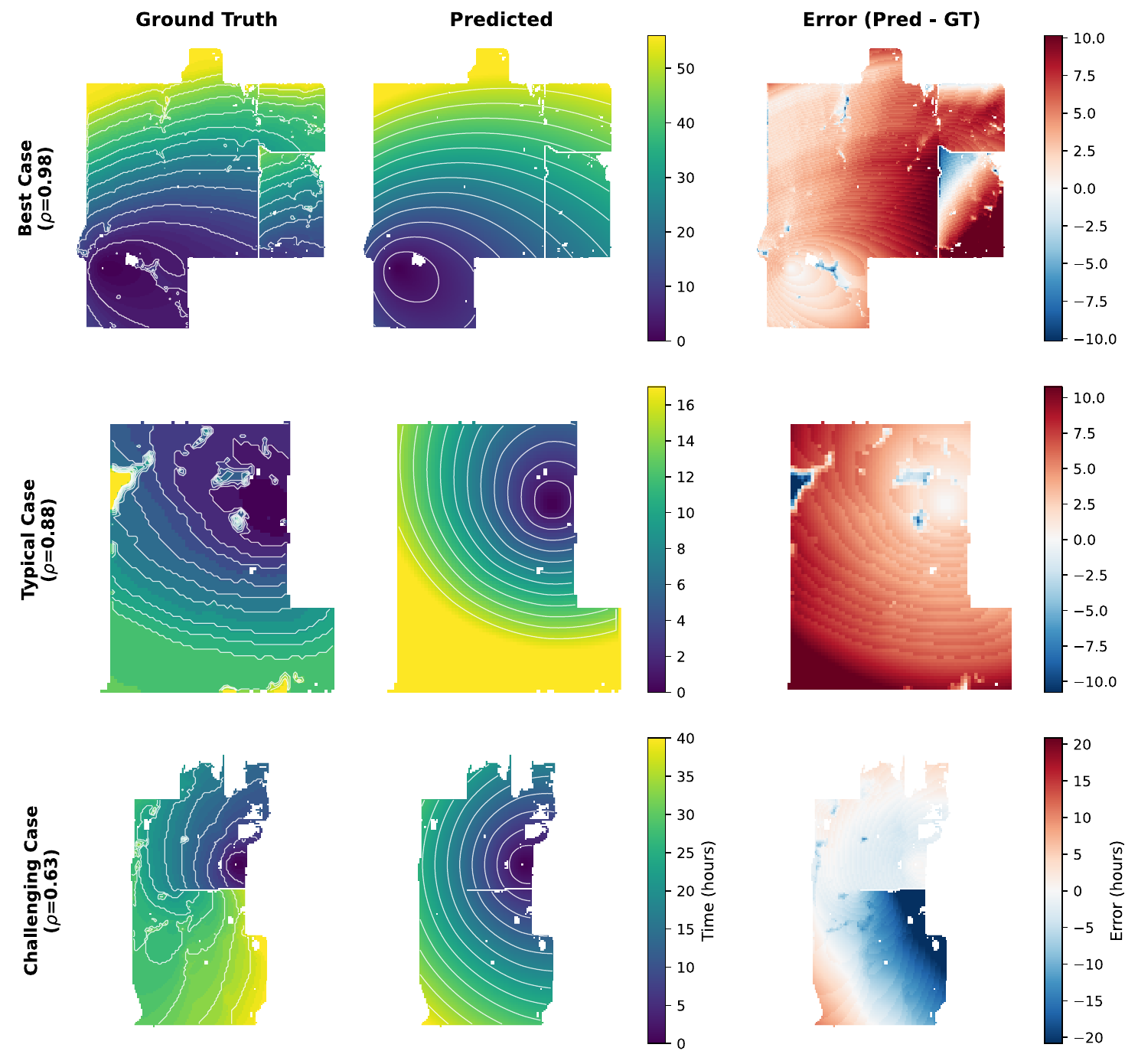}
\caption{Qualitative results on scene 0024\_02655 showing ground truth arrival times (left), predicted arrival times (center), and prediction error (right) for three fires representing best ($\rho=0.98$), typical ($\rho=0.88$), and challenging ($\rho=0.63$) cases. Color scales indicate arrival time in hours from ignition; error maps use a diverging scale where red indicates overprediction (fire predicted to arrive later than observed) and blue indicates underprediction.}
\label{fig:qualitative_within}
\end{figure*}

Figure~\ref{fig:qualitative_within} complements the aggregate trends in Figure~\ref{fig:per_scene_summary} by providing qualitative insight into model behavior across the performance spectrum. In the best case ($\rho=0.98$), the model accurately captures fire spreading radially from the ignition point, with errors confined to a small region where terrain features cause deviation from smooth propagation. The typical case ($\rho=0.88$) reveals a characteristic pattern: the model predicts smooth, roughly elliptical wavefronts while the ground truth exhibits irregular edges shaped by local fuel and terrain heterogeneity. The resulting error shows systematic overprediction in regions where fire spread faster than the learned metric field anticipates. The challenging case ($\rho=0.63$) illustrates the model's limitations: fires with complex, elongated shapes driven by strong directional effects are approximated by smoother patterns, leading to substantial timing errors. These examples suggest that while the model successfully learns average propagation characteristics from covariates, capturing fine-grained spatial variation remains an open challenge.

\subsection{Cross-Scene Evaluation: Geographic Transfer}
\label{sec:cross_scene}

Practical wildfire modeling requires generalization to entirely new 
landscapes, not just new ignition points within known terrain. We evaluate 
this by training on fires from 11 scenes and testing on 4 held-out scenes. 
Cross-scene generalization requires learning universal relationships between covariates and propagation physics: the model must learn that a $20^\circ$ slope has similar effects on fire spread whether it appears in scene 0001 or scene 0024.

\textbf{Training Protocol.} To handle varying scene dimensions ($132 \times 190$ to $574 \times 770$ pixels), we extract $256 \times 256$ patches with stride 128, yielding 18,676 training patches across 11 scenes. Patches containing disconnected burned regions are handled by placing source masks at each component's minimum arrival time pixel. Evaluation uses full scene resolution to obtain accurate metrics. We select four test scenes spanning different difficulty levels and grid sizes; full training details and test scene selection criteria appear in Appendix~\ref{app:cross_scene_training}.

\begin{figure*}[t]
\centering
\includegraphics[width=0.8\textwidth]{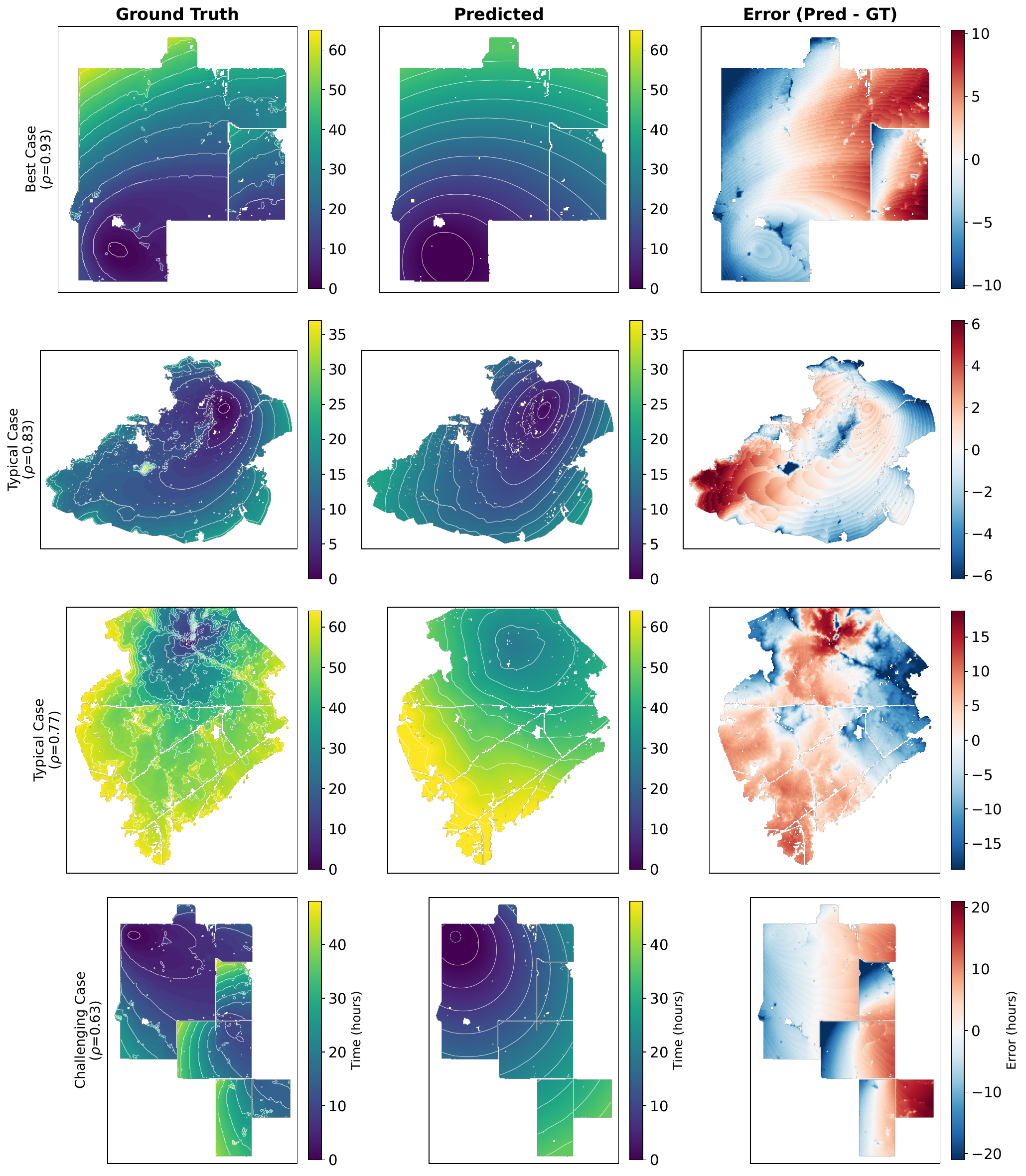}
\caption{Cross-scene wildfire propagation results on held-out geographic regions not seen during training. Each row shows a different fire from the Sim2Real-Fire dataset: ground truth arrival times derived from simulated fire spread (left), predicted arrival times from our learned Randers-Finsler metric (center), and pointwise prediction error (right). Contour lines indicate fire front positions at equal time intervals. The best case ($\rho=0.93$, scene 0024) accurately captures the radial spread pattern and directional bias imposed by terrain and wind. The typical cases ($\rho=0.83$, scene 0023; $\rho=0.77$, scene 0005) reproduce the dominant propagation direction and overall fire shape, though the predicted fronts are smoother than the irregular ground truth perimeters shaped by local fuel heterogeneity. The challenging case ($\rho=0.63$, scene 0024) shows a large fire where the model underpredicts spread in the eastern sector, likely due to unresolved local wind channeling effects. Predictions are affine-rescaled to the ground truth range for visual comparison; correlation is computed on unscaled values. Error maps use a diverging scale (red: overprediction, blue: underprediction).}
\label{fig:qualitative_cross_scene}
\end{figure*}

\begin{table}[!t]
\centering
\caption{Cross-scene generalization results. The model is trained on 11 scenes and tested on 4 held-out scenes.}
\label{tab:cross_scene_results}
\resizebox{\columnwidth}{!}{%
\begin{tabular}{lccccc}
\toprule
\textbf{Scene} & \textbf{Grid Size} & \textbf{Fires} & \textbf{Correlation} & \textbf{Rel.\ RMSE} & \textbf{IoU@50} \\
\midrule
0005\_00725 & $408 \times 345$ & 725 & $0.675 \pm 0.174$ & $0.388 \pm 0.149$ & 0.198 \\
0014\_00426 & $132 \times 190$ & 426 & $0.725 \pm 0.123$ & $0.360 \pm 0.155$ & 0.289 \\
0023\_00995 & $568 \times 579$ & 995 & $0.804 \pm 0.172$ & $1.078 \pm 0.934$ & 0.251 \\
0024\_02655 & $565 \times 531$ & 2655 & $0.783 \pm 0.189$ & $0.205 \pm 0.146$ & 0.369 \\
\midrule
\textbf{Overall} & -- & 4801 & $0.766 \pm 0.184$ & $0.427 \pm 0.560$ & 0.295 \\
\bottomrule
\end{tabular}}
\end{table}
Figure~\ref{fig:qualitative_cross_scene} provides qualitative examples from each of the four held-out test scenes, showing ground truth arrival times, rescaled predictions, and pointwise error maps. The model captures overall fire shape and radial spread patterns across diverse terrains, though predictions tend toward smoother wavefronts than the ground truth, consistent with the eikonal low-pass effect discussed in Section~\ref{sec:conclusion}.

The per-scene breakdown reveals heterogeneous generalization. Scene 0023\_00995 achieves the highest correlation (0.804), suggesting its terrain characteristics are well-represented in the training distribution, though it also exhibits the highest relative RMSE (1.078), indicating less accurate absolute timing predictions. Scene 0005\_00725 proves most challenging (correlation 0.675), consistent with its low within-scene performance. Scene 0024\_02655 demonstrates strong generalization (correlation 0.783, IoU@50 0.369), approaching its within-scene performance despite the model never observing this terrain during training.

The mean correlation drops from 0.824 (within-scene) to 0.766 (cross-scene), a relative decrease of 7\%. IoU@50 decreases more substantially from 0.609 to 0.295. This sharper drop reflects IoU's sensitivity to absolute timing errors: correlation is invariant to affine rescaling of predicted arrival times, so a model that captures the correct relative ordering but misjudges overall fire speed will maintain high correlation while producing spatially misaligned burned-area predictions at fixed time thresholds. Training converged over 28 epochs with learning rate reduction at epoch 19, and validation correlation steadily improved from 0.774 to 0.798, demonstrating that patch-based training successfully learns generalizable features.

\subsection{Model Capacity Analysis} \label{sec:model_capacity}

To separate modeling limitations from generalization challenges, we train with zero regularization on single fires and evaluate on the same fire, measuring the minimum achievable error. We compare per-pixel optimization (125K--1.7M parameters depending on grid size) against our neural network encoder (fixed 116K parameters). Table~\ref{tab:capacity} summarizes results across fires from all 19 scenes.
\begin{table}[t]
\centering
\caption{Model capacity diagnostic. Both strategies train on individual fires with zero regularization and evaluate on the training fire itself.}
\label{tab:capacity}
\resizebox{\columnwidth}{!}{%
\begin{tabular}{lccccc}
\toprule
\textbf{Strategy} & \textbf{Parameters} & \textbf{Mean $\pm$ Std (\%)} & \textbf{Median (\%)} & \textbf{95\% CI (\%)} & \textbf{Corr.} \\
\midrule
Per-pixel & 125K--1.7M & 15.4 $\pm$ 16.7 & 7.4 & [11.3, 20.2] & 0.87 \\
Neural network & 116K (fixed) & \textbf{9.4 $\pm$ 7.4} & \textbf{6.3} & \textbf{[7.5, 11.5]} & \textbf{0.89} \\
\bottomrule
\end{tabular}}%
\end{table}

The neural network achieves lower mean error despite having fewer parameters, with nearly disjoint 95\% confidence intervals. This counterintuitive result arises because per-pixel optimization is severely underdetermined on small fires (1,000 observations constraining $>$1M parameters), causing catastrophic failures that skew the mean. On large fires where per-pixel optimization is well-constrained, it achieves the lowest errors (1.7--2.3\% RMSE), demonstrating that the Randers-Finsler physics model itself imposes no fundamental accuracy limit. The encoder's convolutional architecture provides implicit regularization that enables consistent convergence across diverse fire geometries. With median single-fire error around 6--7\%, the primary challenge lies in learning generalizable covariate-to-metric mappings rather than model expressiveness; per-fire breakdowns appear in Appendix~\ref{app:capacity_analysis}. Additional experiments on temporal band decomposition and real wildfire evaluation appear in Appendices~\ref{app:band_training} and~\ref{app:real_fire_evaluation}.

\section{Conclusion}\label{sec:conclusion} We have presented a differentiable framework for the Randers-Finsler eikonal equation that enables gradient-based learning of spatially-varying propagation parameters from arrival time observations. The framework combines a fast sweeping forward solver with an implicit differentiation backward pass that computes parameter gradients by solving a sparse triangular adjoint system. Discrete stencil selection creates non-differentiability only on measure-zero curves, with gradients remaining stable under perturbation.

Validation on synthetic problems establishes correctness and efficiency across the pipeline. The forward solver achieves sub-1\% error for isotropic, anisotropic, and Randers-Finsler configurations. Gradient verification confirms pointwise-exact agreement with finite differences at interior points. For inverse problems, the framework recovers isotropic metrics to 5.6\% error and drift fields to under 3\% error, with multi-source configurations reducing errors by 45\% through complementary geometric constraints. Application to wildfire propagation on the Sim2Real-Fire dataset demonstrates practical utility, achieving mean correlations of 0.824 for within-scene generalization and 0.766 for cross-scene transfer to unseen geographic regions. Model capacity analysis reveals that convolutional encoders outperform per-pixel optimization despite having fewer parameters, establishing that the primary challenge lies in learning generalizable covariate-to-metric mappings rather than model expressiveness.

\textbf{Limitations and Future Directions.} The current implementation operates on Cartesian grids, limiting direct application to curved surfaces or unstructured meshes. Extension to triangulated surfaces would replace the eight-stencil system with triangle-fan neighborhoods, with the adjoint structure remaining similar.

\textbf{Non-differentiability at stencil boundaries.}
The discrete upwind stencil introduces piecewise-smooth behavior: at each grid node, one of eight triangular stencils is selected based on causality, and the active stencil can change discontinuously under infinitesimal parameter perturbation.  As shown in Section~\ref{sec:gradient_validation}, these stencil boundaries form measure-zero curves in parameter space, so the set of problematic points has zero Lebesgue measure.  In practice, the implicit-differentiation backward pass differentiates through the active stencil, yielding an element of the Clarke subdifferential rather than a classical gradient.  This is sufficient for first-order stochastic optimization (Adam, SGD), which we use throughout; however, it may pose challenges for methods that rely on precise Hessian information or second-order curvature, such as L-BFGS or natural gradient methods.  Empirically, we observe stable convergence in all inverse-problem and wildfire experiments, with gradient perturbation tests (Section~\ref{sec:gradient_validation}) confirming that a 1\% metric perturbation produces only 0.48\% mean gradient variation at interior points.

\textbf{Loss of high-frequency spatial variation.}
The eikonal equation inherently acts as a low-pass filter on the metric field: arrival times integrate the metric along characteristics, so localized high-frequency variations in propagation speed are averaged out over wavefront paths.  This effect is compounded by the neural encoder architecture, which maps covariates to metric fields through convolutional layers that further smooth the output.  As a result, the learned wavefronts approximate irregular fire perimeters well in terms of overall shape and timing (mean correlation 0.824) but miss fine-grained spatial features such as spotting, local terrain channeling, or abrupt fuel transitions (Figure~\ref{fig:qualitative_within}).  This is not primarily a limitation of the gradient computation or the optimizer, but rather a structural property of the forward model: the eikonal equation produces smooth wavefronts by construction, and recovering sharp spatial discontinuities would require either sub-grid parameterizations, hybrid physics-learned architectures that combine eikonal propagation with local correction models, or alternative PDE formulations (e.g., level-set methods) that can represent topological changes in the fire front.

Additional future work includes extension to three dimensions, uncertainty quantification through Bayesian methods, and formal convergence analysis connecting stencil boundaries to Clarke subdifferential theory.

\textbf{Code and Data Availability.} Our implementation is publicly available at \url{https://github.com/BarakGahtan/differentiable-eikonal-wildfire/releases/tag/V1}, with an archived release on Zenodo (DOI: \href{https://doi.org/10.5281/zenodo.18792225}{10.5281/zenodo.18792225}).

\textbf{Acknowledgments.} This project has received funding from the European Research Council (ERC) under the European Union's Horizon 2020 research and innovation programme (grant agreement No 863839)


\printbibliography     

\newpage
\appendix

\newpage
\onecolumn
\appendix

\section{\textbf{Solver Details: Fast Sweeping and Implicit Differentiation}}
\label{app:solver_details}
We solve the Randers--Finsler eikonal equation on a Cartesian grid with a fast sweeping method. The scheme exploits \emph{causality}: each node’s arrival time depends only on neighboring nodes with smaller values (upwind donors), so information propagates outward from sources. Alternating sweep directions ensures that updates are communicated correctly across the grid regardless of wavefront geometry.

\subsection{Stencil Geometry and Local Updates}
\begin{figure}[ht]
    \centering
    \begin{tikzpicture}[scale=0.85]
        \begin{scope}[shift={(0,0)}]
            \draw[gray!20, step=1] (-1.5,-1.5) grid (1.5,1.5);
            \fill[black] (0,0) circle (4pt);
            \node[below right, xshift=2pt, font=\footnotesize] at (0,0) {$T_0$};
            \fill[blue!70] (-1,1) circle (3pt);
            \node[above left, font=\scriptsize] at (-1,1) {0};
            \fill[blue!70] (-1,0) circle (3pt);
            \node[left, font=\scriptsize] at (-1,0) {1};
            \fill[blue!70] (-1,-1) circle (3pt);
            \node[below left, font=\scriptsize] at (-1,-1) {2};
            \fill[blue!70] (0,-1) circle (3pt);
            \node[below, font=\scriptsize] at (0,-1) {3};
            \fill[blue!70] (1,-1) circle (3pt);
            \node[below right, font=\scriptsize] at (1,-1) {4};
            \fill[blue!70] (1,0) circle (3pt);
            \node[right, font=\scriptsize] at (1,0) {5};
            \fill[blue!70] (1,1) circle (3pt);
            \node[above right, font=\scriptsize] at (1,1) {6};
            \fill[blue!70] (0,1) circle (3pt);
            \node[above, font=\scriptsize] at (0,1) {7};
            \draw[gray!50, thick] (0,0) -- (-1,1);
            \draw[gray!50, thick] (0,0) -- (-1,0);
            \draw[gray!50, thick] (0,0) -- (-1,-1);
            \draw[gray!50, thick] (0,0) -- (0,-1);
            \draw[gray!50, thick] (0,0) -- (1,-1);
            \draw[gray!50, thick] (0,0) -- (1,0);
            \draw[gray!50, thick] (0,0) -- (1,1);
            \draw[gray!50, thick] (0,0) -- (0,1);
            \draw[gray!30] (-1,1) -- (-1,0) -- (-1,-1) -- (0,-1) -- (1,-1) -- (1,0) -- (1,1) -- (0,1) -- cycle;
            \node[below, font=\small] at (0,-1.8) {(a)};
        \end{scope}
        
        \begin{scope}[shift={(4.5,0)}]
            \fill[blue!15] (0,0) -- (-1.2,1.2) -- (-1.2,0) -- cycle;
            \fill[gray!30] (-1.2,1.2) circle (2.5pt);
            \fill[gray!30] (-1.2,0) circle (2.5pt);
            \fill[gray!30] (0,1.2) circle (2.5pt);
            \fill[gray!30] (1.2,0) circle (2.5pt);
            \fill[gray!30] (0,-1.2) circle (2.5pt);
            \fill[black] (0,0) circle (4pt);
            \node[right, xshift=2pt, font=\footnotesize] at (0,0) {$T_0$};
            \fill[blue!70] (-1.2,1.2) circle (3pt);
            \node[above left, font=\footnotesize] at (-1.2,1.2) {$T_1$};
            \fill[blue!70] (-1.2,0) circle (3pt);
            \node[left, font=\footnotesize] at (-1.2,0) {$T_2$};
            \draw[->, very thick, red!70!black] (0,0) -- (-1.1,1.1);
            \node[red!70!black, above, font=\footnotesize] at (-0.4,0.7) {$\vm_1$};
            \draw[->, very thick, red!70!black] (0,0) -- (-1.1,0);
            \node[red!70!black, below, font=\footnotesize] at (-0.55,0) {$\vm_2$};
            \draw[->, thick, green!50!black, dashed] (0,0) -- (-0.8,0.5);
            \node[green!50!black, below right, font=\footnotesize] at (-0.7,0.45) {$\nabla T$};
            \draw[blue!50, thick] (-1.2,1.2) -- (-1.2,0);
            \node[below, font=\small] at (0,-1.8) {(b)};
        \end{scope}
    \end{tikzpicture}
    \caption{Stencil geometry for fast sweeping. (a)~Eight neighbors numbered counter-clockwise from upper-left. Stencil $k$ uses neighbors $k$ and $(k+1) \bmod 8$. (b)~Triangular stencil showing displacement vectors $\vm_1$, $\vm_2$ from central node $T_0$ to neighbors $T_1$, $T_2$. The two-point update is valid when wavefront direction $\nabla T$ lies within the shaded triangle.}
    \label{fig:stencil}
\end{figure}
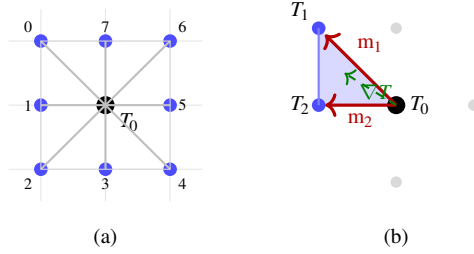

At each interior grid node, we consider eight triangular stencils formed by pairs of adjacent neighbors (Figure~\ref{fig:stencil}a). Each stencil $k \in \{0,\ldots,7\}$ uses two consecutive neighbors indexed as $i=1,2$ with displacement vectors $\vm_i \in \mathbb{Z}^2$ and arrival times $T_i$ (Figure~\ref{fig:stencil}b). We attempt a two-point update using both neighbors; if the geometric validity test fails, we fall back to one-point updates along individual edges.

\textbf{Two-Point Update.} The two-point update solves a quadratic system derived from the discretized eikonal equation. We construct the stencil matrix $\mM = [\vm_1 \mid \vm_2]$ and compute the stencil metric $\mE = \mM^\top \mG \mM$. Define drift-adjusted times $S_i = T_i + \vm_i \cdot \vb$ for $i=1,2$, and let $\vs = [S_1, S_2]^\top$. The candidate arrival time $T_0$ satisfies
\begin{equation}
    \vu^\top \mQ \vu = 1, \qquad \vu = \vs - T_0 \vone, \qquad \mQ = \mE^{-1},
\end{equation}
where $\vone = [1, 1]^\top$. Expanding this quadratic yields an explicit formula for $T_0$ in terms of neighbor values and local metric parameters. We select the larger root to ensure causality, requiring $T_0 > \max(T_1, T_2)$. The update is valid only if both donors satisfy $T_i < T_0$ (causality) and the barycentric coordinates $\boldsymbol{\lambda} = \mQ(T_0 \vone - \vs)$ satisfy $\lambda_1, \lambda_2 \geq 0$ (geometric validity).

\textbf{One-Point Fallback.} When the two-point update fails the geometric validity test, we fall back to one-point updates along individual edges. For a neighbor at displacement $\vm_i$ with arrival time $T_i$, the update formula is Equation~\ref{eq:one_point} representing direct propagation along the edge direction. At each node, we evaluate all applicable stencils and take the minimum over valid candidates.

The solver initializes $T = 0$ at source nodes and $T = \infty$ elsewhere, then performs alternating sweeps through the grid. Each complete iteration consists of four directional sweeps: processing columns left-to-right, then rows top-to-bottom, then columns right-to-left, and finally rows bottom-to-top. Within each sweep, we update nodes sequentially, considering all eight triangular stencils and selecting the minimum valid update. The algorithm converges when $\max_i |T_i^{(k+1)} - T_i^{(k)}| < 10^{-6}$, typically within 2--3 iterations independent of grid size, yielding $O(n)$ complexity per iteration for $n = N \times N$ grid points. We now provide pseudocode for the full sweep schedule and update loop used in the forward solve.

\subsection{Forward Solve Pseudocode}
\begin{algorithm}[!ht]
\small
\caption{Fast Sweeping Forward Solve}
\label{alg:forward}
\begin{algorithmic}[1]
\Require Metric field $\mG$, drift field $\vb$, source mask $\mathcal{S}$
\Ensure Arrival time field $T$
\State $T_i \gets 0$ for $i \in \mathcal{S}$; $T_i \gets \infty$ otherwise
\Repeat
    \For{each sweep direction $d \in \{\rightarrow, \downarrow, \leftarrow, \uparrow\}$}
        \For{each grid point $(i,j)$ in order determined by $d$}
            \For{each triangular stencil $k \in \{0,\ldots,7\}$}
                \State $(\vm_1, \vm_2) \gets$ neighbor displacements for stencil $k$
                \If{$T_1, T_2 < T_{i,j}$ (causality check)}
                    \State $t^{(2)} \gets \textsc{TwoPointUpdate}(T, \mG, \vb, \vm_1, \vm_2)$
                    \If{geometric validity test fails}
                        \State $t^{(1)}_1 \gets \textsc{OnePointUpdate}(T, \mG, \vb, \vm_1)$
                        \State $t^{(1)}_2 \gets \textsc{OnePointUpdate}(T, \mG, \vb, \vm_2)$
                        \State $t^{(2)} \gets \min(t^{(1)}_1, t^{(1)}_2)$
                    \EndIf
                    \State $T_{i,j} \gets \min(T_{i,j}, t^{(2)})$
                \EndIf
            \EndFor
        \EndFor
    \EndFor
\Until{$\max_{i,j} |T^{\text{new}}_{i,j} - T^{\text{old}}_{i,j}| < \epsilon$}
\State \Return $T$
\end{algorithmic}
\end{algorithm}

\subsection{Jacobian Details and Backward Pass}

\textbf{The Jacobian} $\mJ = \partial R / \partial T$ has sparse lower-triangular structure when nodes are ordered by increasing arrival time. For two-point updates with donors at displacements $\vm_1, \vm_2$, the Jacobian entries are
\begin{equation}
    \frac{\partial R}{\partial T_0} = -2 \cdot \vone^\top \mQ \vu, 
    \qquad
    \frac{\partial R}{\partial T_k} = 2 \cdot \ve_k^\top \mQ \vu,\quad k\in\{1,2\},
    \label{eq:jacobian_planar}
\end{equation}
where $\ve_k$ is the $k$-th standard basis vector in $\mathbb{R}^2$.

\textbf{Stencil Boundaries and Non-Smoothness.} The solution operator $T(\theta)$ is piecewise smooth: within regions where the active stencil configuration remains fixed, implicit differentiation yields classical derivatives. At boundaries where the minimizing stencil switches, $T$ remains continuous but non-differentiable. At such points, our method differentiates through the active (winning) stencil, yielding one element of the Clarke subdifferential~\cite{clarke1990optimization}---the convex hull of all limiting gradients. This suffices for convergence under stochastic subgradient descent~\cite{davis2020stochastic}. Smooth alternatives such as soft-min relaxations or gradient averaging under parameter jitter could eliminate these boundary discontinuities entirely, but we found them unnecessary given the measure-zero nature of stencil boundaries and the empirical stability observed in Section~\ref{sec:gradient_validation}.

\begin{algorithm}[ht]
\caption{Implicit Differentiation Backward Pass}
\label{alg:backward}
\begin{algorithmic}[1]
\Require Converged $T$, loss gradient $\vg = \partial \mathcal{L}/\partial T$, parameters $\mG, \vb$
\Ensure Parameter gradients $\partial \mathcal{L}/\partial \mG$, $\partial \mathcal{L}/\partial \vb$
\For{each interior node $i$}
    \State Identify active stencil $k^*_i$ and update type (planar/edge)
    \State Compute Jacobian entries $\mJ_{ii}, \mJ_{ij}$ for donors $j$ via Eq.~\ref{eq:jacobian_planar}
\EndFor
\State Sort nodes by decreasing $T$: permutation $\pi$ with $T_{\pi(1)} \geq T_{\pi(2)} \geq \cdots$
\For{$k = 1$ to $N$}
    \State $i \gets \pi(k)$
    \State $\lambda_i \gets \frac{1}{\mJ_{ii}} \left( \vg_i - \sum_{j: i \in \text{donors}(j)} \mJ_{ji} \lambda_j \right)$
\EndFor
\For{each interior node $i$}
    \State Compute $\partial R_i/\partial \mG$ and $\partial R_i/\partial \vb$ for the active update at $i$ (two-point: Eq.~\ref{eq:two_point}; one-point: Eq.~\ref{eq:one_point})
\EndFor
\State \Return $\partial \mathcal{L}/\partial \mG$, $\partial \mathcal{L}/\partial \vb$
\end{algorithmic}
\end{algorithm}

\noindent\textbf{Explicit residuals and closed-form partial derivatives.}
Each interior node $i$ selects an \emph{active} local update (either a two-point stencil or a one-point edge). We define the corresponding scalar residual $R_i(T,\theta)=0$ and differentiate it w.r.t.\ the local parameters $\theta=(\mG,\vb)$ at node $i$.

\textbf{Two-point active stencil.}
Assume the active update at the current node $T_0$ uses two donors with displacements $\vm_1,\vm_2$ and arrival times $T_1,T_2$. Define the stencil matrix $\mM=[\vm_1\mid \vm_2]$, the stencil metric $\mE=\mM^\top \mG \mM$, and $\mQ=\mE^{-1}$. As in Eq.~\ref{eq:two_point}, define drift-adjusted donor times
\begin{equation}
S_k = T_k + \vm_k\cdot\vb,\quad k\in\{1,2\},
\qquad
\vs = [S_1,S_2]^\top,
\qquad
\vu = \vs - T_0\vone.
\end{equation}
The corresponding two-point residual is
\begin{equation}
R^{(2)}(T,\theta) = \vu^\top \mQ \vu - 1 = 0.
\end{equation}
Differentiating this residual yields the closed-form partial derivatives
\begin{equation}
\frac{\partial R^{(2)}}{\partial \vb}
= 2\,\mM\,\mQ\,\vu,
\qquad
\frac{\partial R^{(2)}}{\partial \mG}
= -\,\mM\,\mQ\,\vu\,\vu^\top\,\mQ\,\mM^\top,
\label{eq:two_point_partials}
\end{equation}
where $\partial R^{(2)}/\partial \mG$ is a $2\times 2$ symmetric matrix (and can be accumulated using the symmetric parameterization used in the implementation).

\textbf{One-point active edge.}
Let $\vm_i$ denote the (active) donor displacement and $T_i$ the corresponding donor arrival time. Define
\begin{equation}
r = T_i + \vm_i\cdot\vb - T_0,
\qquad
e = \vm_i^\top \mG \vm_i,
\end{equation}
so that the one-point update Eq.~\ref{eq:one_point} enforces $r=\sqrt{e}$. The corresponding residual is
\begin{equation}
R^{(1)}(T,\theta)= r^2 - e = 0.
\end{equation}
The closed-form partial derivatives are
\begin{equation}
\frac{\partial R^{(1)}}{\partial \vb}
= 2r\,\vm_i,
\qquad
\frac{\partial R^{(1)}}{\partial \mG}
= -\,\vm_i\,\vm_i^\top.
\label{eq:one_point_partials}
\end{equation}

In Algorithm~\ref{alg:backward}, the line ``Compute $\partial R_i/\partial \mG$ and $\partial R_i/\partial \vb$'' instantiates Eq.~\ref{eq:two_point_partials} when the active update at node $i$ is a valid two-point stencil, and Eq.~\ref{eq:one_point_partials} when the active update is a one-point fallback. The per-node gradients are then accumulated via the adjoint contraction $-\lambda_i\,\partial R_i/\partial \theta$ (Eq.~\ref{eq:param_grad}).

\section{Learning Framework Details: Projections and Regularization}
\label{app:learning_details}

\textbf{Metric Representation.} The network outputs five scalar fields at each grid location: the symmetric metric tensor components $(g_{11}, g_{12}, g_{22})$ stored as a tensor of shape $(B, 3, M, N)$, and the drift vector $(b_1, b_2)$ stored as $(B, 2, M, N)$. The metric tensor is reconstructed as $\mG = \begin{pmatrix} g_{11} & g_{12} \\ g_{12} & g_{22} \end{pmatrix}$.

\textbf{Feasibility Projections.} The metric tensor $\mG$ must be symmetric positive-definite, and the drift vector $\vb$ must satisfy the Randers feasibility condition $\|\vb\|_{\mG^{-1}} < 1$ to ensure well-posed propagation. We enforce these constraints through differentiable projection layers applied after the network output. For $\mG$, we compute eigenvalues via the trace-discriminant formula $\lambda_{\pm} = \frac{1}{2}\bigl((g_{11} + g_{22}) \pm \sqrt{(g_{11} - g_{22})^2 + 4g_{12}^2}\bigr)$, clamp them to $[\epsilon, \lambda_{\max}]$, compute the principal angle $\theta = \frac{1}{2}\mathrm{atan2}(2g_{12}, g_{11} - g_{22})$, and reconstruct $\mG = \mU \mD \mU^\top$ with the clamped eigenvalues. For $\vb$, we first clip the Euclidean norm as a safety measure, then compute the $\mG^{-1}$-norm via $\|\vb\|^2_{\mG^{-1}} = (b_1^2 g_{22} - 2 b_1 b_2 g_{12} + b_2^2 g_{11}) / \det(\mG)$ and rescale if it exceeds the threshold. These projections are differentiable almost everywhere, allowing gradients to flow through during training.

\textbf{Total Variation Regularization.} We employ total variation regularization to promote piecewise-smooth parameter fields. For the drift vector field, this takes the standard form $\mathcal{R}(\vb) = \sum_{i,j} \|\nabla \vb_{i,j}\|_2$. For the metric tensor field, we consider two formulations. The Frobenius norm variant computes $\mathcal{R}(\mG) = \sum_{i,j} \|\nabla \mG_{i,j}\|_F$, treating $\mG$ as a collection of independent components with the off-diagonal weighted by 2 to account for symmetry. The Log-Euclidean variant respects the Riemannian geometry of the positive-definite cone by measuring distances in the tangent space: $\mathcal{R}(\mG) = \sum_{i,j} \|\nabla \log \mG_{i,j}\|_F$, where the matrix logarithm is computed in closed form using the same eigenvalue decomposition. The Log-Euclidean formulation is rotation-invariant and geometrically principled, whereas the Frobenius variant is often more numerically stable in practice.

\newpage
\onecolumn
\newpage
\onecolumn
\section{\textbf{Wildfire Dataset and Training Details}}
\label{app:wildfire_details} 

This appendix provides additional details on the Sim2Real-Fire dataset, cross-scene training protocol, and model capacity analysis that complement the experiments in Section~\ref{sec:wildfire}.

\subsection{Dataset Details}
\label{app:dataset_details}

The Sim2Real-Fire dataset~\cite{li2024sim2real} comprises over one million simulated wildfire scenarios generated using three established fire spread simulators: FARSITE~\cite{finney1998farsite}, WFDS~\cite{meerpoel2023modeling}, and WRF-SFIRE~\cite{mandel2011coupled}. Each scenario includes five aligned data modalities capturing the environmental factors that govern fire propagation. The topography maps provide elevation, slope, and aspect information derived from LANDFIRE at 30-meter resolution. Vegetation maps encode existing vegetation type, cover, and height according to the USNVC classification system. Fuel maps characterize surface fuel distribution, canopy properties, and fuel disturbance patterns essential for combustion modeling. Weather data recorded at hourly intervals include temperature, relative humidity, precipitation, wind speed, wind direction, and cloud cover. Finally, satellite-view images with annotated fire region masks provide the spatiotemporal ground truth for fire progression.

The simulation data were generated by varying ignition locations and controllable parameters such as spread speed and scope, producing approximately 200 distinct fire evolution sequences per environmental configuration. This diversity ensures that models must learn generalizable fire dynamics rather than memorizing specific scenarios. The temporal resolution of one hour and spatial resolution of 30 meters align well with operational fire management requirements while providing sufficient detail for learning fine-grained propagation patterns. This dataset is well-suited for evaluating physics-informed learning: the simulations encode physical principles of fire propagation consistent with our eikonal framework, the dense temporal sampling enables rigorous arrival time prediction assessment, and the scale (15 scenes with 426--3,070 fires each) provides sufficient data for both within-scene and cross-scene generalization experiments.

\subsection{Cross-Scene Training Protocol}
\label{app:cross_scene_training}

\textbf{Patch Extraction.} Scenes in the Sim2Real-Fire dataset vary substantially in grid dimensions, ranging from $132 \times 190$ to $574 \times 770$ pixels. Training on full-resolution scenes would require either padding to a common size (wasting computation on empty regions) or processing scenes individually (preventing efficient batching). We address this through a patch-based strategy that extracts fixed-size $256 \times 256$ patches from each fire event using a sliding window with stride 128 pixels. Patches with fewer than 100 burned pixels are discarded as they provide insufficient training signal, yielding 18,676 training patches across the 11 training scenes.

\textbf{Handling Disconnected Regions.} When a patch boundary intersects a fire's burned region, the patch may contain multiple disconnected burned areas---for example, if fire spread around a lake or ridge. For each connected component, we locate the pixel with minimum arrival time (where fire first entered that region) and place a $3 \times 3$ source mask around it, allowing the eikonal solver to handle patches with multiple entry points. Within each patch, arrival times are shifted so that source pixels have $T = 0$, since the physics we are learning concerns relative spread patterns rather than absolute ignition times.

\textbf{Training vs.\ Evaluation Resolution.} Training operates on fixed-size patches while evaluation uses full scene resolution. During training, the patch-based approach enables efficient batching across fires with different grid sizes (batch size 64 across 4 GPUs). During validation and testing, we process complete fires at their native resolution to obtain accurate performance metrics. Within the 11 training scenes, fires are split into 70\% for training, 15\% for validation (used for early stopping and learning rate scheduling), and 15\% held out.

\textbf{Test Scene Selection Criteria.} We select four test scenes to represent diverse generalization challenges: scene 0005\_00725 exhibited the lowest within-scene correlation (0.738) in our per-scene experiments; scene 0014\_00426 has the smallest grid ($132 \times 190$), testing whether the model handles different spatial scales; scene 0023\_00995 represents medium difficulty; and scene 0024\_02655 achieved the best within-scene performance and contains the most fires, providing a benchmark for comparing cross-scene to within-scene accuracy.

\subsection{Model Capacity Analysis Details}
\label{app:capacity_analysis}

Table~\ref{tab:capacity_examples} compares both optimization strategies on representative fires spanning different burned area sizes.

\begin{table}[htbp]
\centering
\caption{Per-fire optimization comparison showing encoder advantage (116K parameters, fixed across all fires) on small fires where per-pixel optimization is underdetermined.}
\label{tab:capacity_examples}
\small

\begin{tabular}{lrrrrc}
\toprule
& & \multicolumn{2}{c}{\textbf{Relative Error (\%)}} & & \\
\cmidrule(lr){3-4}
\textbf{Fire ID} & \textbf{Size} & \textbf{NN} & \textbf{Pixel} & \textbf{Grid} & \textbf{Params} \\
\midrule
\multicolumn{6}{l}{\textit{Small fires (encoder advantage):}} \\
0001\_00016 & 1,036 & \textbf{17.4} & 55.1 & 116$\times$287 & 166K \\
0003\_00041 & 1,011 & \textbf{19.1} & 60.3 & 411$\times$429 & 882K \\
0016\_00056 & 1,080 & \textbf{18.8} & 49.9 & 523$\times$612 & 1.6M \\
0017\_00014 & 1,498 & \textbf{10.7} & 40.6 & 310$\times$319 & 494K \\
\midrule
\multicolumn{6}{l}{\textit{Large fires (comparable performance):}} \\
0003\_00009 & 11,538 & 3.6 & \textbf{5.9} & 411$\times$429 & 882K \\
0019\_00022 & 12,023 & 3.2 & \textbf{1.9} & 424$\times$771 & 1.6M \\
0025\_00003 & 11,783 & 3.4 & \textbf{1.7} & 574$\times$502 & 1.4M \\
0024\_00020 & 6,901 & 2.7 & \textbf{2.3} & 565$\times$531 & 1.5M \\
\bottomrule
\end{tabular}
\end{table}

The pattern in Table~\ref{tab:capacity_examples} highlights the difference between the two strategies. Per-pixel optimization is severely underdetermined when the burned area is small relative to the parameter count: a fire burning 1,000 pixels on a $500 \times 500$ grid provides only 1,000 observations to constrain $1.25 \times 10^6$ parameters, a ratio exceeding 1,000:1. Under such conditions, the optimization landscape contains vast regions of near-equivalent solutions, making convergence to a good minimum unlikely. We observe this directly in the training dynamics: per-pixel optimization on small fires begins with loss values 10--100$\times$ higher than the encoder and often stagnates at poor local minima, achieving over 50\% relative RMSE on fires where the encoder reaches 17--19\%.

Conversely, when the burned area constitutes a substantial fraction of the domain, per-pixel optimization achieves the lowest errors in our experiments. Fires 0019\_00022, 0025\_00003, and 0024\_00020 demonstrate that per-pixel optimization can reach 1.7--2.3\% relative RMSE when sufficiently constrained, compared to 2.7--3.4\% for the encoder. This demonstrates that the Randers-Finsler physics model itself imposes no fundamental accuracy limit; the median per-pixel performance of 7.4\% reflects optimization difficulty rather than modeling capacity.

The encoder's convolutional architecture implicitly enforces spatial smoothness and learns to predict metric fields from local terrain features, providing an inductive bias that regularizes the solution space. This enables consistent convergence across diverse fire geometries, as reflected in the substantially lower variance (standard deviation of 7.4\% versus 16.7\%). Notably, the encoder uses a fixed 116K parameters regardless of whether the grid is $116 \times 287$ or $574 \times 502$, whereas per-pixel optimization scales linearly with domain size. The encoder thus represents a practical sweet spot: sufficient capacity to capture fire dynamics while maintaining enough structure for reliable optimization.

Table~\ref{tab:capacity_iou} evaluates spatial overlap using Intersection over Union (IoU) at different fractions of each fire's duration.

\begin{table}[htbp]
\centering
\caption{Mean IoU ($\pm$ std) at different time thresholds, measuring spatial overlap between predicted and observed burned regions.}
\label{tab:capacity_iou}
\begin{tabular}{lcccc}
\toprule
\textbf{Strategy} & \textbf{IoU@25} & \textbf{IoU@50} & \textbf{IoU@75} & \textbf{IoU@100} \\
\midrule
Per-pixel & 0.64 $\pm$ 0.32 & 0.72 $\pm$ 0.32 & 0.80 $\pm$ 0.27 & \textbf{1.00 $\pm$ 0.01} \\
Neural network & \textbf{0.66 $\pm$ 0.27} & \textbf{0.77 $\pm$ 0.23} & \textbf{0.85 $\pm$ 0.21} & 0.97 $\pm$ 0.05 \\
\bottomrule
\end{tabular}
\end{table}

\newpage
\onecolumn
\section{Temporal Band Decomposition Training}
\label{app:band_training}

Training neural network parameters via full-fire MSE loss faces a fundamental challenge: errors in early propagation compound throughout the simulation, creating both optimization difficulties (vanishing gradients through long eikonal solves) and attribution problems (the loss cannot distinguish early errors from late ones). We address this through temporal band decomposition, which partitions each fire's evolution into overlapping temporal segments and trains on each segment independently.

For a fire with maximum arrival time $T_{\max}$, we define bands as intervals $[t_k, t_k + w \cdot T_{\max}]$ where $w$ is the band width (as a fraction of $T_{\max}$) and bands are spaced by stride $s \cdot T_{\max}$. Crucially, each band solve uses the ground-truth fire perimeter at $t_k$ as its source region rather than the original ignition point. This means pixels where $T_{\text{gt}} \leq t_k$ are treated as already burned ($T_{\text{pred}} = 0$), and the eikonal equation propagates outward from this known boundary. The loss is computed only over pixels within the band, where $t_k < T_{\text{gt}} \leq t_k + w \cdot T_{\max}$, measuring how well the predicted metric field explains local propagation given perfect knowledge of the fire's prior extent.

This formulation offers several advantages. First, gradient signal flows through shorter temporal horizons, improving optimization stability. Second, later bands benefit from large source regions that accelerate solver convergence. Third, the decomposition provides a curriculum-like effect where the model receives supervision at multiple temporal scales simultaneously. However, band training evaluates a fundamentally different task than full-fire prediction: it measures one-step propagation accuracy given oracle boundary conditions, not multi-step forecasting from ignition alone.

We evaluated band training ($w=0.5$, $s=0.25$, yielding approximately 3 bands per fire) on two scenes with 1200 training fires each. On scene 0024\_02655 with a $565 \times 531$ grid ($\sim$300K pixels), models achieved test correlation $0.805 \pm 0.174$ and band-IoU $0.823 \pm 0.129$. On scene 0001\_02614 with a smaller $116 \times 287$ grid ($\sim$33K pixels), performance was lower at correlation $0.777 \pm 0.109$ and band-IoU $0.709 \pm 0.141$. The larger grid provides more spatial context and terrain diversity within a single scene, potentially explaining the performance gap. Scaling training data from 300 to 1200 fires yielded only marginal improvements ($\Delta\text{corr} = +0.003$), and increasing model capacity 8-fold (116K to 962K parameters) produced no improvement, confirming that the bottleneck lies outside data quantity and architectural capacity. These results suggest that increasing the number of bands (smaller $w$, finer temporal resolution) would similarly fail to improve performance, as the limiting factor appears to be feature informativeness or the expressiveness of the Randers-Finsler parameterization rather than optimization dynamics.

To evaluate real-fire transfer, we tested both single-scene band-trained models on 43 real wildfires from the Sim2Real-Fire dataset. The scene 0024 model (larger grid) achieved Pearson correlation $0.584 \pm 0.166$, Spearman correlation $0.696 \pm 0.221$, and mean IoU $0.351 \pm 0.190$. The scene 0001 model (smaller grid) performed worse with Pearson $0.541 \pm 0.189$, Spearman $0.690 \pm 0.228$, and IoU $0.307 \pm 0.193$. For comparison, the full-fire trained cross-scene model (10 scenes, no band decomposition) achieved Pearson $0.588 \pm 0.172$, Spearman $0.695 \pm 0.226$, and IoU $0.342 \pm 0.198$ on the same test set. The scene 0024 band-trained model matches real-fire transfer performance despite seeing only one-tenth the geographic diversity during training, suggesting that band decomposition may be more data-efficient for learning generalizable propagation dynamics when training on a sufficiently large and representative scene. However, the scene 0001 model underperforms, indicating that transfer quality depends on both training scene selection and grid size. Notably, Spearman correlation remains consistent across all three models ($\approx 0.69$), suggesting that rank ordering of arrival times is learned robustly while absolute timing calibration varies with training configuration.

\newpage
\onecolumn
\section{Evaluation on Real Wildfire Data}
\label{app:real_fire_evaluation}

To if whether the learned Randers metric generalizes beyond simulation, we evaluate our method on 50 real wildfires from the Sim2Real-Fire dataset observed via satellite imagery across diverse landscapes in the western United States. This experiment tests two questions: if a model trained directly on real observations can learn meaningful spread parameters from sparse supervision, and if a model trained entirely on FARSITE simulations transfers zero-shot to real fire dynamics.

\textbf{Real Fire Data Characteristics.} Each real fire shares the same five covariate modalities as the synthetic scenarios---topography, fuel, vegetation, and weather from LANDFIRE and USGS at identical spatial resolution---enabling direct application of models trained on simulations without domain adaptation of the input features.

The critical difference lies in the supervision signal. Synthetic fires provide continuous arrival time fields at sub-hourly resolution from FARSITE's deterministic solver, yielding dense per-pixel ground truth. Real fires are observed through satellite imagery that provides only 2--8 binary burned-area masks per fire, spaced days to weeks apart depending on revisit schedules and cloud cover. The precise moment each pixel burned is unknown; we observe only that burning occurred between consecutive acquisitions. Ground truth arrival times are therefore constructed by assigning each pixel the date when it first appears burned, introducing temporal quantization at the satellite revisit interval. Beyond observational sparsity, real fires exhibit phenomena absent from FARSITE simulations: spotting events create disconnected burned regions that violate the continuous wavefront assumption, and firefighting suppression efforts artificially halt spread in certain directions.

\textbf{Evaluation Metrics.} The sparse temporal observations require an evaluation strategy different from the synthetic experiments, where dense ground truth arrival times allow direct pointwise comparison. For each real fire, the satellite record provides $K$ dated binary masks $\{(d_0, M_0), (d_1, M_1), \ldots, (d_K, M_K)\}$, where $M_k$ indicates the burned area at acquisition date~$d_k$. The earliest observation $M_0$ serves as the ignition region: we initialize the eikonal source at all pixels where $M_0 > 0$ and set $T_{\text{pred}} = 0$ there. The solver then produces a continuous predicted arrival time field across the domain.

To compare this prediction against the remaining $K$ observations, we define $t_k = d_k - d_0$ as the elapsed time in days since first observation and compute IoU at each subsequent date:
\begin{equation}
\text{IoU}_{t_k} = \frac{|\hat{B}_{t_k} \cap M_k|}{|\hat{B}_{t_k} \cup M_k|}, \quad k = 1, \ldots, K, \nonumber
\end{equation}
where $\hat{B}_{t_k} = \{(x,y) : T_{\text{pred}}(x,y) \leq t_k\}$ is the predicted burned region at elapsed time~$t_k$. This metric directly evaluates if the predicted wavefront reaches the correct spatial extent by each observation date, without requiring per-pixel arrival time ground truth. We report the mean IoU across all $K$ dates for each fire.

As secondary metrics, we construct a quantized ground truth arrival time map by assigning each pixel the elapsed time $t_k$ of the earliest mask in which it appears burned. Pearson and Spearman rank correlations between these quantized values and the predicted arrival times measure temporal ordering accuracy. Note that correlation is invariant to affine transformations of the predicted times, making it robust to any unit mismatch between the solver output and the day-scale observation times. IoU, by contrast, requires that the predicted arrival times are on a compatible scale with the elapsed days, which we find to hold empirically for these fires. Five fires where the solver produced degenerate solutions, likely caused by spotting or suppression-induced gaps that prevent the wavefront from reaching the observed burned region, are excluded from the statistics.

\textbf{Training Protocol.} We evaluate two models on the real fires. The first is trained directly on real fire data: 50 fires split into 35 training, 7 validation, and 8 test fires (32 training and 5 validation after filtering fires exceeding GPU memory). Grid sizes vary from $241 \times 296$ to $1648 \times 2051$ pixels, processed at native resolution with batch size one. The loss is MSE between predicted and quantized ground truth arrival times within the burned region. The second model is the cross-scene model from Section~\ref{sec:cross_scene}, trained on 11 synthetic scenes containing thousands of FARSITE-simulated fires with dense arrival time supervision. This model is applied directly to all 50 real fires without any fine-tuning, testing pure simulation-to-reality transfer. Both models use the same architecture (5-layer CNN encoder with 64 hidden channels) and identical covariate preprocessing.

\textbf{Results.} Table~\ref{tab:real_fire_results} summarizes both evaluations. Because the real-trained model requires a held-out test set, its metrics are computed on 8 fires unseen during training, whereas the simulation-trained model is evaluated on all 43 valid fires since none were used for its training. Despite this difference in evaluation scope, the comparison is informative: even restricting attention to the 8 fires common to both evaluations, the simulation-trained model achieves higher correlation (0.61 vs.\ 0.50) and IoU (0.31 vs.\ 0.16).

\begin{table}[t]
\centering
\caption{Evaluation on real wildfires from Sim2Real-Fire. The simulation-trained model (cross-scene, Section~\ref{sec:cross_scene}) is applied zero-shot without fine-tuning. Five degenerate fires are excluded from both evaluations.}
\label{tab:real_fire_results}
\begin{tabular}{lccc}
\toprule
\textbf{Model} & \textbf{Pearson $\rho$} & \textbf{Spearman $r_s$} & \textbf{Mean IoU} \\
\midrule
Trained on real fires (8 test) & $0.503 \pm 0.240$ & $0.517 \pm 0.261$ & $0.159 \pm 0.131$ \\
Trained on simulations (43 fires test) & $0.588 \pm 0.172$ & $0.695 \pm 0.226$ & $0.342 \pm 0.198$ \\
\bottomrule
\end{tabular}
\end{table}

This result reflects two advantages of simulation-based training. First, the synthetic model was trained on thousands of fires with dense per-pixel arrival time supervision, whereas the real-trained model had access to only 32 fires with coarse temporal labels. Second, the continuous arrival time fields from FARSITE provide a much stronger training signal than binary masks at irregular intervals, enabling the model to learn finer-grained relationships between covariates and propagation speed. The shared covariate modalities between synthetic and real data: identical topographic, fuel, and vegetation layers from the same LANDFIRE and USGS sources, facilitate this transfer.

The best-performing fires under the simulation-trained model (0002 with $\rho = 0.89$, 0003 with $\rho = 0.82$) exhibit relatively uniform directional spread well-captured by the eikonal wavefront model. The most challenging fires (0021 with $\rho = 0.17$, 0018 with $\rho = 0.23$) involve complex dynamics such as multiple fronts or suppression-induced irregularities that violate continuous wavefront assumptions. Notably, fire 0025 achieves high IoU (0.37) despite low correlation (0.31) across its 14 observation dates, suggesting the model captures overall burned area extent even when fine temporal ordering is poor.

Compared to within-scene synthetic experiments achieving correlation 0.824 and IoU 0.609, the gap to real fire performance reflects the inherent difficulty of real-world prediction: noisy covariates, unmodeled phenomena, and coarse supervision. Nevertheless, the strong simulation-to-real transfer demonstrates that covariate-to-metric mappings learned from physics-based simulations capture physically meaningful relationships that generalize to actual fire behavior. These results point toward a practical workflow in which models are pre-trained on abundant simulation data and subsequently fine-tuned on sparse real observations, though we leave this fine-tuning step to future work.

\newpage
\onecolumn
\section{Additional Sim2Real-Fire Results}
\label{app:sim2real_additional}

Table~\ref{tab:per_scene_results} reports the full per-scene within-scene generalization metrics across all 15 scenes. This table is moved from the main text for readability; Figure~\ref{fig:per_scene_summary} in the main paper provides a compact summary of the same results.

\begin{table}[htbp]
\centering
\caption{Within-scene generalization results on the Sim2Real-Fire dataset. Each model is trained on fires from a single scene and evaluated on held-out fires from the same scene. This protocol tests whether the learned metric tensor generalizes to new ignition points within a known landscape.}
\label{tab:per_scene_results}
\resizebox{\textwidth}{!}{%
\begin{tabular}{lcccccccccc}
\toprule
Scene & Grid Size & Fires & Train/Val/Test & Correlation & Rel.\ RMSE & MAE & IoU@25 & IoU@50 & IoU@75 & IoU@100 \\
\midrule
0001\_02614 & $116 \times 287$ & 2614 & 1829/392/393 & $0.780 \pm 0.111$ & $0.226 \pm 0.099$ & $12.32 \pm 6.07$ & $0.426 \pm 0.207$ & $0.556 \pm 0.215$ & $0.678 \pm 0.181$ & $0.937 \pm 0.102$ \\
0002\_00714 & $205 \times 229$ & 714 & 500/106/108 & $0.845 \pm 0.099$ & $0.191 \pm 0.092$ & $9.92 \pm 5.46$ & $0.510 \pm 0.202$ & $0.640 \pm 0.202$ & $0.719 \pm 0.188$ & $0.930 \pm 0.081$ \\
0003\_01715 & $411 \times 429$ & 1715 & 1200/257/258 & $0.889 \pm 0.089$ & $0.204 \pm 0.140$ & $11.23 \pm 8.59$ & $0.551 \pm 0.219$ & $0.649 \pm 0.192$ & $0.746 \pm 0.141$ & $0.912 \pm 0.096$ \\
0004\_01191 & $417 \times 770$ & 1191 & 833/178/180 & $0.818 \pm 0.206$ & $0.241 \pm 0.208$ & $9.76 \pm 7.33$ & $0.551 \pm 0.217$ & $0.658 \pm 0.188$ & $0.797 \pm 0.167$ & $0.900 \pm 0.150$ \\
0005\_00725 & $408 \times 345$ & 725 & 507/108/110 & $0.738 \pm 0.168$ & $0.358 \pm 0.155$ & $20.07 \pm 9.81$ & $0.338 \pm 0.166$ & $0.455 \pm 0.171$ & $0.533 \pm 0.162$ & $0.778 \pm 0.159$ \\
0014\_00426 & $132 \times 190$ & 426 & 298/63/65 & $0.780 \pm 0.152$ & $0.424 \pm 0.295$ & $17.60 \pm 10.91$ & $0.397 \pm 0.209$ & $0.473 \pm 0.202$ & $0.616 \pm 0.203$ & $0.819 \pm 0.185$ \\
0015\_02986 & $473 \times 734$ & 2986 & 2090/447/449 & $0.848 \pm 0.125$ & $0.191 \pm 0.109$ & $10.53 \pm 7.75$ & $0.597 \pm 0.203$ & $0.742 \pm 0.188$ & $0.848 \pm 0.147$ & $0.936 \pm 0.103$ \\
0016\_03070 & $523 \times 612$ & 3070 & 2149/460/461 & $0.795 \pm 0.149$ & $0.311 \pm 0.170$ & $14.61 \pm 8.77$ & $0.418 \pm 0.197$ & $0.493 \pm 0.191$ & $0.617 \pm 0.199$ & $0.824 \pm 0.175$ \\
0017\_02731 & $310 \times 319$ & 2731 & 1912/408/411 & $0.755 \pm 0.144$ & $0.321 \pm 0.148$ & $12.55 \pm 6.06$ & $0.420 \pm 0.194$ & $0.491 \pm 0.193$ & $0.618 \pm 0.187$ & $0.837 \pm 0.161$ \\
0018\_02440 & $284 \times 309$ & 2440 & 1708/366/366 & $0.871 \pm 0.103$ & $0.204 \pm 0.114$ & $9.08 \pm 5.80$ & $0.510 \pm 0.203$ & $0.585 \pm 0.212$ & $0.710 \pm 0.190$ & $0.877 \pm 0.150$ \\
0019\_01316 & $424 \times 771$ & 1316 & 921/197/198 & $0.862 \pm 0.158$ & $0.179 \pm 0.137$ & $8.36 \pm 8.03$ & $0.600 \pm 0.219$ & $0.753 \pm 0.192$ & $0.870 \pm 0.145$ & $0.946 \pm 0.095$ \\
0022\_01630 & $533 \times 396$ & 1630 & 1141/244/245 & $0.867 \pm 0.101$ & $0.276 \pm 0.139$ & $12.07 \pm 6.90$ & $0.522 \pm 0.209$ & $0.649 \pm 0.193$ & $0.745 \pm 0.171$ & $0.898 \pm 0.113$ \\
0023\_00995 & $568 \times 579$ & 995 & 696/149/150 & $0.820 \pm 0.162$ & $0.371 \pm 0.275$ & $12.88 \pm 7.76$ & $0.454 \pm 0.214$ & $0.558 \pm 0.216$ & $0.683 \pm 0.221$ & $0.860 \pm 0.227$ \\
0024\_02655 & $565 \times 531$ & 2655 & 1858/398/399 & $0.835 \pm 0.151$ & $0.187 \pm 0.113$ & $7.30 \pm 5.70$ & $0.589 \pm 0.206$ & $0.740 \pm 0.192$ & $0.864 \pm 0.152$ & $0.961 \pm 0.077$ \\
0025\_02019 & $574 \times 502$ & 2019 & 1413/302/304 & $0.862 \pm 0.135$ & $0.213 \pm 0.126$ & $7.82 \pm 5.97$ & $0.593 \pm 0.198$ & $0.699 \pm 0.188$ & $0.799 \pm 0.165$ & $0.908 \pm 0.132$ \\
\midrule
\textbf{Mean $\pm$ Std} & -- & -- & -- & $\mathbf{0.824 \pm 0.044}$ & $\mathbf{0.260 \pm 0.076}$ & $\mathbf{12.49 \pm 4.81}$ & $\mathbf{0.496 \pm 0.082}$ & $\mathbf{0.609 \pm 0.099}$ & $\mathbf{0.719 \pm 0.093}$ & $\mathbf{0.879 \pm 0.063}$ \\
\bottomrule
\end{tabular}%
}
\end{table}

\newpage
\onecolumn
\section{Forward Solver Validation} \label{app:forward_solver}
This section provides supplementary validation for the fast sweeping solver for the Randers--Finsler eikonal equation. The main text (Section~\ref{sec:forward_validation}) reports the primary convergence and accuracy results; here we include additional anisotropic/heterogeneous cases and the runtime scaling table that was moved from the main text.

\begin{table}[t]
\centering
\caption{Iteration counts and wall-clock time per problem instance for the fast sweeping solver (Numba-optimized).}
\small
\label{tab:complexity}
\begin{tabular}{ccc}
\toprule
Grid Size & Iterations & Time (ms) \\
\midrule
$50 \times 50$   & 2 & 2.3 \\
$100 \times 100$ & 2 & 5.8 \\
$200 \times 200$ & 2 & 27.2 \\
$400 \times 400$ & 2 & 102.9 \\
$800 \times 800$ & 2 & 445.1 \\
\bottomrule
\end{tabular}
\end{table}

\textbf{Anisotropic Metric Validation.} The solver's ability to handle anisotropic propagation is critical for wildfire modeling where terrain and fuel properties induce directional speed variations. We validate this capability through two experiments with known analytical solutions.

For the uniform diagonal metric $\mG = \text{diag}(a^2, b^2)$ with $a = 2.0$ and $b = 0.5$, the analytical solution produces elliptical wavefronts with semi-axes proportional to $1/a$ and $1/b$. On a $200 \times 200$ grid, the solver achieves a relative $L^2$ error of 0.18\% against the exact solution, with maximum pointwise error of 0.53\%. Figure~\ref{fig:A2_anisotropic} illustrates the computed arrival time field with metric ellipse overlays demonstrating the expected anisotropic propagation pattern.

The rotated anisotropic case tests the handling of off-diagonal metric components. We construct $\mG = \mR^\top \mLambda \mR$ with eigenvalues $\lambda_1 = 4.0$ and $\lambda_2 = 0.25$ rotated by $\theta = 45^\circ$, yielding $g_{12} = 1.875$ at the domain center. The solver maintains comparable accuracy with relative $L^2$ error of 0.14\%, confirming correct treatment of the full symmetric positive-definite metric tensor. Results are visualized in Figure~\ref{fig:A3_rotated}.

\textbf{Drift Field Effects.} The Randers component of our formulation introduces a drift vector field $\vb(\vx)$ that biases wavefront propagation. We verify this behavior with a constant drift $\vb = (0.3, 0)$ applied to an isotropic metric. The resulting arrival times exhibit the expected asymmetry: measuring at equal distances from the source along the drift direction, we obtain $T_{-30} = 39.0$ and $T_{+30} = 21.0$, confirming the expected upwind/downwind asymmetry induced by $\vb$. The measured asymmetry is consistent with the theoretical dependence on the drift magnitude $\|\vb\|=0.3$. Figure~\ref{fig:A4_drift} shows the asymmetric wavefronts and the horizontal arrival time profile.

The combined anisotropic-drift experiment validates the full Randers--Finsler formulation with $\lambda_1 = 2.0$, $\lambda_2 = 0.5$, $\theta = 30^\circ$, and $\mathbf{b} = (0.2, 0.1)$. Since no closed-form solution exists for this case, we employ Richardson extrapolation comparing solutions on $100 \times 100$ and $400 \times 400$ grids. After appropriate rescaling to account for different grid spacings, the relative $L^2$ difference is 0.45\%, indicating consistent convergence behavior for the complete equation. Figure~\ref{fig:A5_combined} displays both solutions and their difference map.

\textbf{Heterogeneous Media.} Wildfire propagation involves spatially varying material properties requiring robust handling of metric discontinuities and smooth variations. The piecewise constant metric experiment places an interface at $x = N/2$ separating regions with $\mG = \mI$ (fast) and $\mG = 4\mI$ (slow). The solver correctly captures the $2\times$ slowdown across the interface, with measured propagation speed in the slow region matching the expected value of 2.0 grid units per unit time. Figure~\ref{fig:A6_piecewise} shows the wavefront refraction at the material interface.

For smoothly varying media, we prescribe $g_{11}(\vx) = g_{22}(\vx) = 1 + 0.5\sin(2\pi x/N)$. Rather than comparing to an unavailable analytical solution, we verify that the computed arrival times satisfy the eikonal equation $|\boldsymbol{\nabla} T|_\mG = 1$ throughout the interior. The mean residual $\big| \|\boldsymbol{\nabla} T\|_\mG - 1 \big|$ is 0.22\% with maximum 1.3\%, confirming that the solver accurately captures the spatially varying propagation speeds. Figure~\ref{fig:A7_varying} visualizes the metric field and eikonal residual distribution.

\textbf{Multiple Ignition Points.} Wildfire scenarios frequently involve multiple ignition sources whose fire fronts eventually merge. The solver handles this naturally through the initialization $T = 0$ at all source locations. With three sources placed at $(N/4, N/4)$, $(N/4, 3N/4)$, and $(3N/4, N/2)$, the computed solution matches the analytical minimum-distance field $T(\vx) = \min_i \|\vx - \vx_i\|$ with relative $L^2$ error of 0.20\%. Figure~\ref{fig:A8_multisource} displays the Voronoi-like structure of the arrival time field with source locations marked.

\textbf{Summary.} All nine validation experiments confirm the correctness and efficiency of the fast sweeping implementation. The solver achieves sub-1\% relative errors across isotropic, anisotropic, and combined Randers--Finsler configurations while maintaining the expected $O(N^2)$ computational complexity.

\newpage
\onecolumn
\subsection{Supplementary figures A}
\begin{figure}[htbp]
\centering
\includegraphics[width=0.5\columnwidth]{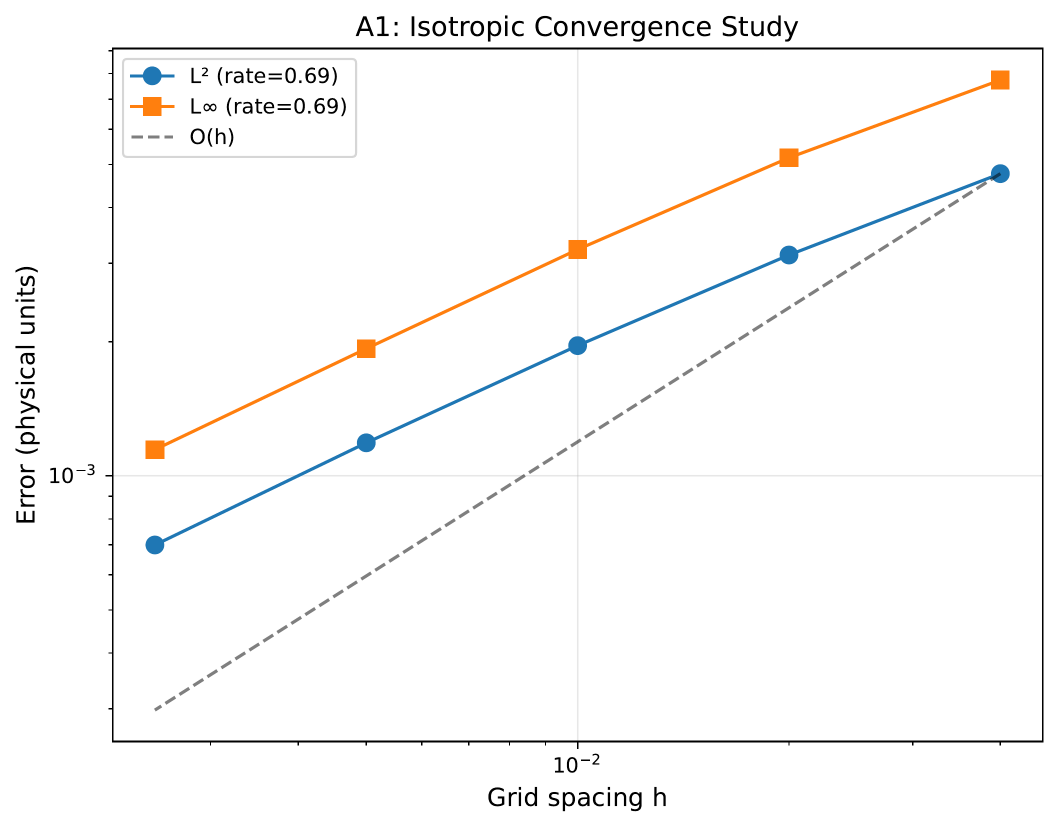}
\caption{Convergence study for the isotropic eikonal equation. Both $L^2$ and $L^\infty$ errors decrease with grid refinement at rate $\approx 0.69$, consistent with first-order accuracy.}
\label{fig:A1_convergence}
\end{figure}

\begin{figure*}[htbp]
\centering
\includegraphics[width=\textwidth]{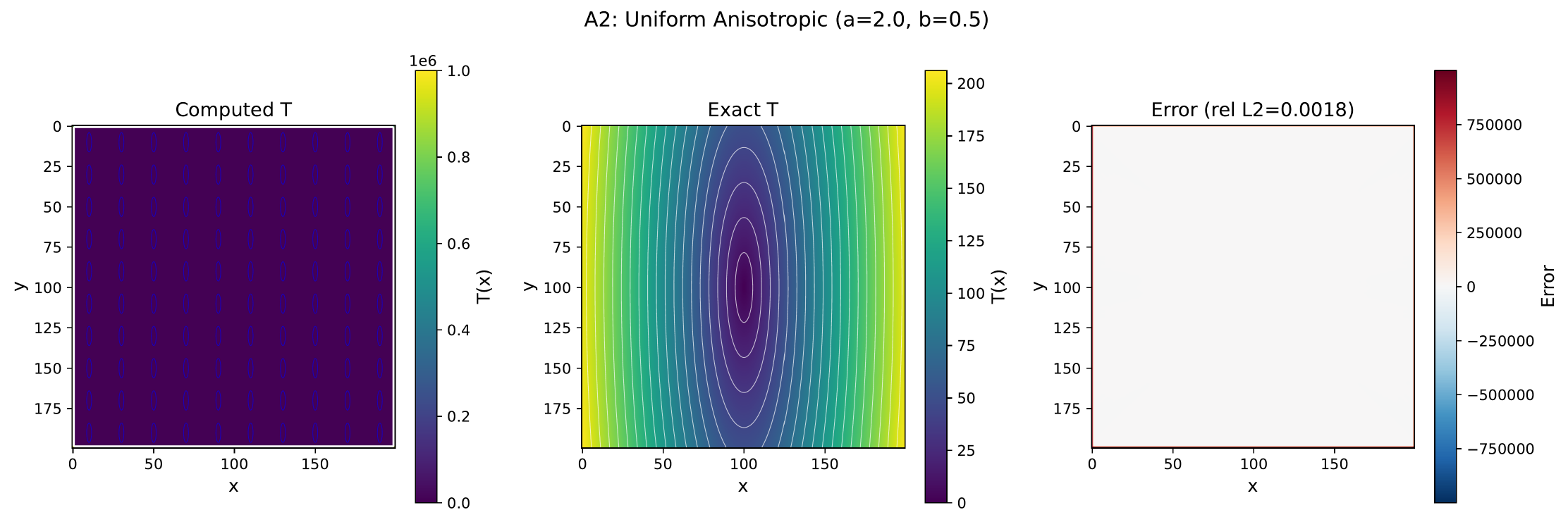}
\caption{Uniform anisotropic metric validation with $G = \text{diag}(4, 0.25)$. Left: computed arrival time with metric ellipses showing local propagation speeds. Center: analytical solution. Right: pointwise error (relative $L^2 = 0.18\%$).}
\label{fig:A2_anisotropic}
\end{figure*}

\begin{figure}[t!]
\centering
\includegraphics[width=\textwidth]{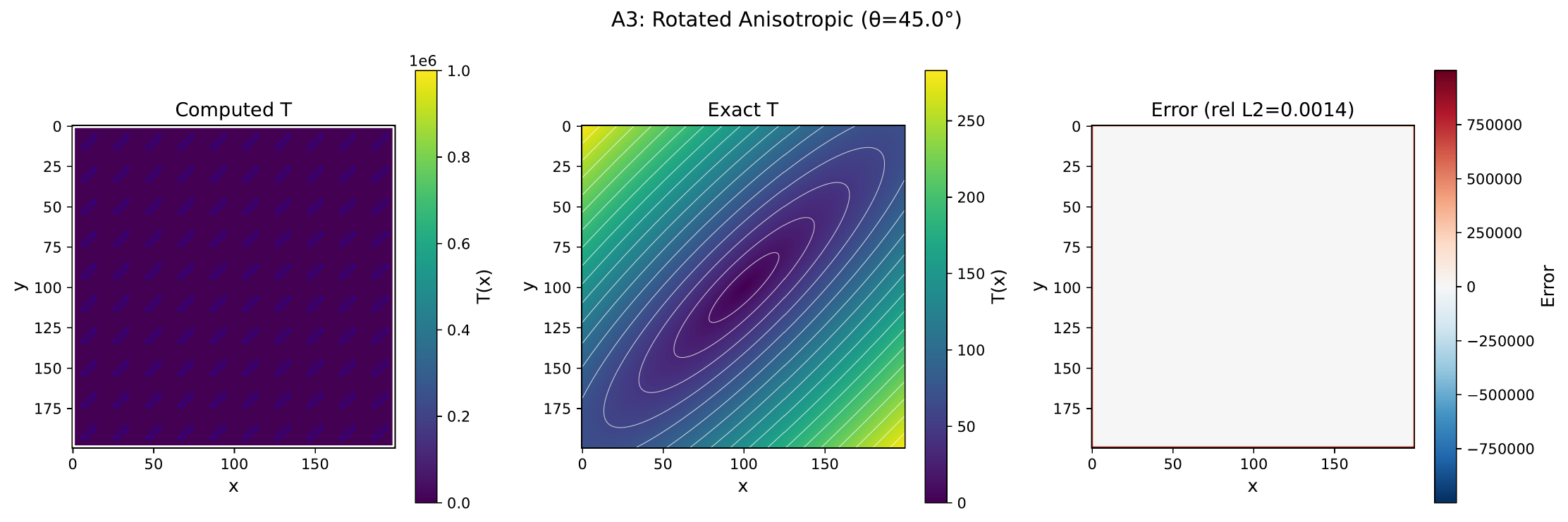}
\caption{Rotated anisotropic metric with $\theta = 45^\circ$, demonstrating correct handling of off-diagonal components $g_{12} \neq 0$. The solver achieves relative $L^2$ error of 0.14\%.}
\label{fig:A3_rotated}
\end{figure}

\begin{figure*}[htbp]
\centering
\includegraphics[width=\textwidth]{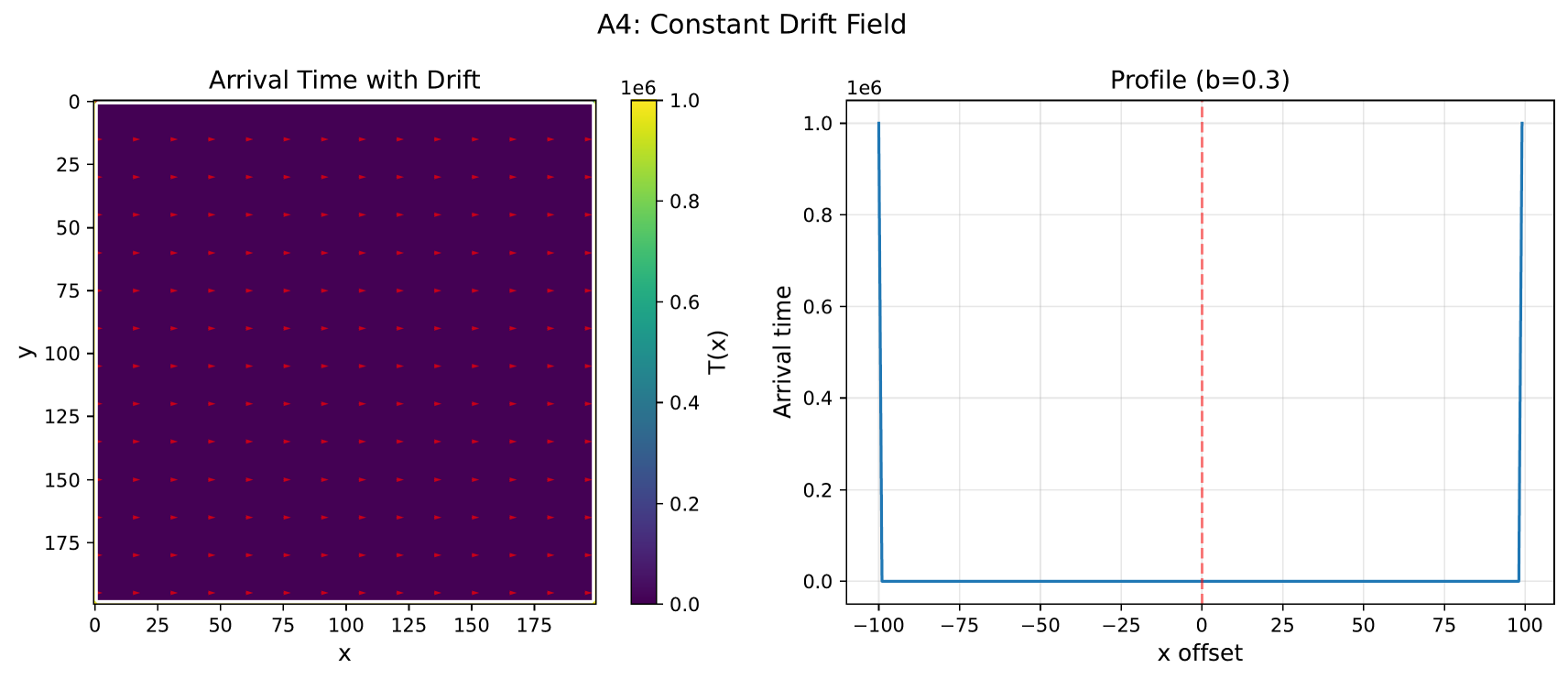}
\caption{Constant drift field validation with $\vb = (0.3, 0)$. Left: arrival time field showing asymmetric wavefronts elongated against the drift direction. Right: horizontal profile through source demonstrating the expected asymmetry ratio of 0.30.}
\label{fig:A4_drift}
\end{figure*}

\begin{figure*}[htbp]
\centering
\includegraphics[width=\textwidth]{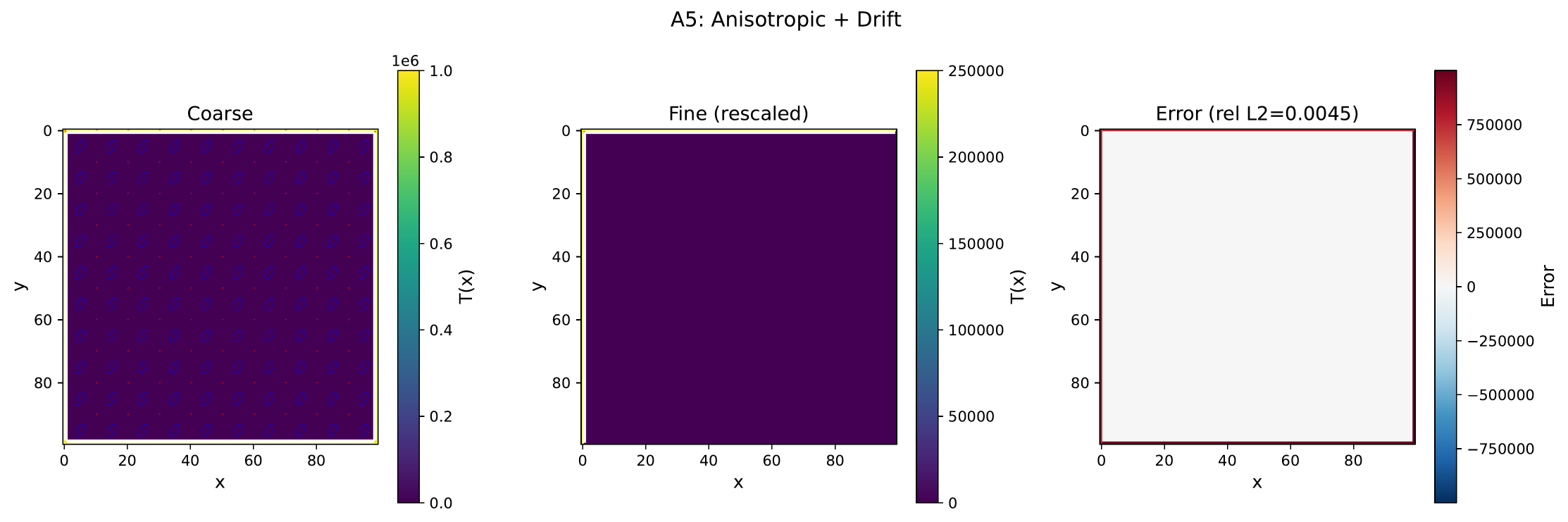}
\caption{Combined anisotropic metric and drift field validation. Left: coarse grid solution with metric ellipses and drift vectors overlaid. Center: fine grid reference solution (rescaled). Right: difference map showing 0.45\% relative $L^2$ error.}
\label{fig:A5_combined}
\end{figure*}

\begin{figure*}[t]
\centering
\includegraphics[width=1\textwidth]{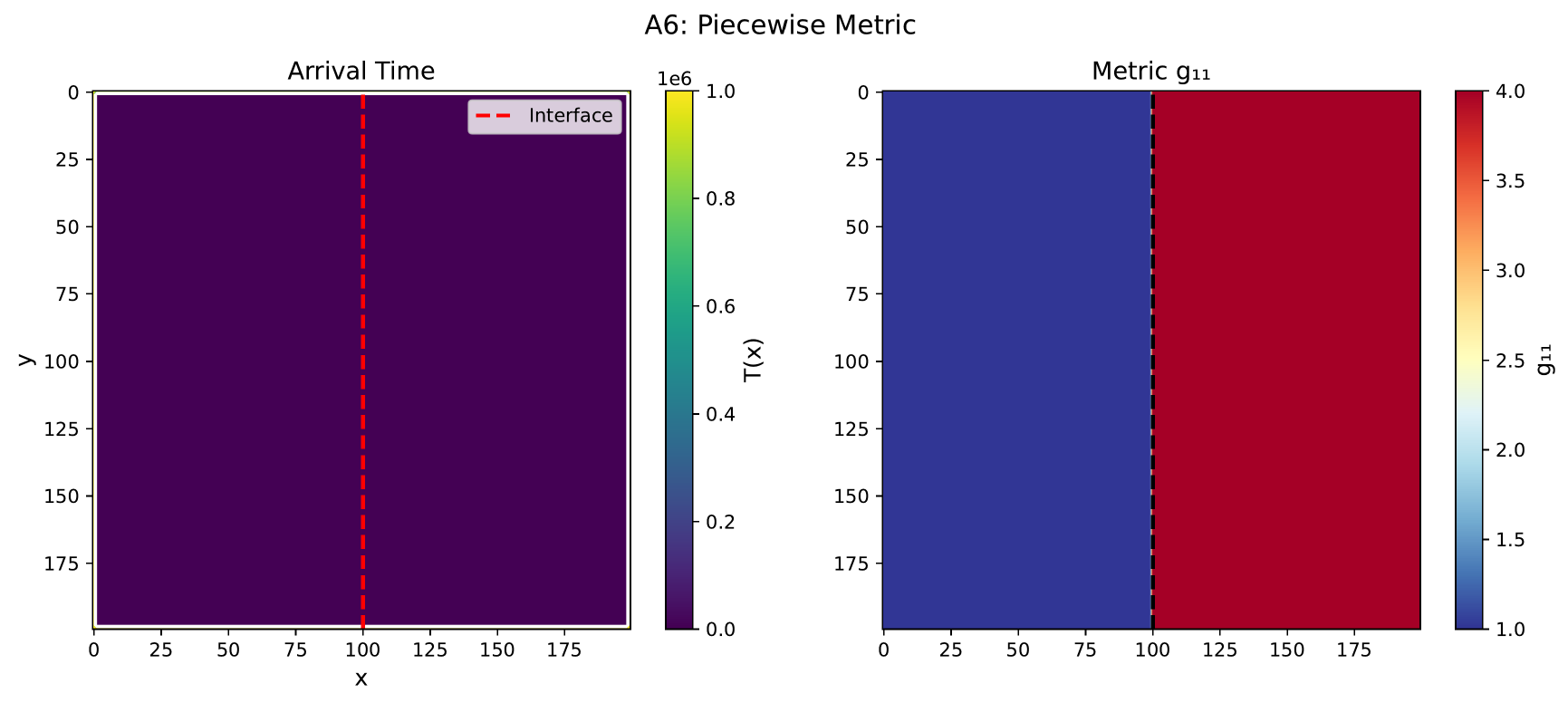}
\caption{Piecewise constant metric with interface at $x = N/2$. Left: arrival time field showing wavefront refraction. Right: metric field $g_{11}$ with interface location marked.}
\label{fig:A6_piecewise}
\end{figure*}

\begin{figure}[t]
\centering
\includegraphics[width=\textwidth]{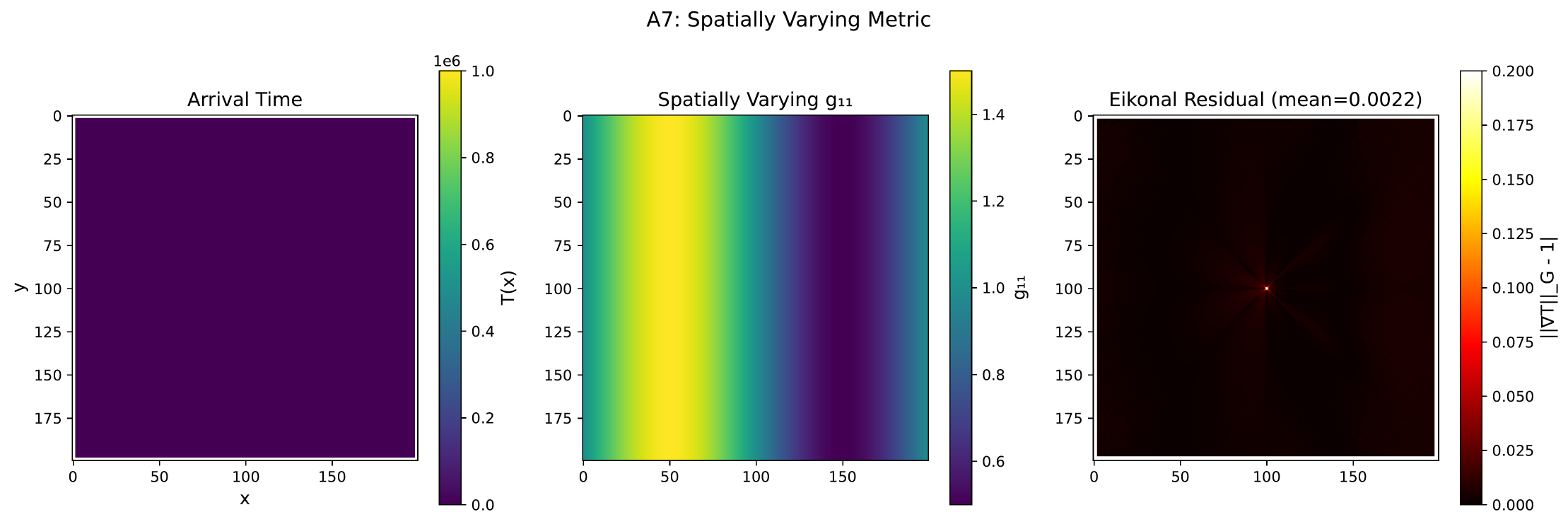}
\caption{Spatially varying metric validation. Left: computed arrival time. Center: metric field $g_{11}(\mathbf{x}) = 1 + 0.5\sin(2\pi x/N)$. Right: eikonal residual $\big| \|\nabla T\|_G - 1 \big|$ with mean 0.22\%.}
\label{fig:A7_varying}
\end{figure}

\begin{figure*}[htbp]
\centering
\includegraphics[width=\textwidth]{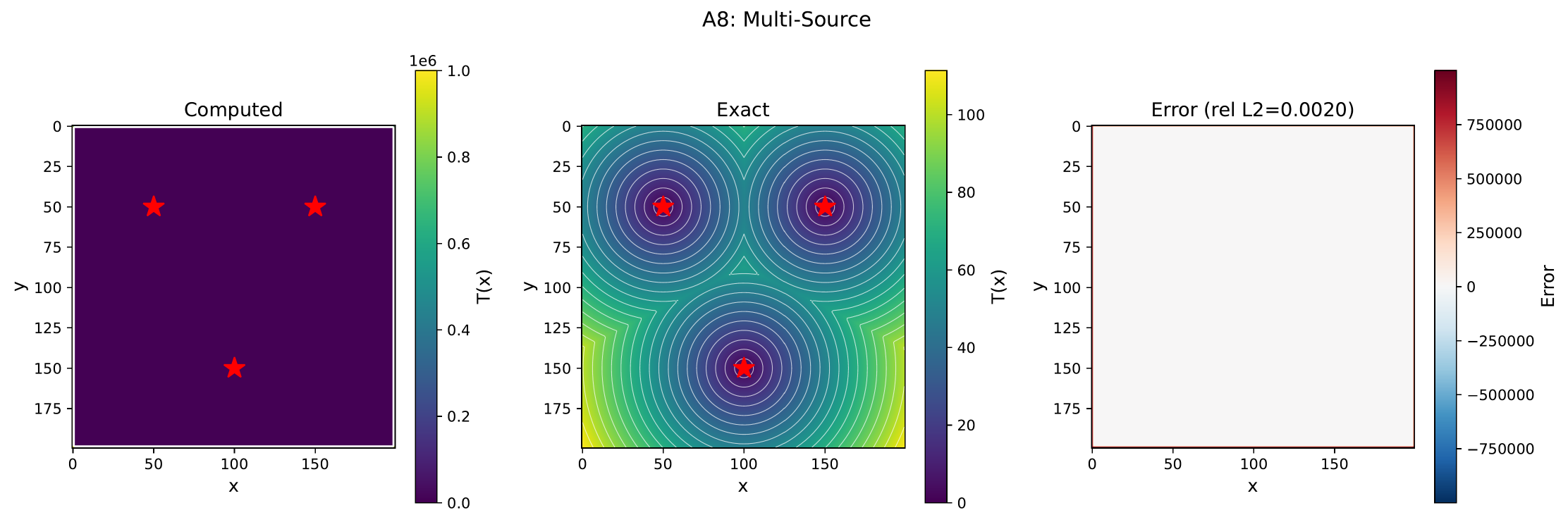}
\caption{Multi-source validation with three ignition points (red stars). Left: computed arrival time showing Voronoi-like structure. Center: analytical minimum-distance solution. Right: pointwise error (relative $L^2 = 0.20\%$).}
\label{fig:A8_multisource}
\end{figure*}

\begin{figure}[htbp]
\centering
\includegraphics[width=\textwidth]{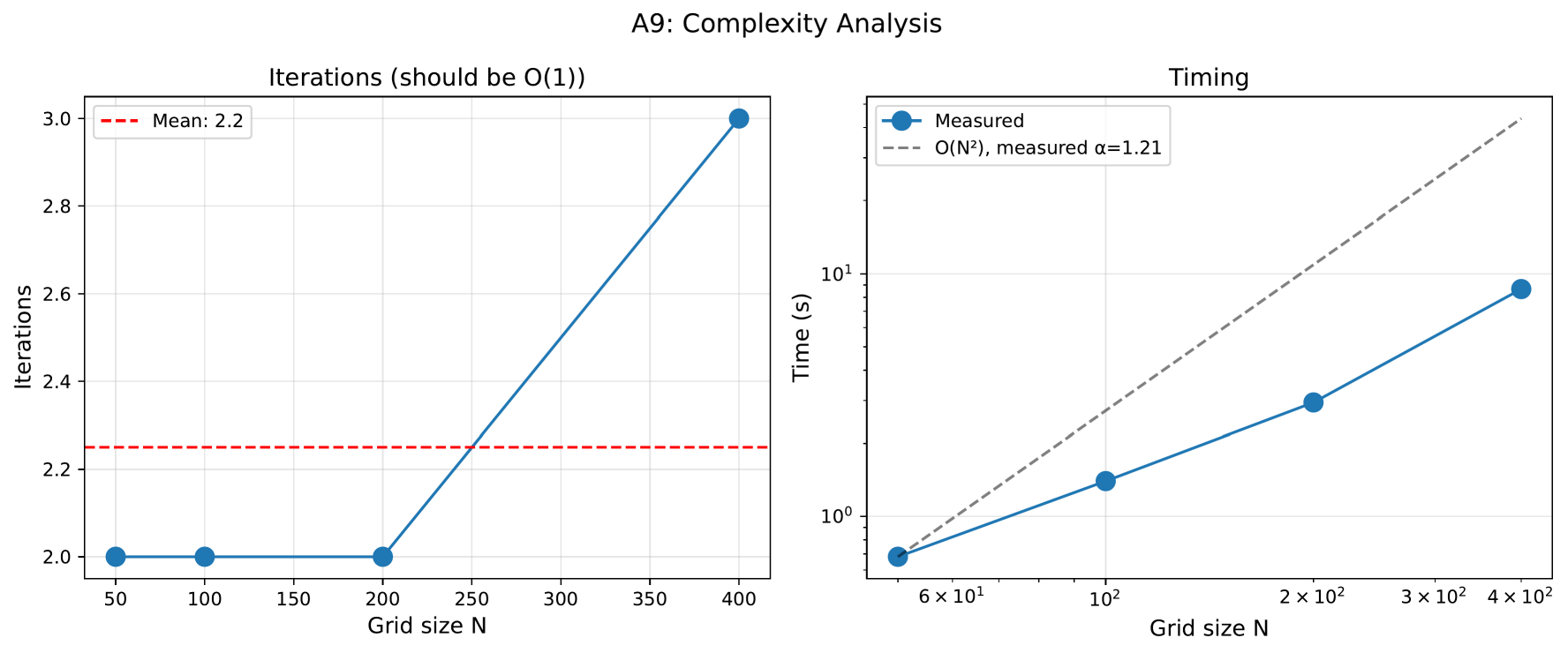}
\caption{Computational complexity analysis. Left: iteration count versus grid size, demonstrating $O(1)$ iterations. Right: wall-clock time versus total grid points $n = N^2$ on log-log scale, with measured scaling exponent $\alpha \approx 1$, confirming $O(n)$ complexity.}
\label{fig:A9_complexity}
\end{figure}

\newpage
\onecolumn
\section{\textbf{Gradient Verification}}
\label{app:gradient_verification}

This section provides additional visualizations and experiments validating the correctness of gradients computed via implicit differentiation of the fast sweeping solver. Core results including pointwise accuracy, stencil boundary analysis, and stability findings are presented in Section~\ref{sec:gradient_validation}.

\begin{figure}[htbp]
\centering
\includegraphics[width=\textwidth]{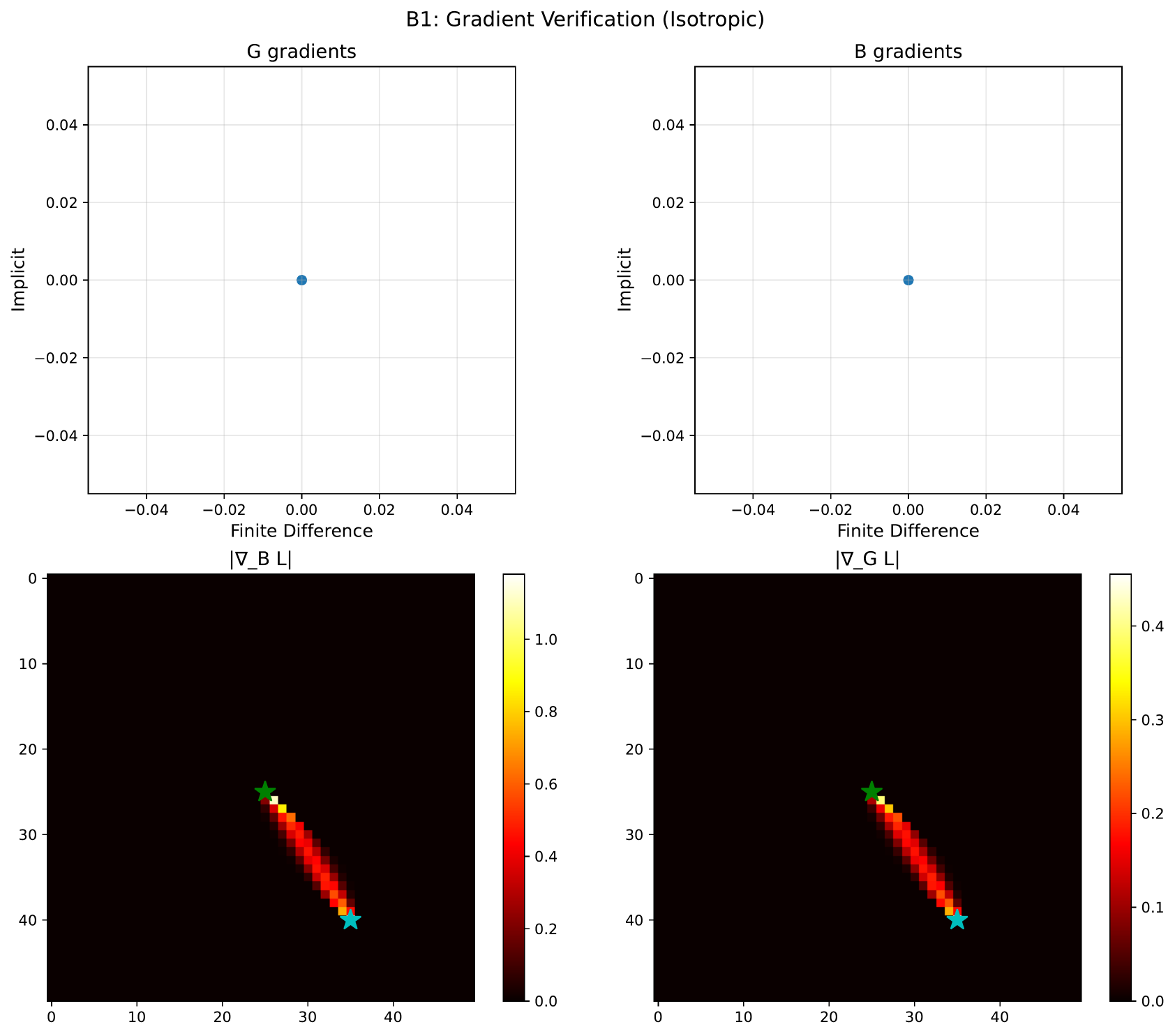}
\caption{\textbf{Isotropic gradient verification.} Scatter plots comparing finite difference (FD) and implicit differentiation gradients for metric $\mG$ and drift $\vb$ components. Perfect alignment along the identity line confirms 100\% accuracy.}
\label{fig:S10}
\end{figure}

\begin{figure}[htbp]
\centering
\includegraphics[width=\textwidth]{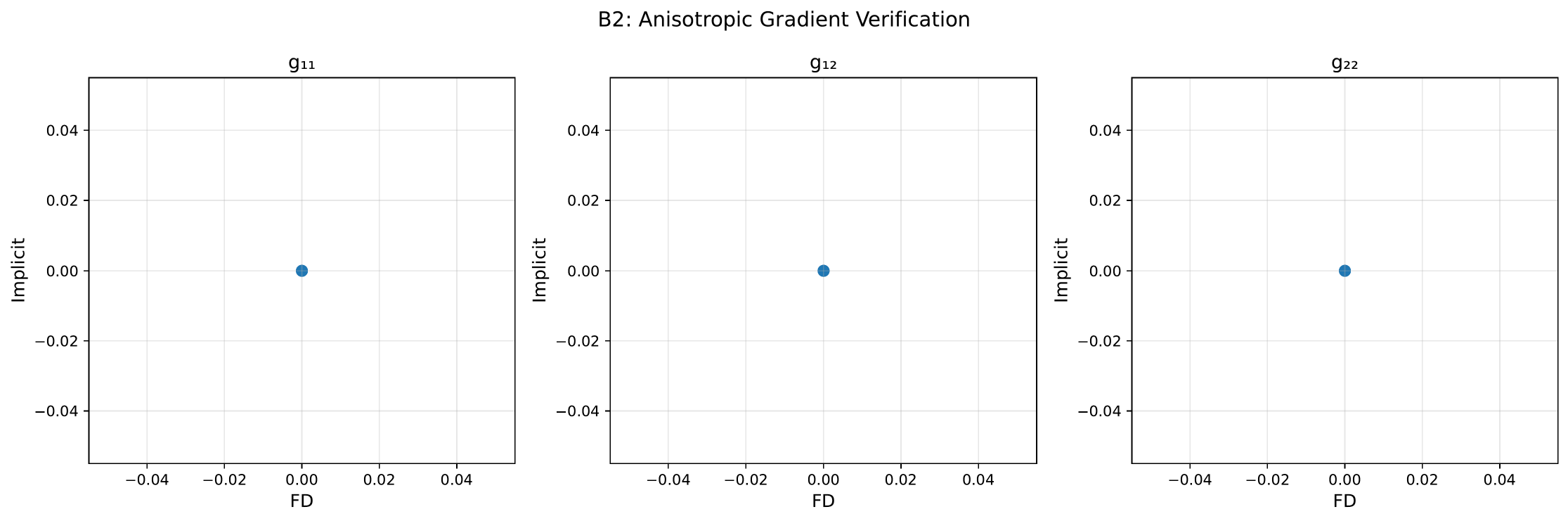}
\caption{\textbf{Anisotropic gradient verification.} All three metric components ($g_{11}$, $g_{12}$, $g_{22}$) achieve perfect agreement between FD and implicit gradients at 15 randomly sampled points.}
\label{fig:S11}
\end{figure}

\begin{figure}[htbp]
\centering
\includegraphics[width=0.8\textwidth]{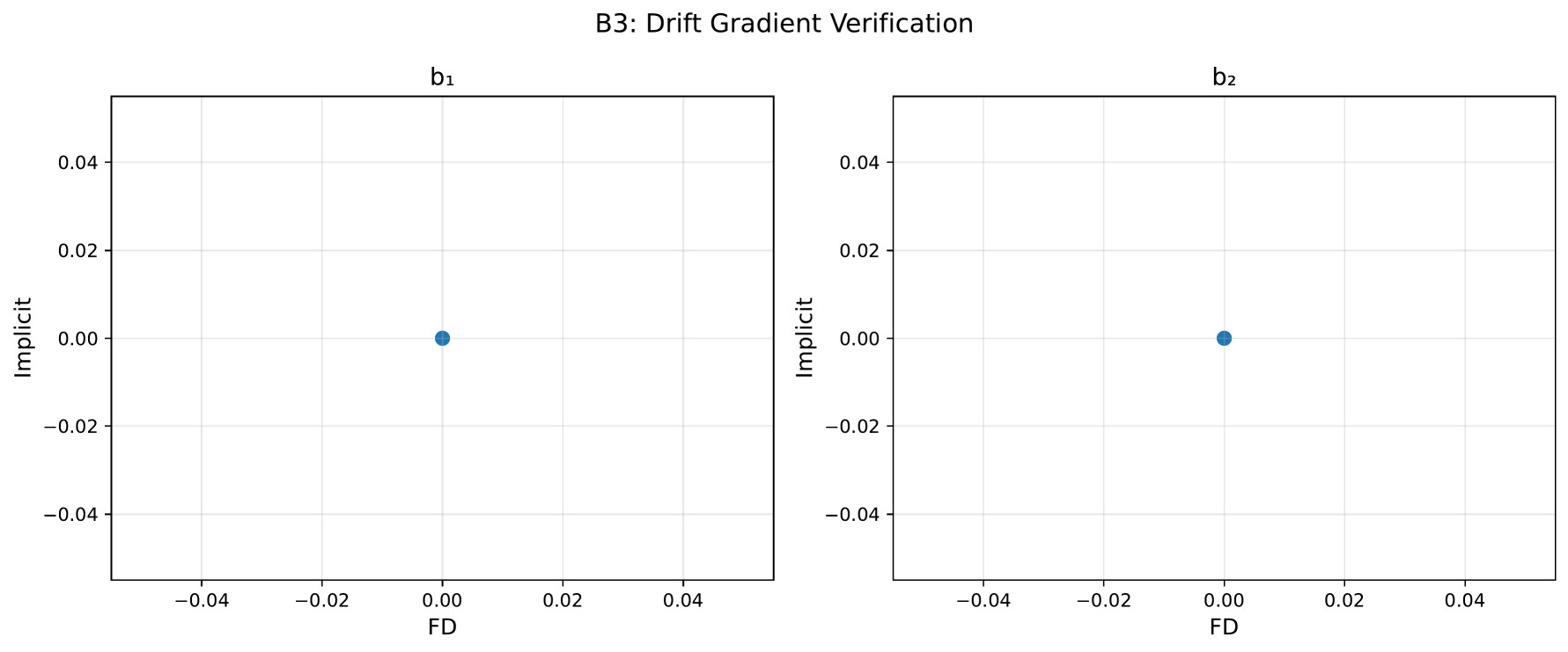}
\caption{\textbf{Drift gradient verification.} Both drift components ($b_1$, $b_2$) show exact agreement between FD and implicit differentiation.}
\label{fig:S12}
\end{figure}

\textbf{Pointwise Gradient Accuracy.} We verify gradient correctness by comparing implicit differentiation against central finite differences at randomly sampled interior points. For the isotropic case with $\mG = \mI$ and $\vb = \vzero$, we test both metric and drift gradients. The anisotropic case with rotated metric ($\lambda_1 = 2.0$, $\lambda_2 = 0.5$, $\theta = 30^\circ$) tests all three metric components $g_{11}$, $g_{12}$, and $g_{22}$. We also test drift gradients with $\vb = (0.15, 0.08)$ for both $b_1$ and $b_2$ components. Supplementary Figures~\ref{fig:S10}--\ref{fig:S12} display scatter plots comparing finite difference and implicit gradients, demonstrating perfect alignment along the identity line.

\begin{figure}[htbp]
\centering
\includegraphics[width=\textwidth]{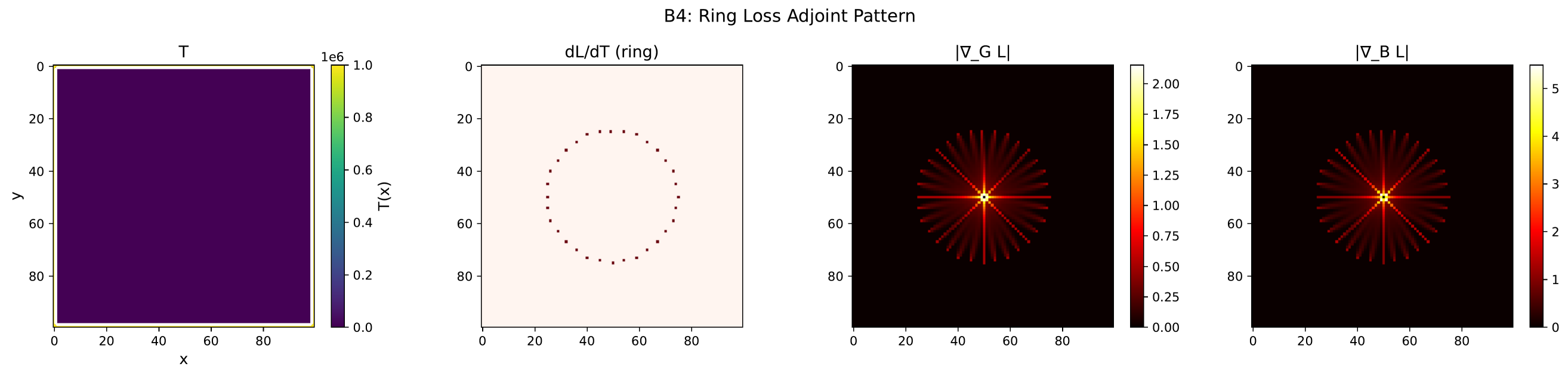}
\caption{\textbf{Adjoint propagation pattern.} With observations placed on a ring around the source, gradient magnitudes propagate inward, consistent with adjoint equation structure.}
\label{fig:S13}
\end{figure}

\begin{figure}[htbp]
\centering
\includegraphics[width=\textwidth]{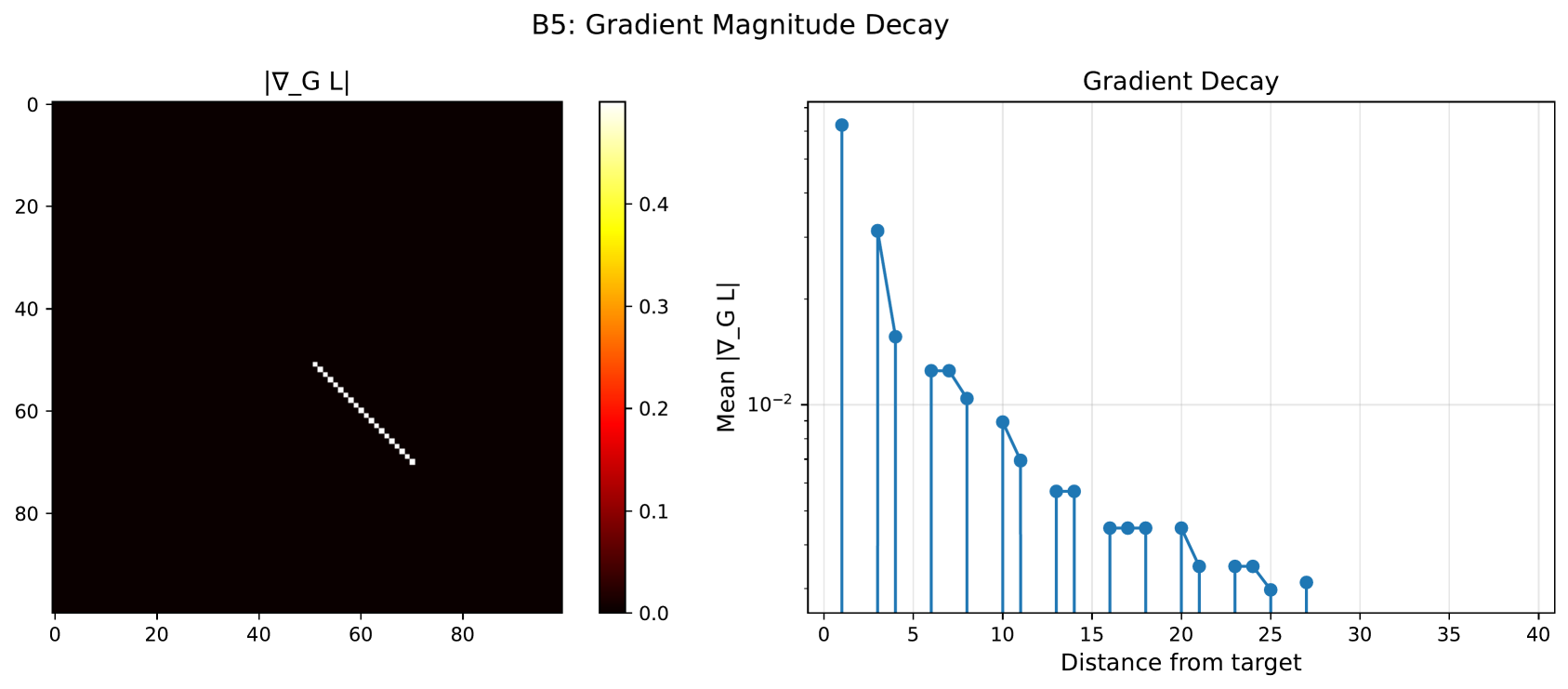}
\caption{\textbf{Gradient magnitude decay.} The gradient $|\nabla_G \mathcal{L}|$ decreases with distance from the single observation point, reflecting localized parameter influence.}
\label{fig:S14}
\end{figure}

\textbf{Adjoint Propagation.} We examine the spatial structure of gradient fields to verify correct adjoint propagation. Placing observations on a ring of radius 25 around the source, the resulting gradient magnitudes $|\nabla_\mG \mathcal{L}|$ and $|\nabla_{\vb} \mathcal{L}|$ propagate inward from the observation locations toward the source, consistent with the adjoint equation structure (Supplementary Figure~\ref{fig:S13}). Testing gradient magnitude decay from a single target point confirms that $|\nabla_\mG \mathcal{L}|$ decreases with distance from the observation, as expected from the localized influence of parameter perturbations on arrival times (Supplementary Figure~\ref{fig:S14}).

\begin{figure}[htbp]
\centering
\includegraphics[width=\textwidth]{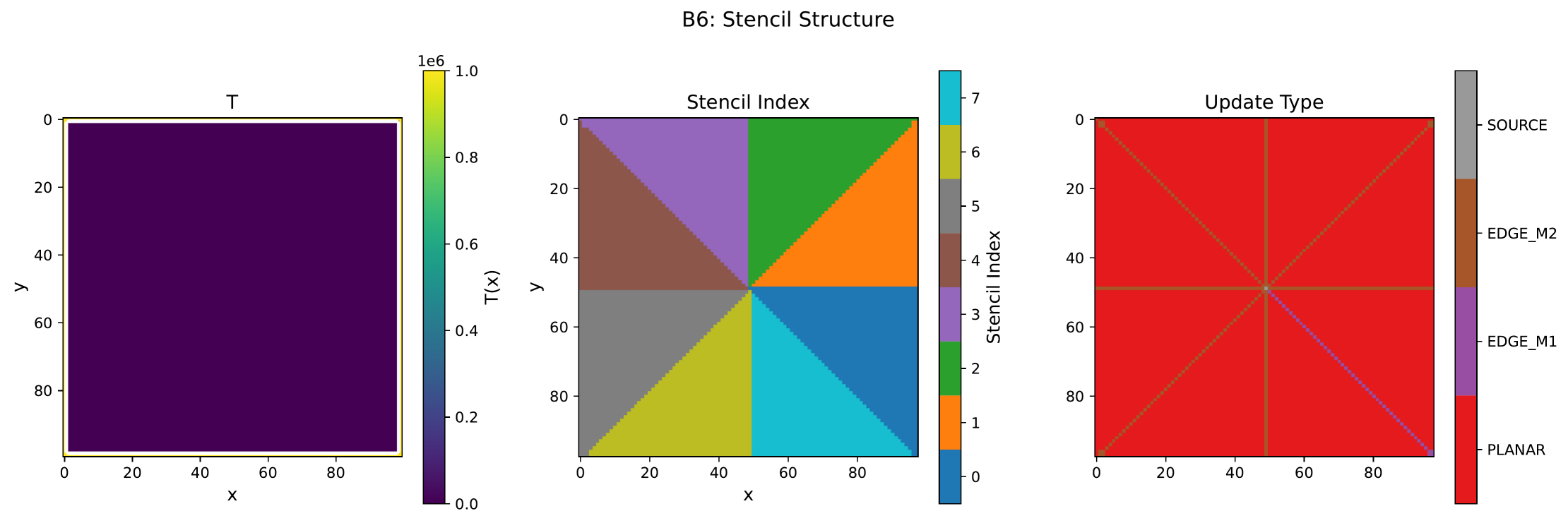}
\caption{\textbf{Stencil map visualization.} Left: active stencil index (0--7) at each grid point. Right: update type distribution showing 96\% planar (two-point) updates.}
\label{fig:S15}
\end{figure}

\textbf{Stencil Structure Analysis.} The fast sweeping method selects one of eight triangular stencils at each grid point based on the local solution structure. We visualize this stencil map to understand the solver's behavior. The eight stencils are distributed approximately uniformly across the domain (Supplementary Figure~\ref{fig:S15}), with stencil boundary pixels tracing characteristic curves emanating from the source (Figure~\ref{fig:stencil_boundaries}).

\textbf{Random Direction Test.} To assess gradient accuracy along arbitrary perturbation directions, we test 100 random unit vectors in the full parameter space. Each direction perturbs all metric components simultaneously, computing the directional derivative via both finite differences and the implicit gradient.

\begin{figure}[htbp]
\centering
\includegraphics[width=\textwidth]{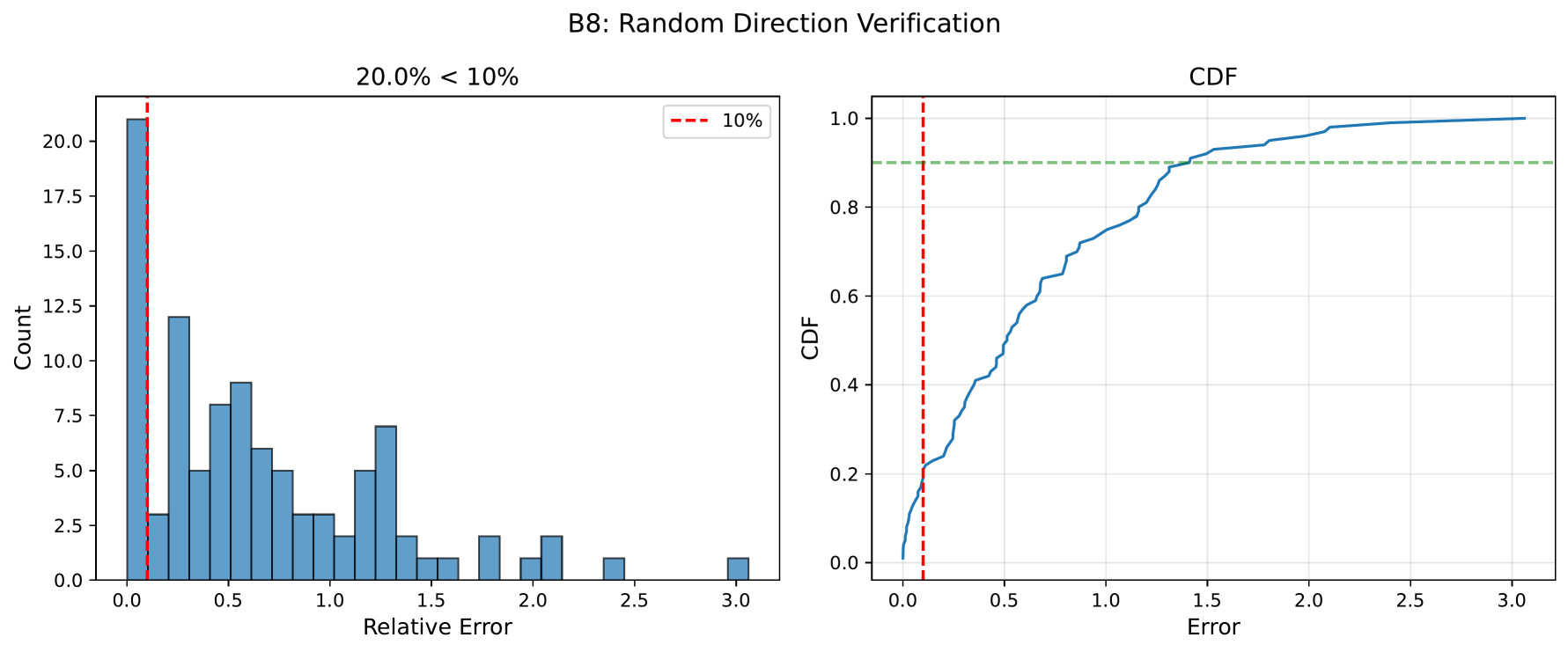}
\caption{\textbf{Random direction test.} Left: histogram of relative errors across 100 random perturbation directions. Right: CDF showing 20\% achieve $<$10\% error. The reduced accuracy reflects stencil boundary crossings.}
\label{fig:S17}
\end{figure}

This reduced accuracy compared to pointwise tests is expected and validates our theoretical analysis: random perturbations almost surely cross at least one stencil boundary somewhere in the domain, causing the finite difference to measure a different (discontinuous) function than the implicit gradient assumes. The success rate reflects directions that happen to avoid significant boundary crossings, while the majority encounter the measure-zero non-differentiability set (Supplementary Figure~\ref{fig:S17}).

\begin{figure}[htbp]
\centering
\includegraphics[width=0.5\textwidth]{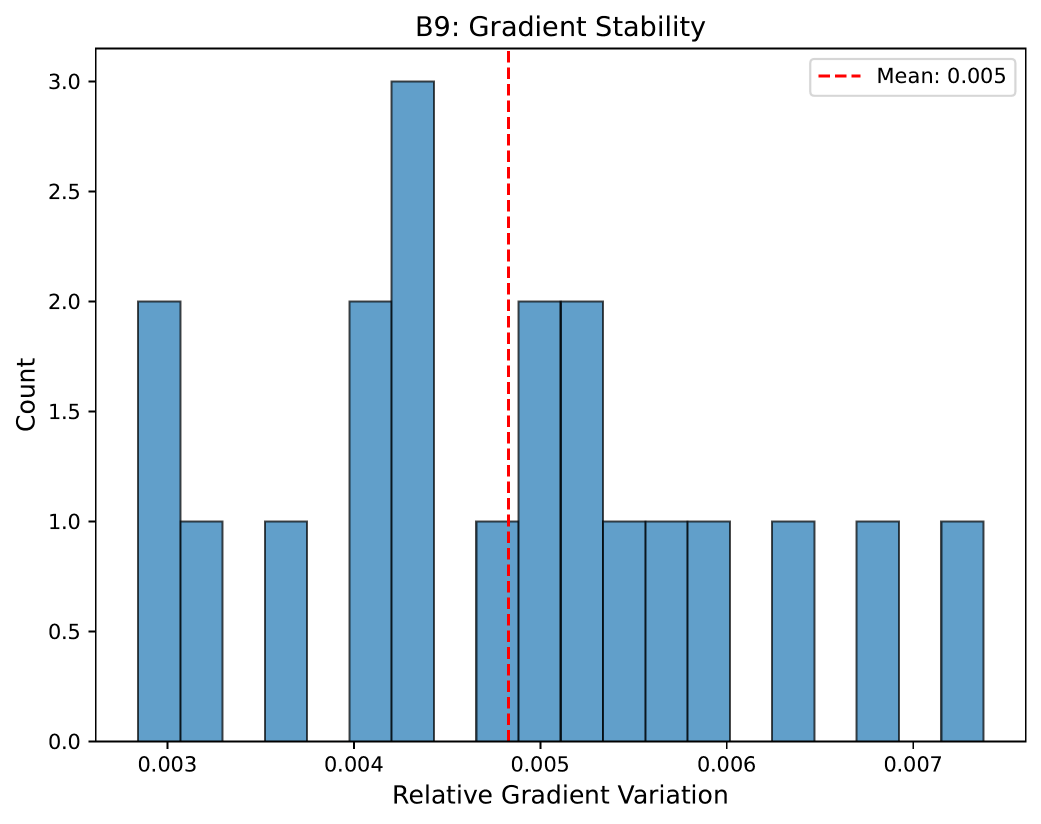}
\caption{\textbf{Gradient stability.} Distribution of gradient variations under 1\% metric perturbations. Mean variation of 0.48\% confirms stability despite stencil switching.}
\label{fig:S18}
\end{figure}

\textbf{Gradient Stability.} Despite stencil boundaries, we verify that gradients remain stable under small parameter perturbations. Adding 1\% Gaussian noise to the metric tensor and recomputing gradients confirms that while the solver's stencil selection may change discretely, the resulting gradient changes are small and bounded (Supplementary Figure~\ref{fig:S18}).

\newpage
\onecolumn
\subsection{Supplementary figures B}

\newpage
\onecolumn
\section{\textbf{Inverse Problem Experiments}}
\label{app:inverse_problem}

This section provides additional inverse problem experiments and visualizations. Core results for isotropic recovery, drift recovery, and multi-source learning are presented in Section~\ref{sec:inverse_validation}. All experiments use Adam optimization with gradient clipping and learning rate scheduling.

\textbf{Isotropic Metric Recovery.} We consider recovering an isotropic metric $\mG = g(\vx)\mI$ from arrival time observations. The true metric is piecewise constant with $g=1$ in the left half and $g=2$ in the right half of an $80 \times 80$ domain, creating a sharp interface at $x=40$.

\begin{figure*}[htbp]
\centering
\includegraphics[width=1\textwidth]{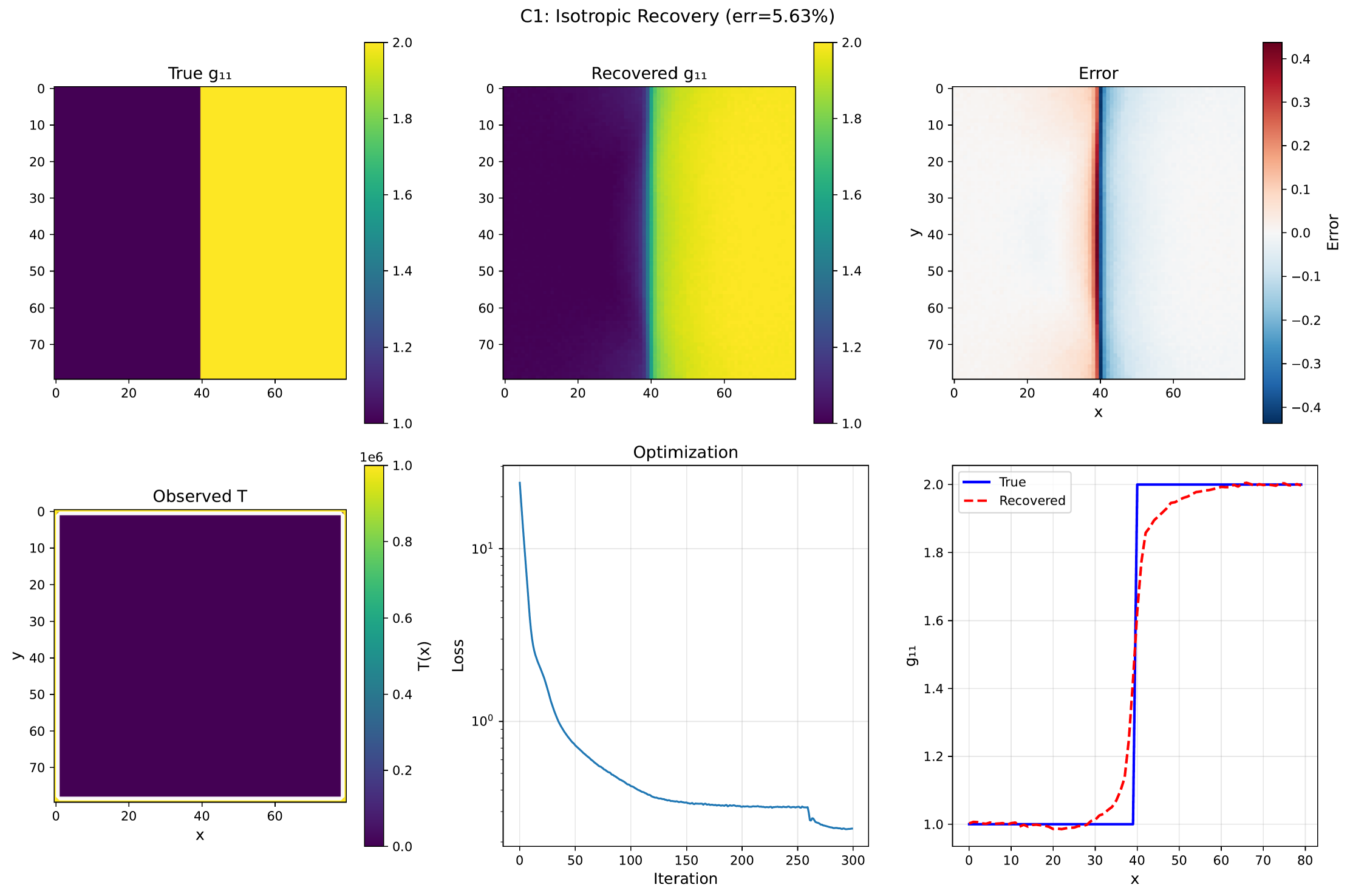}
\caption{Isotropic metric recovery with full observations achieves 5.6\% error. Top: true metric, recovered metric, and pointwise error. Bottom: observed arrival times, loss curve, and profile through the interface.}
\label{fig:isotropic_recovery}
\end{figure*}

With full observations, the recovered metric accurately captures both the values in each region and the location of the interface. The one-dimensional profile reveals slight smoothing at the discontinuity, which is expected since the TV regularization penalizes sharp gradients. The optimization converges smoothly over approximately two orders of magnitude in loss (Figure~\ref{fig:isotropic_recovery}).

\begin{figure}[htbp]
\centering
\includegraphics[width=\textwidth]{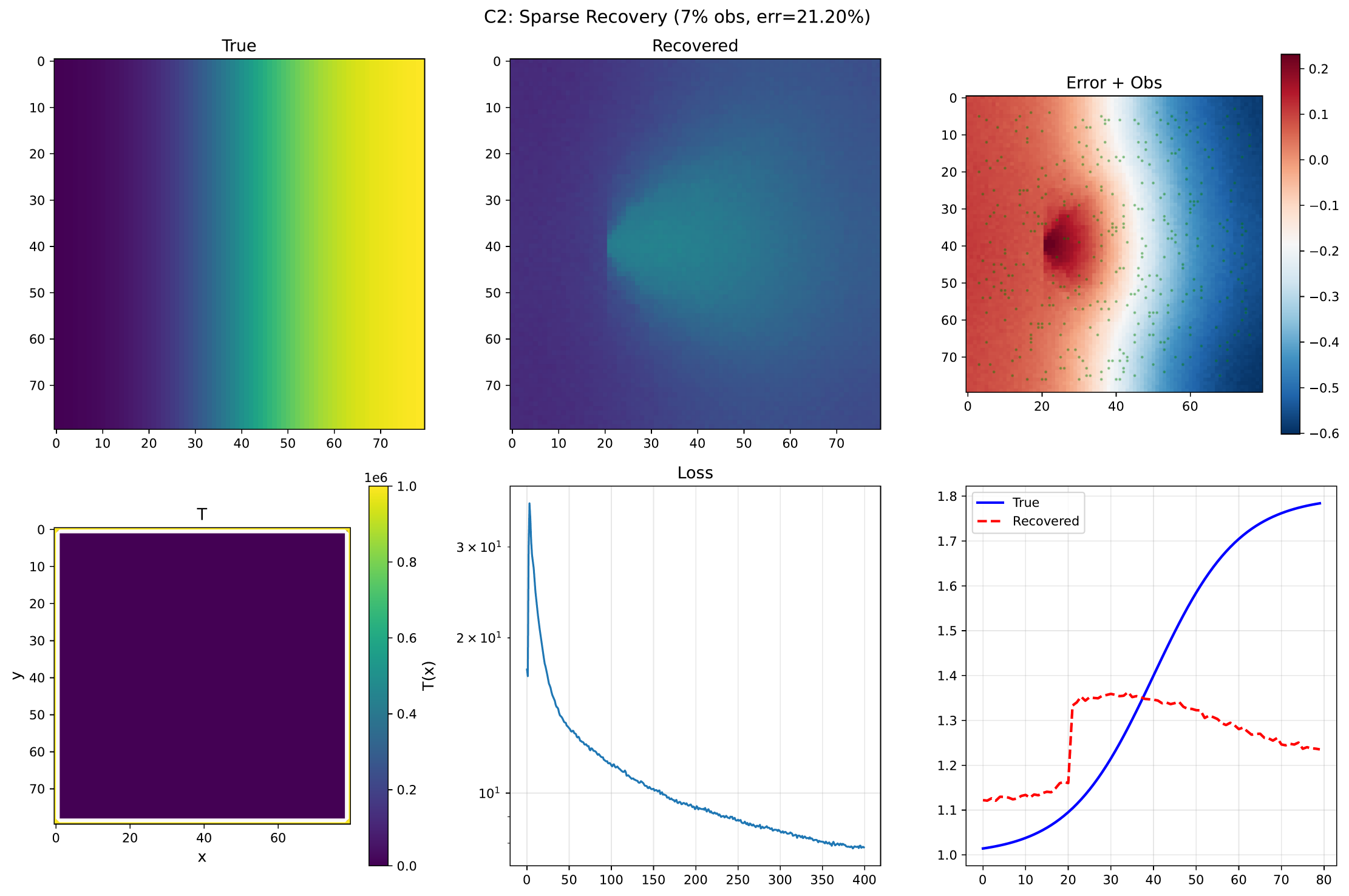}
\caption{\textbf{Isotropic metric recovery with sparse observations (7\%).} Green dots in the error panel indicate observation locations. Despite limited data, the general spatial structure is recovered, though with increased local variation. Recovery error: 21.2\%.}
\label{fig:C2}
\end{figure}

Reducing observations to 7\% of pixels (379 randomly sampled points) demonstrates that the method still recovers the general spatial structure, though with increased local variation in regions far from observation points (Supplementary Figure~\ref{fig:C2}). This result is encouraging for practical applications where satellite observations provide only sparse coverage.

\textbf{Anisotropic Metric Recovery.} Recovering a diagonal anisotropic metric $\mG = \text{diag}(g_{11}, g_{22})$ is substantially harder because the optimizer must distinguish between two independent components from a single scalar observation (arrival time). We test recovery of a metric where both $g_{11}$ and $g_{22}$ have piecewise structure with different values on each side of the domain.

\begin{figure}[htbp]
\centering
\includegraphics[width=\textwidth]{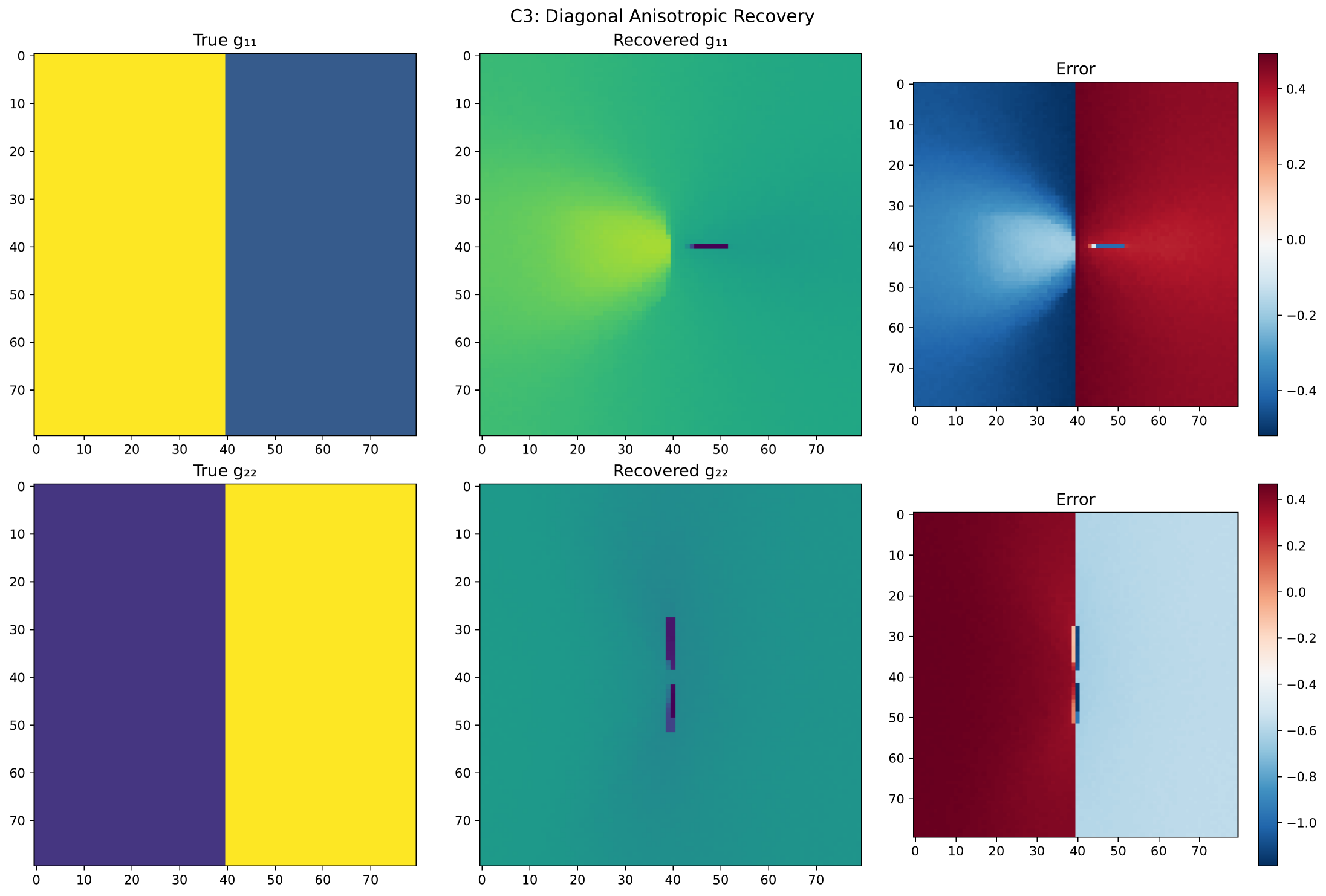}
\caption{\textbf{Diagonal anisotropic metric recovery.} Top row: $g_{11}$ component. Bottom row: $g_{22}$ component. Both components show blurred interfaces compared to the true sharp discontinuities, illustrating the increased difficulty of anisotropic recovery.}
\label{fig:C3}
\end{figure}

The recovery achieves errors of 42\% for $g_{11}$ and 52\% for $g_{22}$. Supplementary Figure~\ref{fig:C3} shows that while the optimizer correctly identifies the presence of an interface and the general magnitude difference between regions, the recovered fields are significantly blurred compared to the true sharp discontinuities. This illustrates a fundamental limitation: anisotropic structure is less constrained by arrival time data than isotropic structure, as multiple metric configurations can produce similar wavefront shapes.

\begin{figure}[htbp]
\centering
\includegraphics[width=\textwidth]{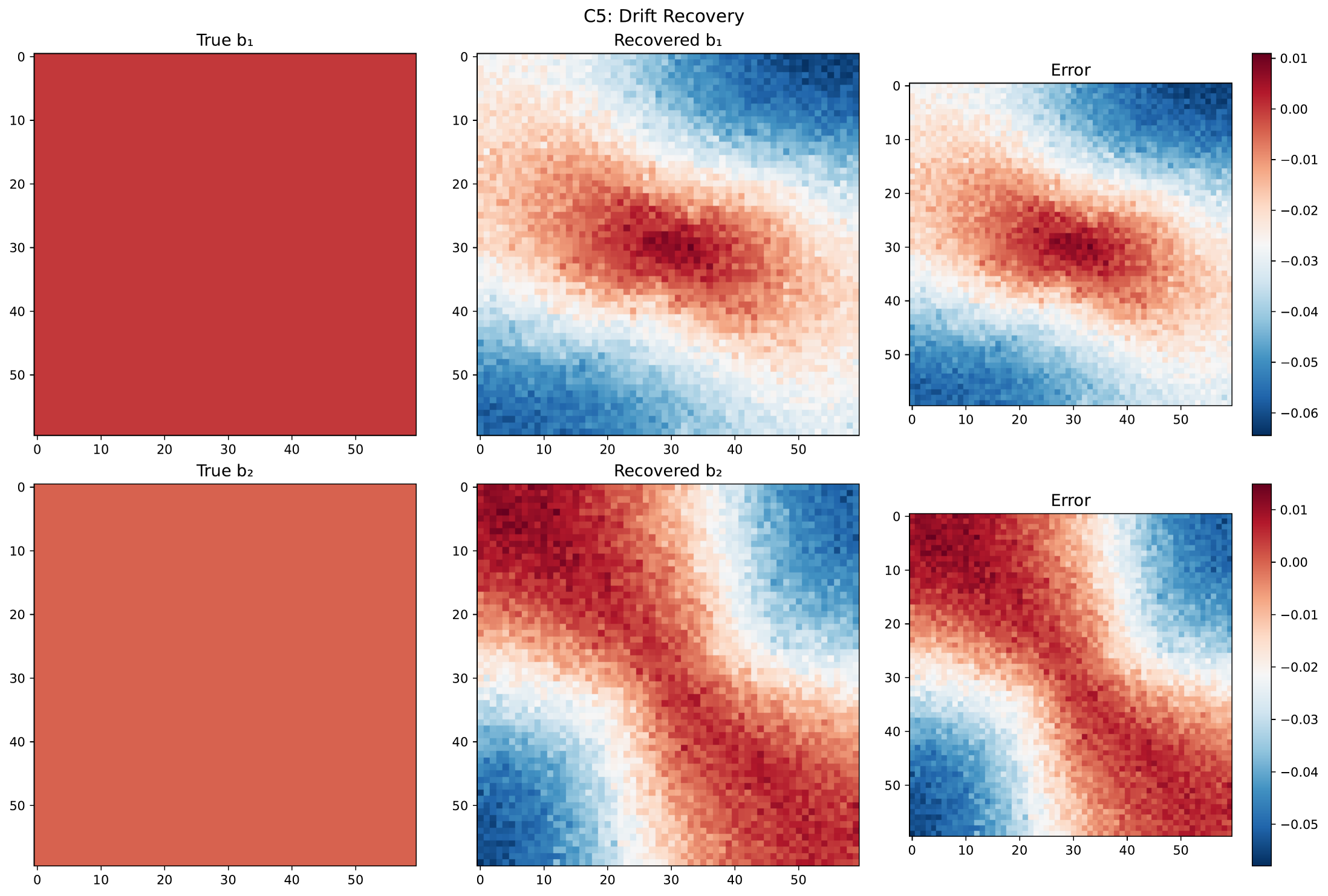}
\caption{\textbf{Drift field recovery.} Top row: $b_1$ component. Bottom row: $b_2$ component. The true constant drift is recovered with high accuracy (errors $<$3\%), with small spatial variations around the correct mean value.}
\label{fig:C5}
\end{figure}

\textbf{Drift Recovery.} Recovering the drift field $\vb$ with known metric $\mG$ tests the sensitivity of arrival times to directional propagation effects. The true drift is spatially constant with $\vb = (0.15, 0.08)$. Supplementary Figure~\ref{fig:C5} shows that while the recovered field exhibits spatial variation around the true constant value, the mean is accurately captured and errors remain within $\pm 0.05$ throughout the domain. The superior performance compared to metric recovery confirms that drift creates a stronger, more distinctive signature in arrival time patterns---specifically, the characteristic asymmetry between upwind and downwind propagation.

\textbf{Regularization Ablation.} Total variation regularization prevents overfitting and promotes piecewise-smooth solutions. We systematically vary the regularization strength $\lambda$ from 0 to 0.5 to identify the optimal value.

\begin{figure}[htbp]
\centering
\includegraphics[width=0.5\textwidth]{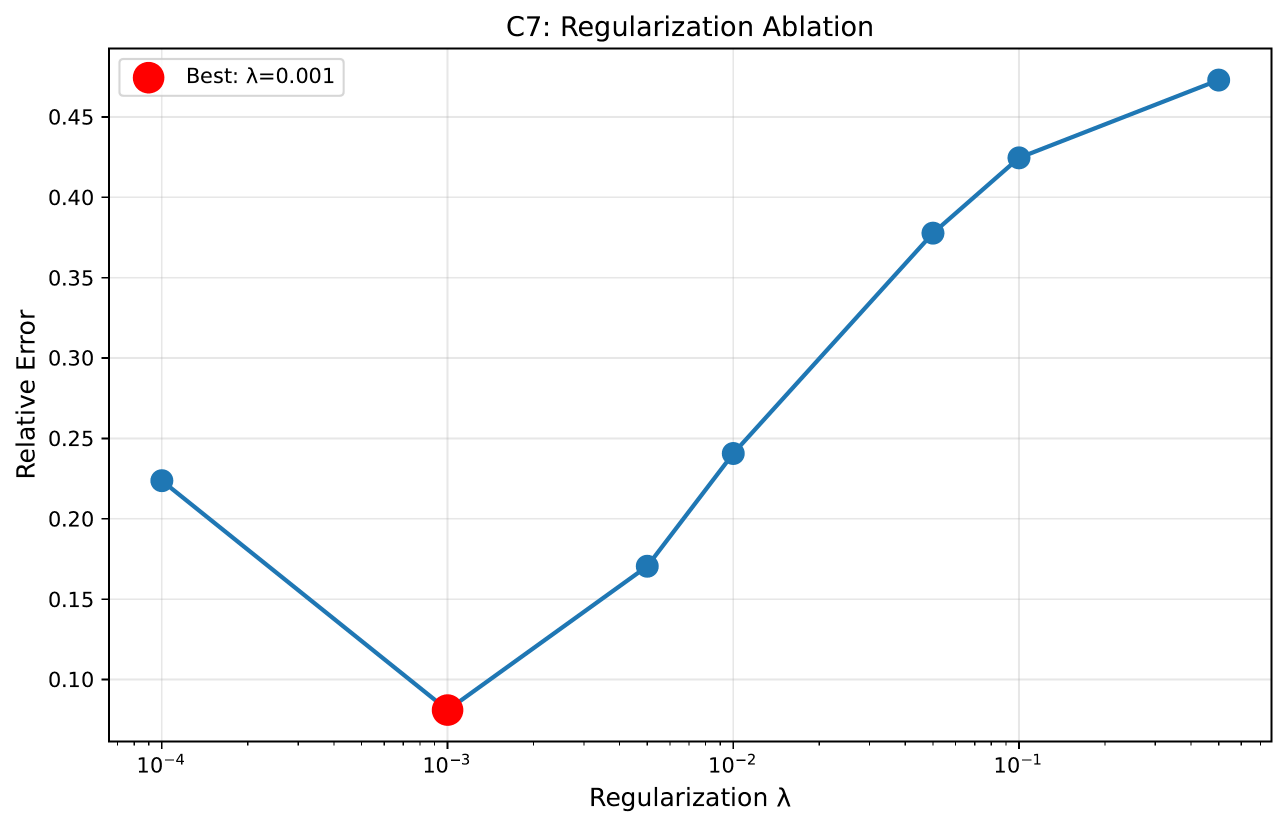}
\caption{\textbf{Regularization strength ablation.} Recovery error versus TV regularization parameter $\lambda$. The U-shaped curve reveals optimal performance at $\lambda=0.001$, with degradation from both under-regularization (overfitting) and over-regularization (excessive smoothing).}
\label{fig:C7}
\end{figure}

Supplementary Figure~\ref{fig:C7} reveals a clear U-shaped relationship between $\lambda$ and recovery error. Without regularization ($\lambda=0$), the optimizer overfits to noise and produces spatially oscillatory solutions. Increasing regularization beyond the optimal point progressively degrades performance, as excessive smoothing prevents recovery of the true spatial structure.

\begin{figure}[htbp]
\centering
\includegraphics[width=0.5\textwidth]{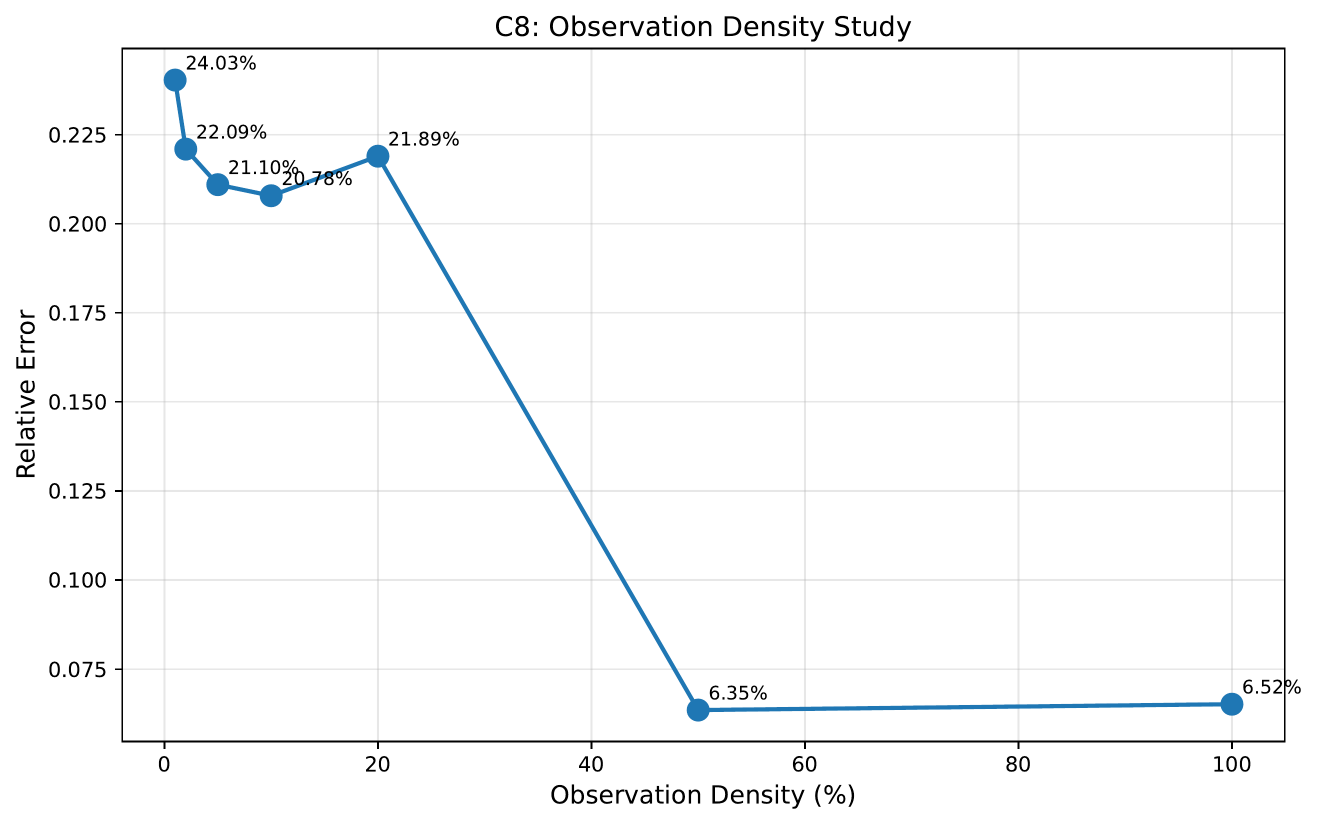}
\caption{\textbf{Observation density study.} Recovery error versus percentage of observed pixels. A phase transition occurs near 50\% density, below which the problem is underdetermined ($\sim$21\% error) and above which performance saturates ($\sim$6\% error).}
\label{fig:C8}
\end{figure}

\textbf{Observation Density Study.} Practical applications often have limited observation coverage. We vary the fraction of observed pixels from 1\% to 100\% to characterize how recovery accuracy depends on data density. Supplementary Figure~\ref{fig:C8} reveals a striking nonlinear relationship with a phase transition near 50\% coverage. Performance is relatively flat between 1\% and 20\% observations. Beyond 50\%, performance saturates. This phase transition suggests a critical density threshold below which the inverse problem is fundamentally underdetermined, and above which additional observations provide diminishing returns.

\begin{figure}[htbp]
\centering
\includegraphics[width=0.5\textwidth]{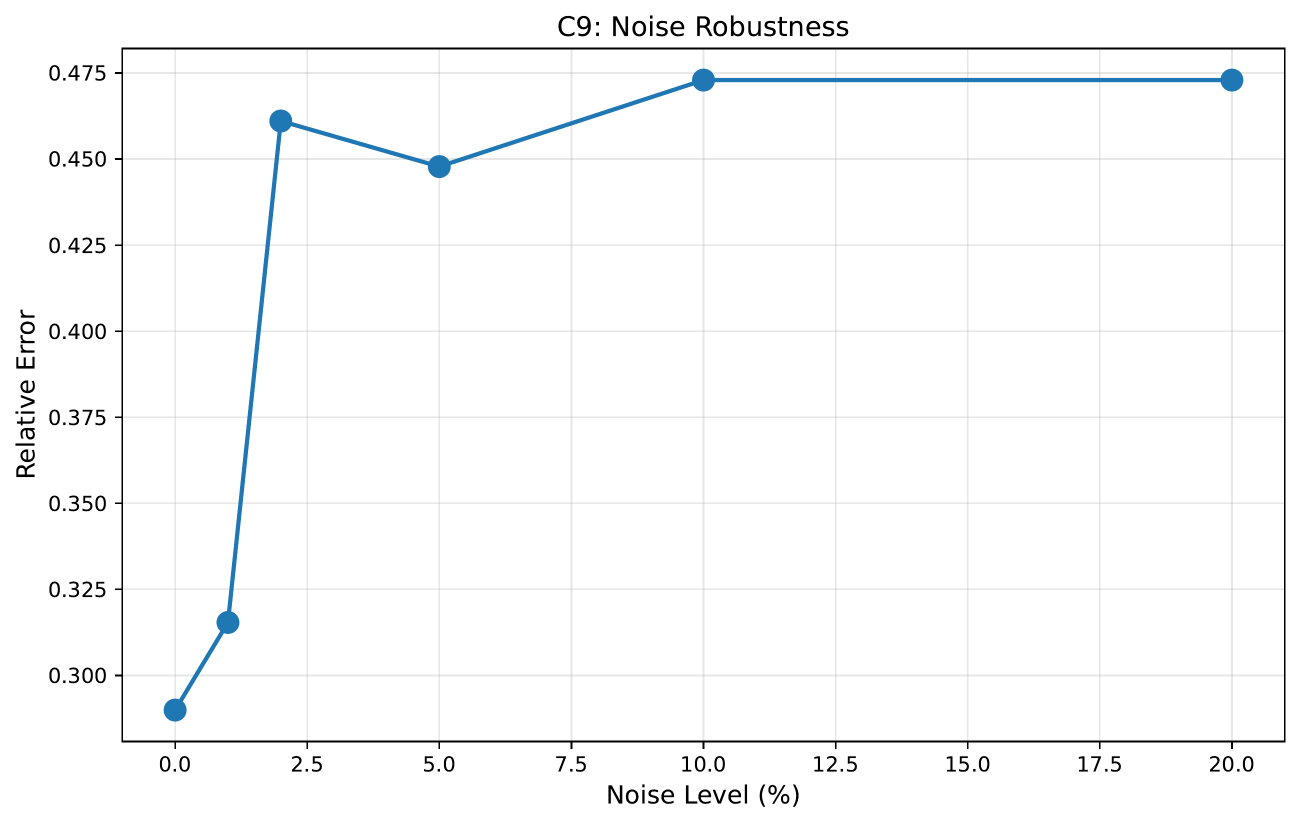}
\caption{\textbf{Noise robustness.} Recovery error versus observation noise level. Performance degrades gracefully, with errors plateauing around 45--47\% for noise levels above 2\%.}
\label{fig:C9}
\end{figure}

\textbf{Noise Robustness.} Real observations contain measurement noise. We add Gaussian noise to the arrival times at levels from 0\% to 20\% of the signal standard deviation. Supplementary Figure~\ref{fig:C9} shows graceful degradation with errors plateauing at higher noise levels, suggesting the regularization effectively prevents overfitting to noisy observations. However, the sensitivity to even small noise (1--2\%) indicates that accurate arrival time measurements are important for precise parameter recovery.

\begin{figure}[htbp]
\centering
\includegraphics[width=0.5\textwidth]{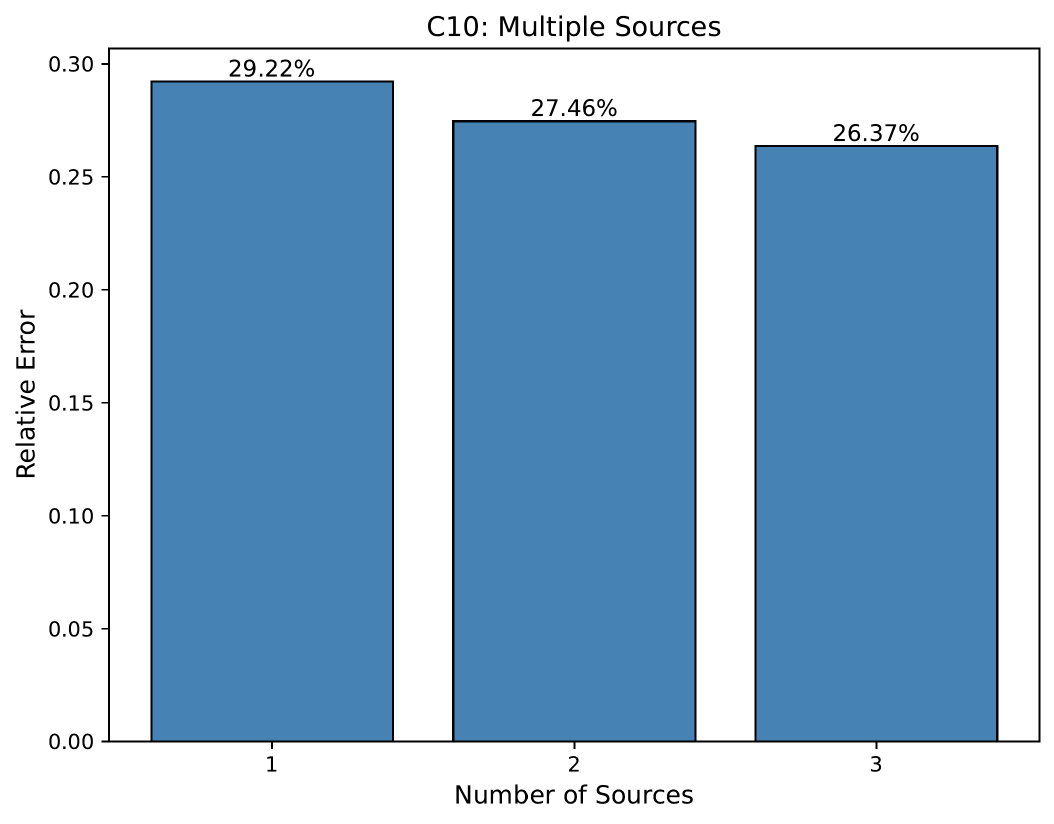}
\caption{\textbf{Multiple ignition sources.} Recovery error decreases modestly from 29.2\% (one source) to 26.4\% (three sources), demonstrating the value of complementary geometric information from multiple wavefront directions.}
\label{fig:C10}
\end{figure}

\textbf{Multiple Ignition Sources.} Multiple ignition points create intersecting wavefronts that probe the domain from different directions. We test whether this additional information improves recovery by varying the number of sources from 1 to 3. Supplementary Figure~\ref{fig:C10} shows modest but consistent improvement. Each additional source provides complementary geometric information, as wavefronts arriving from different directions sample the metric tensor along different characteristic paths.

\textbf{Multi-Source Recovery.} We test if combining data from multiple sources with different ignition locations improves metric recovery over single-source estimation. We simulate 1, 2, 3, and 5 sources on the same domain, each providing sparse (7\%) observations, and jointly optimize a shared metric tensor.

\begin{figure}[htbp]
\centering
\includegraphics[width=0.8\textwidth]{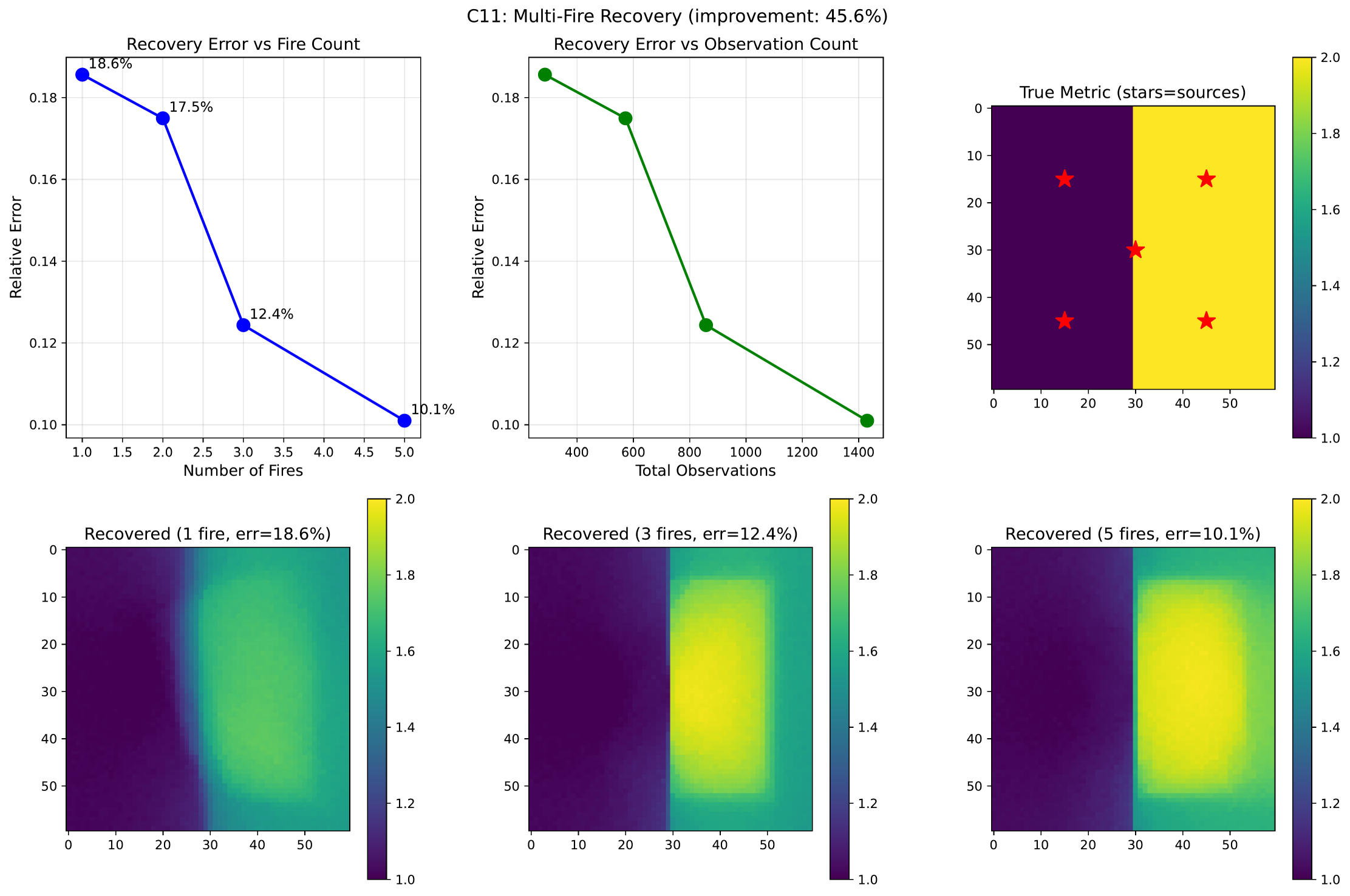}
\caption{\textbf{Multi-fire recovery.} Recovery error decreases from 18.6\% (one fire) to 10.1\% (five fires), a 45.6\% improvement. Different fire geometries provide complementary constraints on the metric tensor beyond simply increasing observation count.}
\label{fig:C11}
\end{figure}

Supplementary Figure~\ref{fig:C11} shows that the improvement exceeds what would be expected from simply increasing observation count. The additional benefit arises because different sources probe the domain from different directions, providing complementary geometric constraints on the metric tensor that redundant observations from a single source cannot.

\newpage
\onecolumn
\subsection{Supplementary figures C}

\newpage
\onecolumn
\section{\textbf{Synthetic Propagation Scenarios}}
\label{app:synthetic_scenarios}

This section demonstrates the forward model's ability to capture realistic propagation patterns driven by spatially-varying metrics and drift fields. These experiments validate that the Randers-Finsler framework produces physically plausible behavior across diverse configurations.

\begin{figure}[htbp]
\centering
\includegraphics[width=\textwidth]{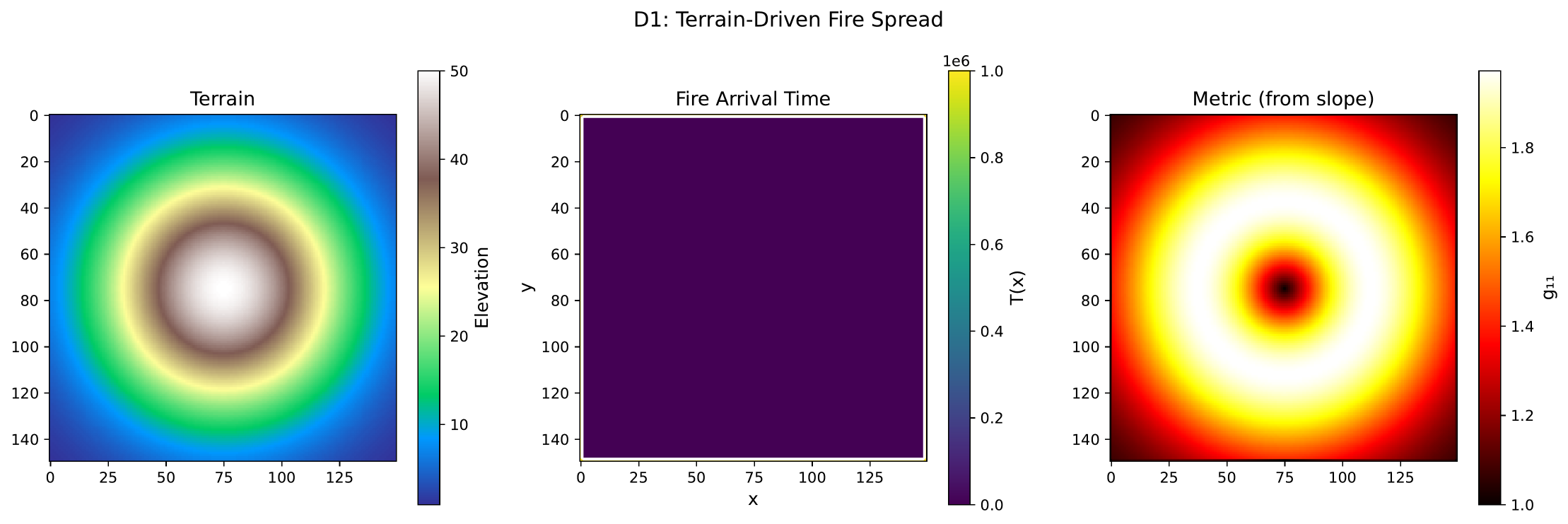}
\caption{\textbf{Terrain-driven fire spread.} Left: Gaussian hill terrain with elevation indicated by color. Center: arrival time field for fire ignited at summit. Right: metric tensor derived from slope magnitude. Fire spreads fastest on flat terrain (summit) and slowest on steep slopes (flanks).}
\label{fig:D1}
\end{figure}

\textbf{Terrain-Driven Spread.} Wavefronts propagate at different speeds depending on local metric values derived from terrain. We model this by constructing the metric tensor from terrain slope: $g(\mathbf{x}) = 1 + \alpha |\boldsymbol{\nabla} z(\mathbf{x})|$ where $z$ is elevation and $\alpha$ controls slope sensitivity. Supplementary Figure~\ref{fig:D1} shows propagation from the peak of a Gaussian hill. The metric is lowest at the summit (flat terrain) and highest on the flanks where slopes are steepest. The resulting arrival time field shows the wavefront initially spreading rapidly from the flat summit, then slowing as it encounters steep slopes.

\begin{figure}[htbp]
\centering
\includegraphics[width=\textwidth]{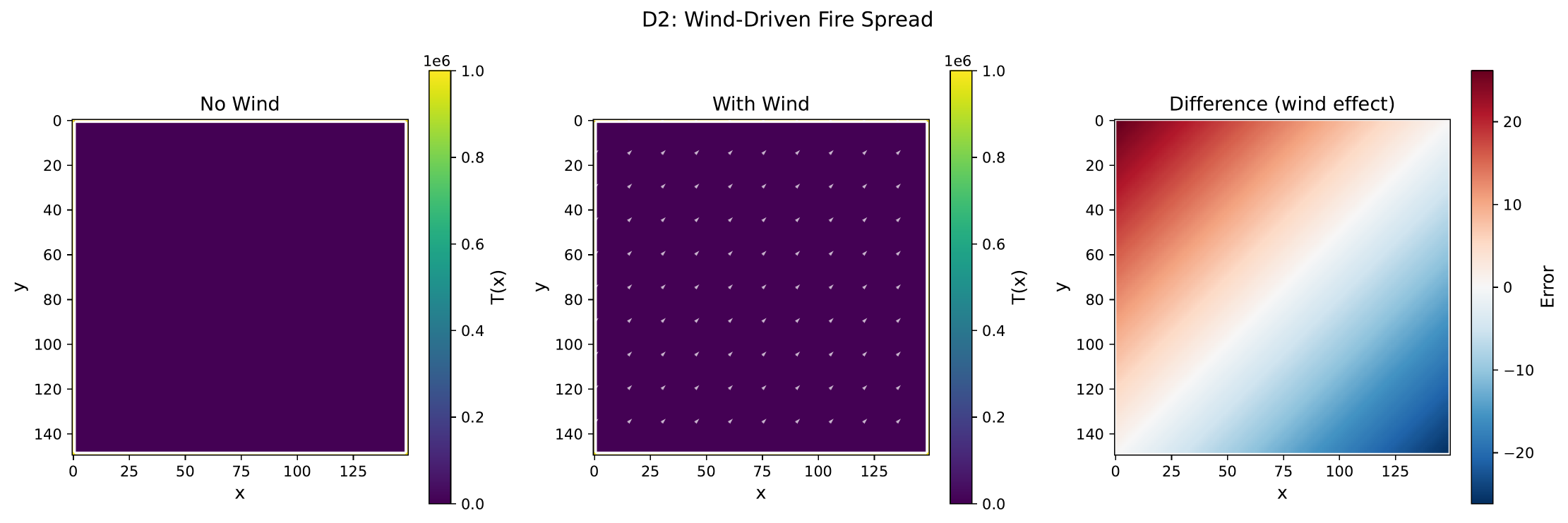}
\caption{\textbf{Wind-driven fire spread.} Left: arrival time without wind (circular spread). Center: arrival time with wind (arrows indicate direction). Right: difference showing wind effect - negative values (blue) indicate faster arrival downwind, positive values (red) indicate delayed arrival upwind.}
\label{fig:D2}
\end{figure}

\textbf{Drift-Driven Spread.} External forces create asymmetric propagation through the drift term $\vb$ in the Randers metric. Supplementary Figure~\ref{fig:D2} compares propagation with and without drift on a flat, homogeneous domain. Without drift, the wavefront spreads in concentric circles from the source. With drift (arrows indicate direction), the wavefront elongates in the drift direction while being suppressed in the opposite direction. The difference map quantifies this asymmetry: arrival times decrease by up to 20 units in the drift direction and increase by similar amounts opposite to drift. This demonstrates the model's ability to capture directional bias in propagation.

\begin{figure}[htbp]
\centering
\includegraphics[width=\textwidth]{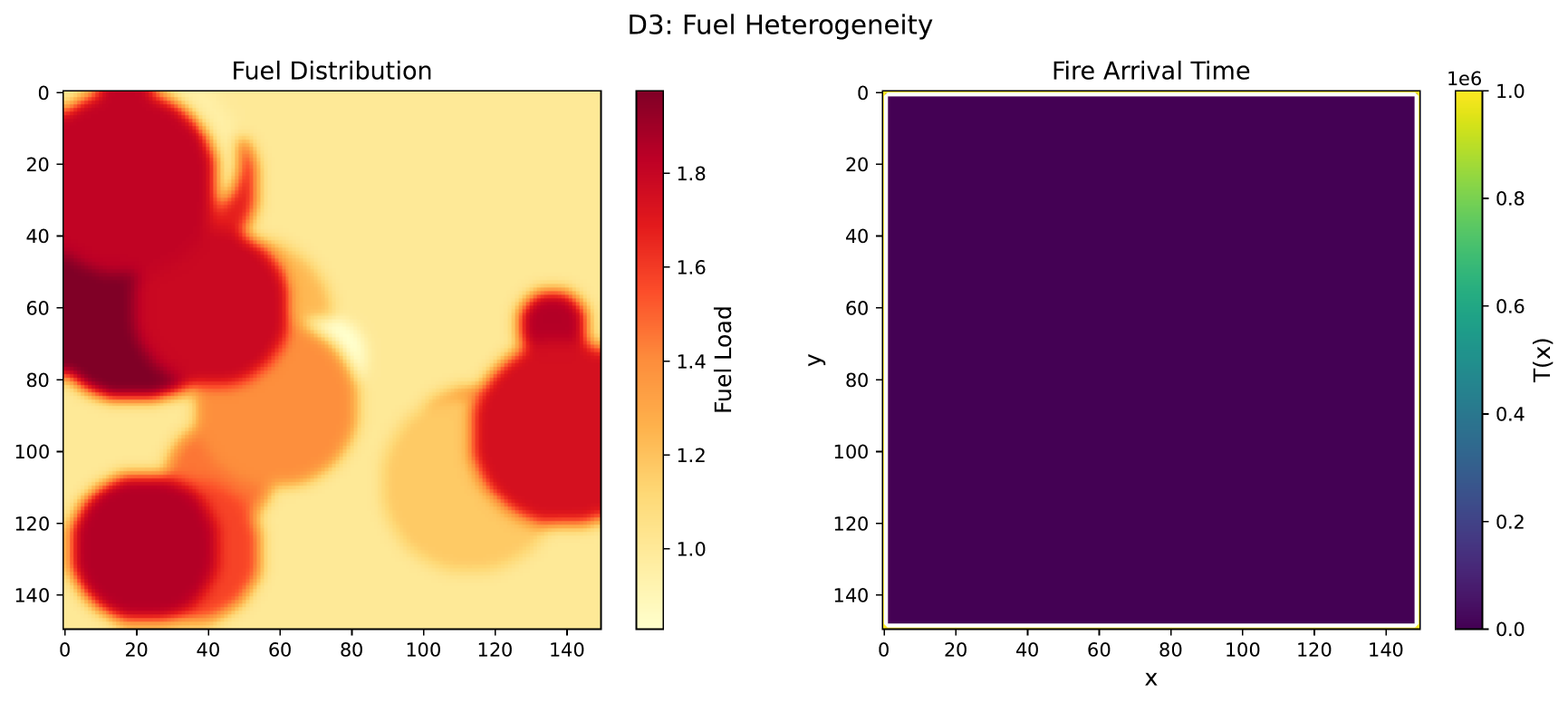}
\caption{\textbf{Fuel heterogeneity effects.} Left: spatial distribution of fuel load with dense patches (red) and light fuel (yellow). Right: arrival time field showing irregular wavefront propagation as fire slows through dense fuel regions.}
\label{fig:D3}
\end{figure}

\textbf{Heterogeneous Media.} Material properties vary spatially in many applications. We model this by assigning the metric $g(\vx)$ proportional to local resistance, so that denser regions produce slower propagation. Supplementary Figure~\ref{fig:D3} shows a domain with randomly distributed high-resistance patches (red regions). The arrival time field reveals how the wavefront deforms around these obstacles: propagation slows through high-resistance regions and accelerates through low-resistance regions, creating irregular, non-circular wavefronts.

\begin{figure}[htbp]
\centering
\includegraphics[width=\textwidth]{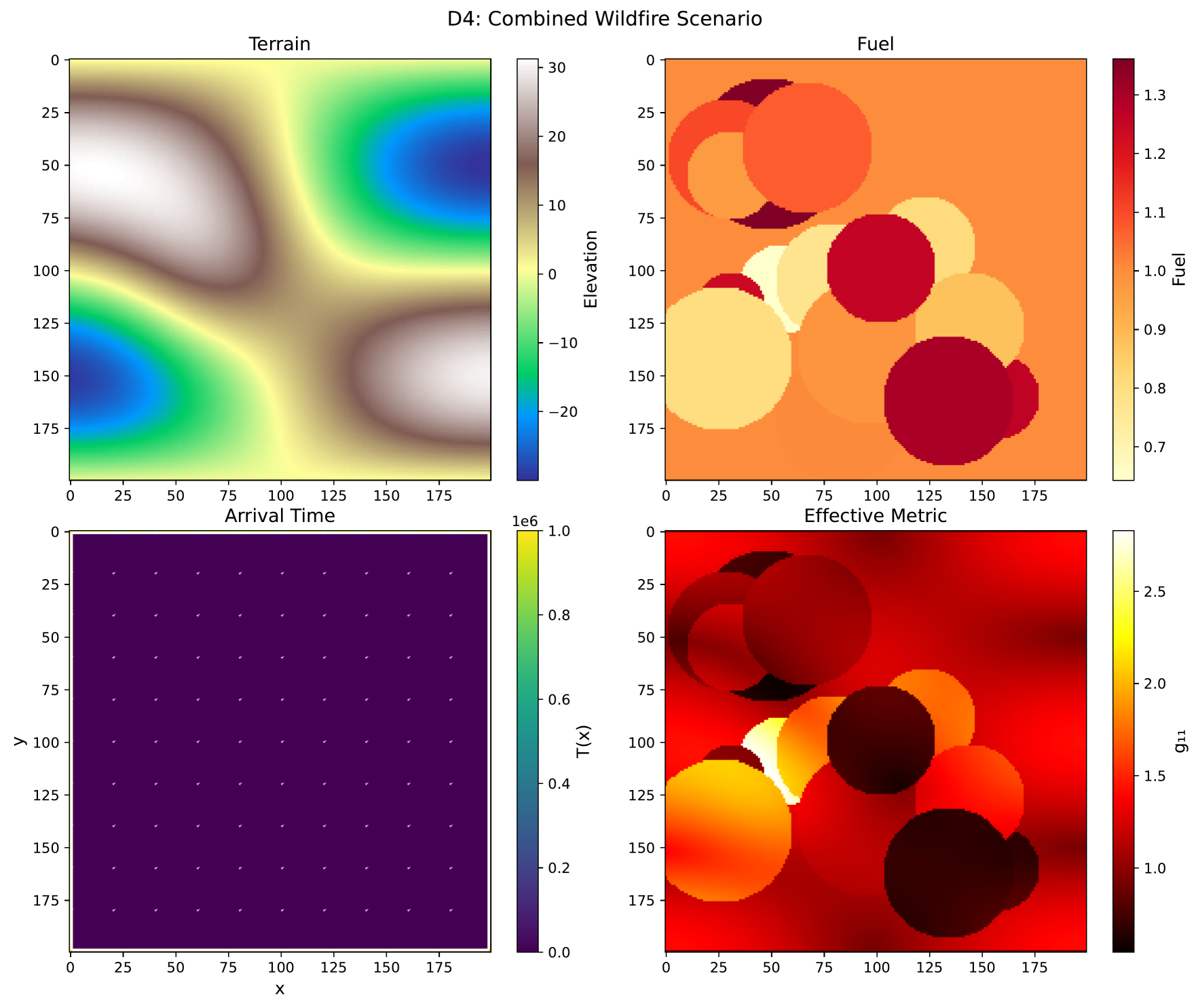}
\caption{\textbf{Combined wildfire scenario.} Top row: terrain (left) and fuel distribution (right). Bottom row: arrival time field (left) and effective metric combining terrain, fuel, and wind effects (right). The complex metric pattern produces realistic heterogeneous fire spread.}
\label{fig:D4}
\end{figure}

\textbf{Combined Scenario.} Real applications experience terrain, drift, and heterogeneity effects simultaneously. Supplementary Figure~\ref{fig:D4} combines all three factors: complex terrain with ridges and valleys, heterogeneous material patches, and uniform drift. The effective metric integrates these influences multiplicatively, producing a complex spatial pattern. The resulting arrival time field shows the wavefront responding to local conditions---slowing on steep slopes and high-resistance regions, accelerating in the drift direction and through low-resistance regions. This demonstrates that the Randers-Finsler framework can represent multi-factor complexity.

\textbf{Wavefront Reconstruction.} A key application is reconstructing continuous wavefronts from sparse observations. Given arrival times at scattered locations, we solve the eikonal equation to interpolate the full arrival time field, from which wavefront contours (iso-arrival curves) can be extracted at any desired time.

\begin{figure}[htbp]
\centering
\includegraphics[width=\textwidth]{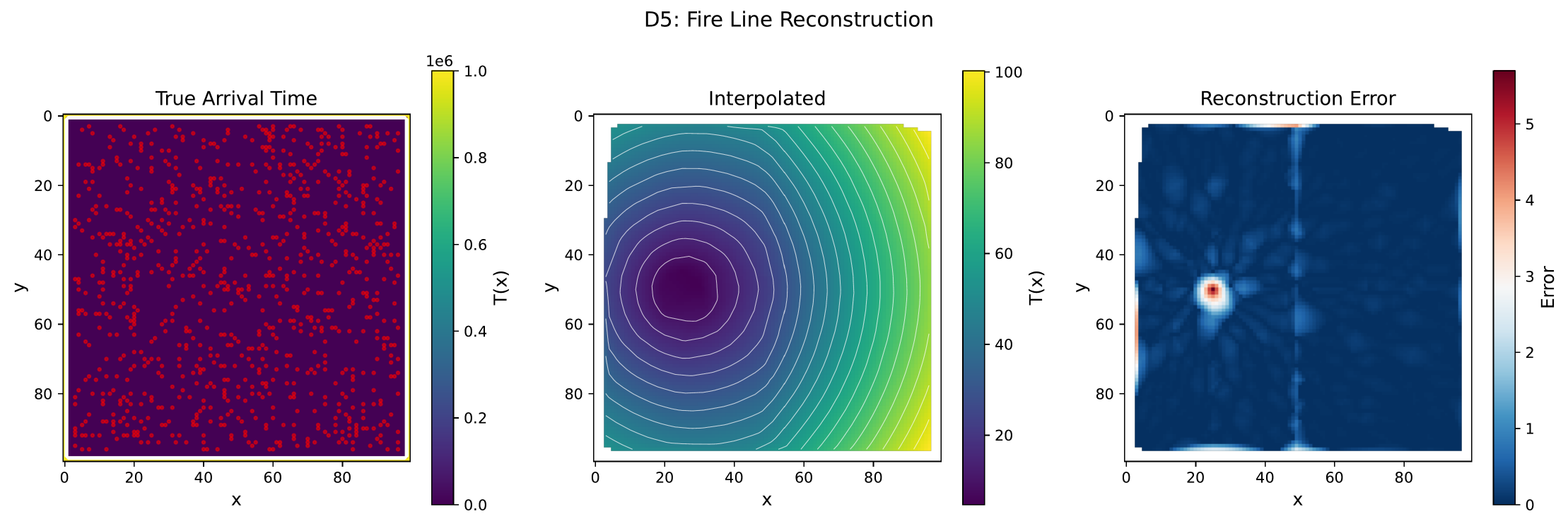}
\caption{\textbf{Fire line reconstruction from sparse observations.} Left: true arrival time with observation locations (red dots). Center: interpolated arrival time field with fire line contours. Right: reconstruction error, concentrated near the source singularity. Overall reconstruction error: 12\%.}
\label{fig:D5}
\end{figure}

Supplementary Figure~\ref{fig:D5} shows reconstruction from 878 randomly distributed observations. The interpolated field accurately recovers the smooth, concentric arrival time pattern, with reconstruction error of 12\%. Error is concentrated near the source where the gradient singularity creates local inaccuracies. Away from the source, errors remain below 1--2 units, demonstrating that the eikonal framework provides physically consistent interpolation that respects the causal structure of wavefront propagation.

\textbf{Parameter Sensitivity.} Understanding how parameter errors propagate to prediction errors is essential for uncertainty quantification. We perturb the metric tensor by adding spatially-correlated Gaussian noise at levels from 0\% to 50\% and measure the resulting error in predicted arrival times.

\begin{figure}[htbp]
\centering
\includegraphics[width=0.8\textwidth]{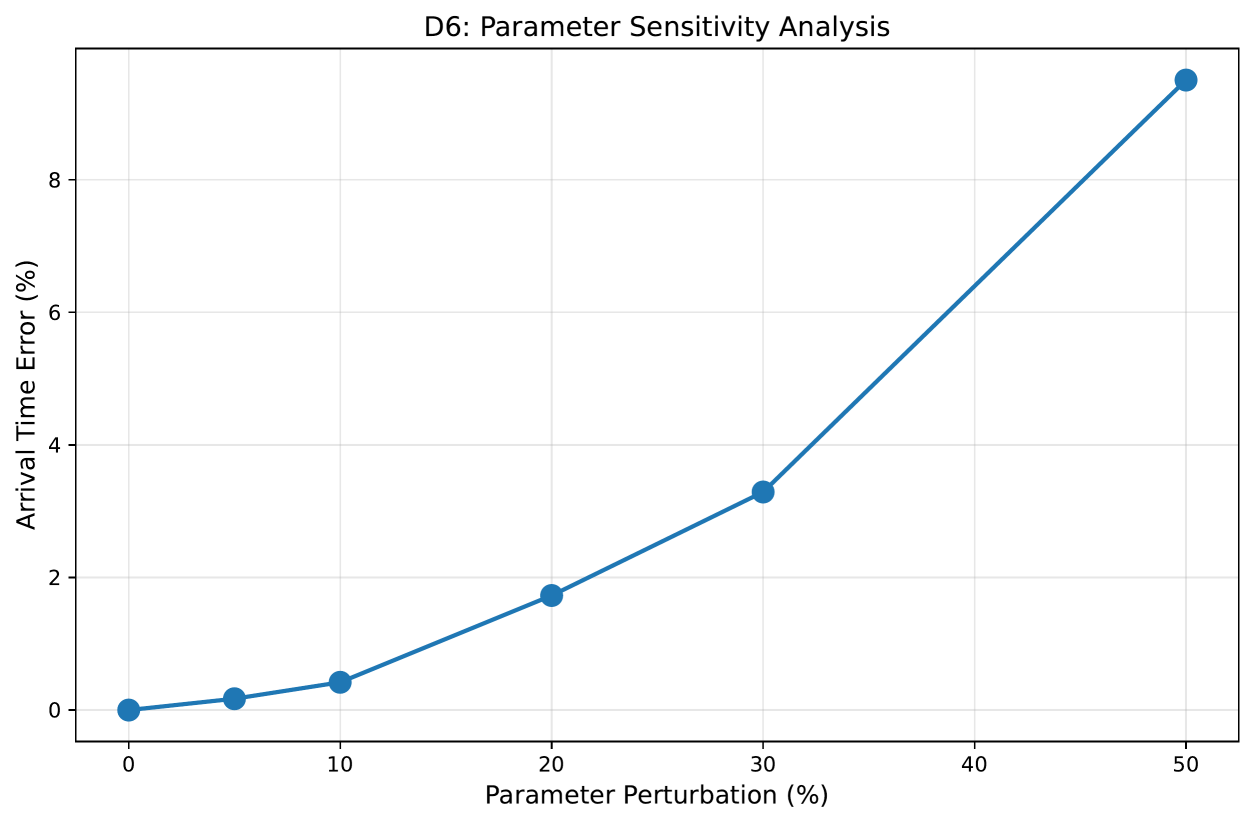}
\caption{\textbf{Parameter sensitivity analysis.} Arrival time error versus metric perturbation level. The near-linear relationship with slope $\approx$0.2 indicates favorable error propagation: 10\% parameter error produces only 0.4\% prediction error.}
\label{fig:D6}
\end{figure}

Supplementary Figure~\ref{fig:D6} shows a nearly linear relationship between parameter perturbation and arrival time error. A 10\% perturbation in the metric produces only 0.4\% error in arrival times, while 30\% perturbation yields 3.3\% error. This favorable error amplification ratio (approximately 1:5) indicates that the forward model is well-conditioned: moderate uncertainty in parameters translates to manageable uncertainty in predictions. This property is crucial for practical applications where input parameters are never known exactly.

\newpage
\onecolumn
\subsection{Supplementary figures D}

\newpage
\onecolumn
\section{\textbf{Computational Comparisons}}
\label{app:comparisons}

This section provides additional computational comparisons. Core efficiency results are presented in Section~\ref{sec:efficiency}.

\textbf{Baseline Selection.} We compare fast sweeping against Jacobi iteration rather than the Fast Marching Method because our focus is on differentiable solving: FMM's heap-based ordering creates parameter-dependent node orderings that complicate the adjoint structure, whereas both Jacobi and fast sweeping process nodes in fixed orders amenable to implicit differentiation. For the backward pass, we compare implicit differentiation against finite differences rather than unrolled automatic differentiation because the latter requires storing all intermediate states across sweeping iterations; for the grid sizes in our wildfire experiments ($300 \times 500$ and larger), this memory requirement exceeded available GPU capacity (48GB).  

\textbf{Fast Sweeping vs Jacobi Iteration.} The eikonal equation can be solved iteratively using either Jacobi iteration (updating all nodes simultaneously) or fast sweeping (alternating directional sweeps). We compare runtime on grids from $50 \times 50$ to $400 \times 400$.

\begin{table}[htbp]
\centering
\caption{Runtime comparison of fast sweeping and Jacobi iteration.}
\label{tab:sweep_vs_jacobi}
\begin{tabular}{cccc}
\toprule
Grid Size & Sweeping (s) & Jacobi (s) & Speedup \\
\midrule
$50 \times 50$ & 1.2 & 22.6 & 18$\times$ \\
$100 \times 100$ & 1.4 & 183.5 & 129$\times$ \\
$200 \times 200$ & 2.8 & 1509.2 & 533$\times$ \\
$400 \times 400$ & 8.7 & 12294.6 & 1419$\times$ \\
\bottomrule
\end{tabular}
\end{table}

\begin{figure}[htbp]
\centering
\includegraphics[width=\textwidth]{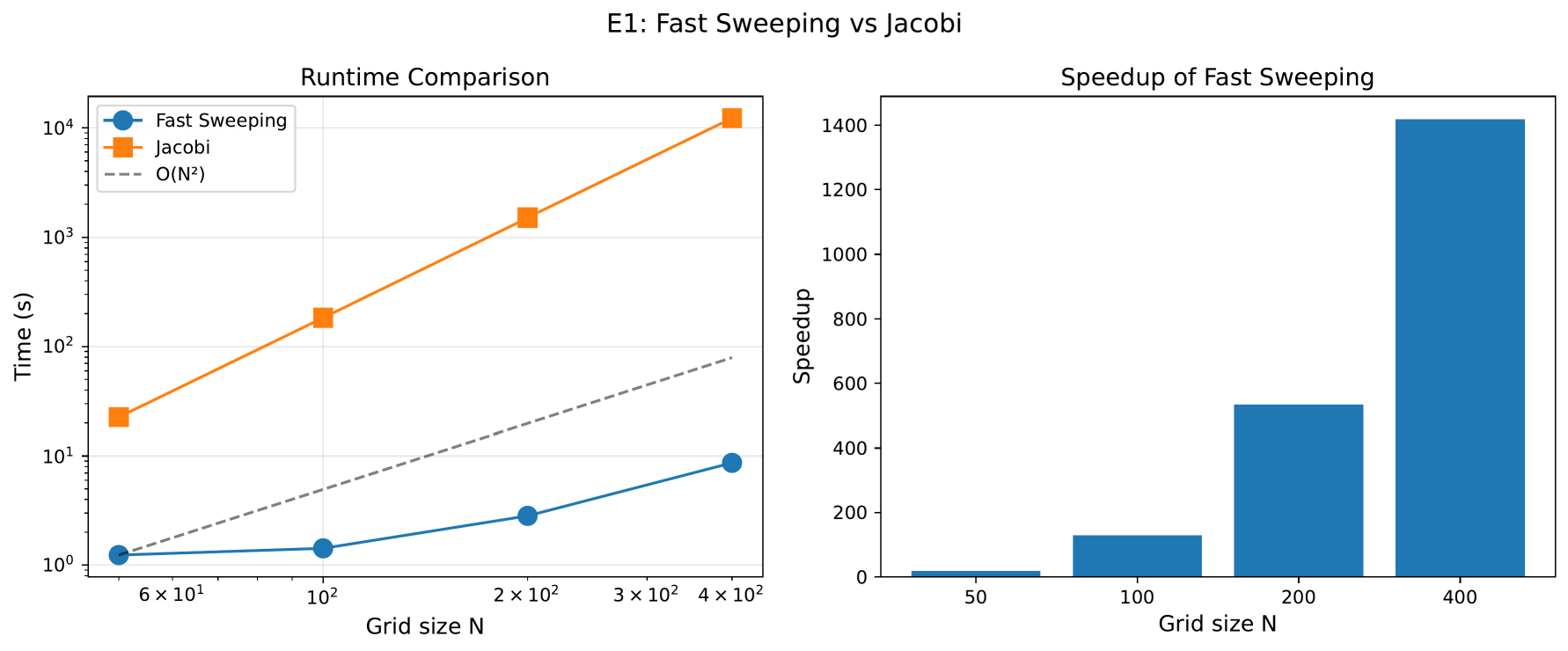}
\caption{\textbf{Fast sweeping vs Jacobi iteration.} Left: runtime comparison on log-log scale showing fast sweeping maintains $O(n)$ scaling while Jacobi approaches $O(n^{3/2})$ for $n = N^2$ grid points. Right: speedup factor increases with grid size, reaching 1400$\times$ at $400 \times 400$.}
\label{fig:E1}
\end{figure}

Supplementary Figure~\ref{fig:E1} and Table~\ref{tab:sweep_vs_jacobi} show the detailed comparison. Fast sweeping converges in 2--3 iterations regardless of grid size because it propagates information along characteristics in each sweep direction. Jacobi iteration requires $O(N)$ iterations for information to traverse an $N \times N$ grid, resulting in $O(n^{3/2})$ total complexity versus $O(n)$ for fast sweeping, where $n = N^2$ is the total number of grid points.

\begin{figure}[htbp]
\centering
\includegraphics[width=0.8\textwidth]{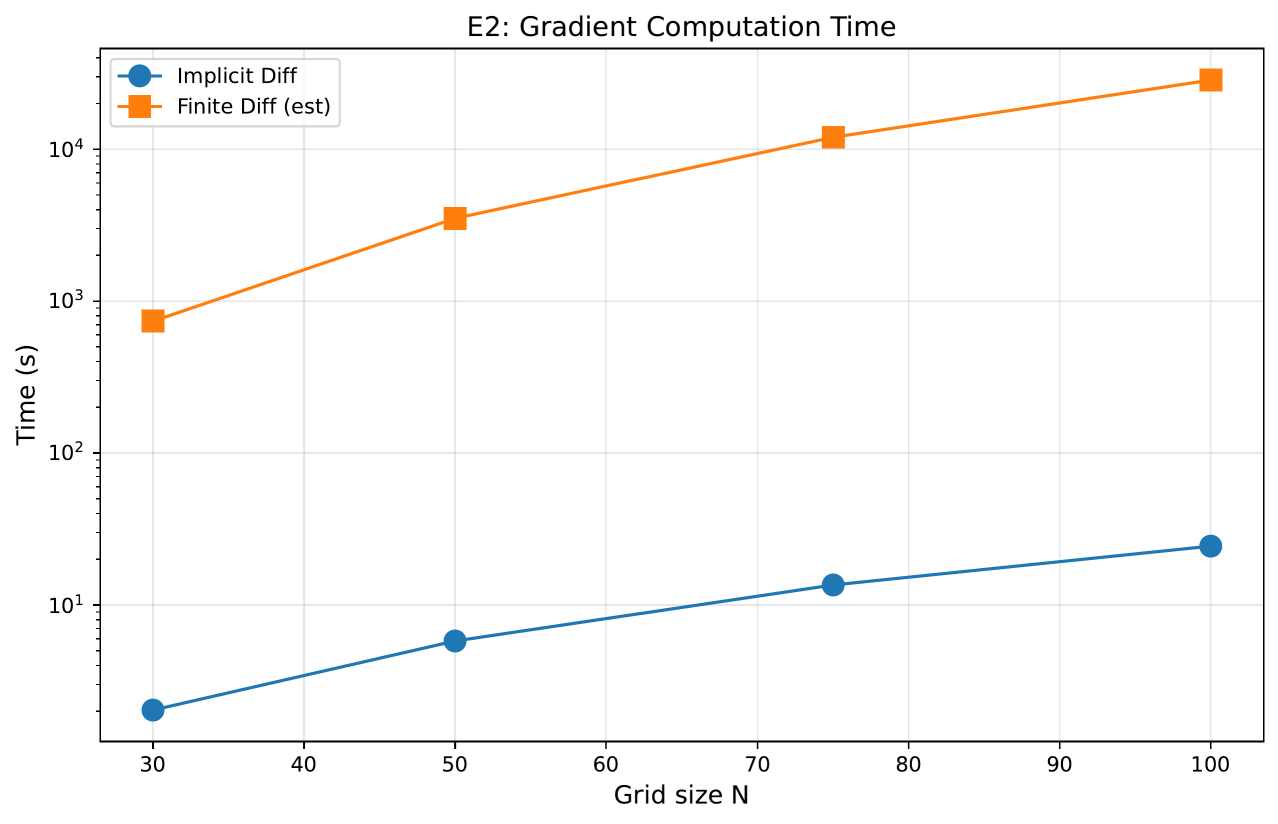}
\caption{\textbf{Implicit differentiation vs finite differences.} Gradient computation time on log scale. Implicit differentiation (blue) achieves 750$\times$ average speedup over finite differences (orange, estimated), with the gap widening at larger grids.}
\label{fig:E2}
\end{figure}

\textbf{Implicit Differentiation vs Finite Differences.} Computing gradients through the eikonal solver can use either implicit differentiation (solving an adjoint system) or finite differences (perturbing each parameter). For a grid with $N^2$ parameters, finite differences requires $O(N^2)$ forward solves, while implicit differentiation requires only one backward pass. Supplementary Figure~\ref{fig:E2} shows the detailed timing comparison.

\begin{figure}[htbp]
\centering
\includegraphics[width=\textwidth]{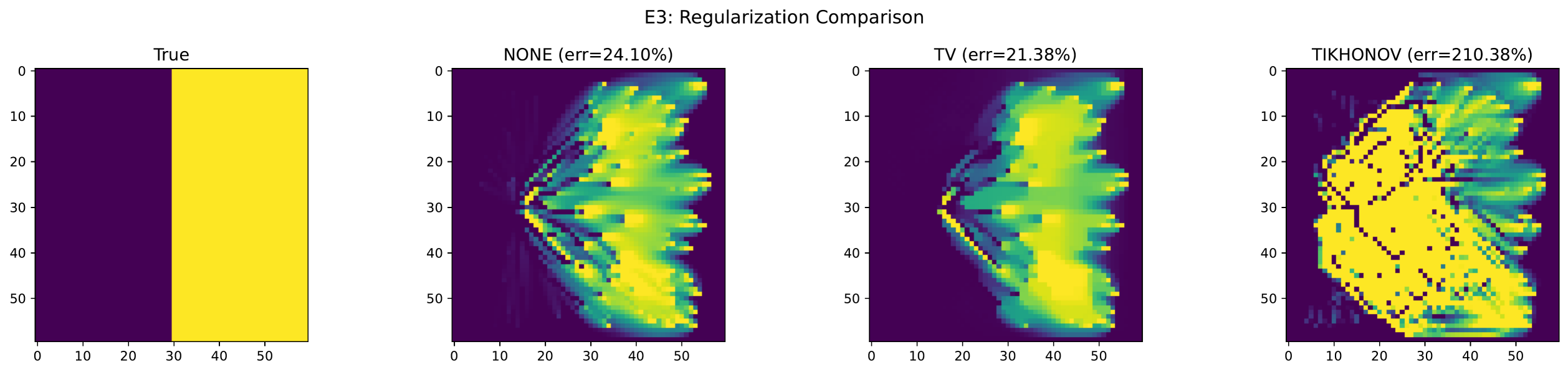}
\caption{\textbf{Regularization comparison.} From left: true metric, no regularization (24\% error), TV regularization (21\% error), Tikhonov regularization (210\% error). TV produces the cleanest recovery; Tikhonov fails by penalizing correct parameter values.}
\label{fig:E3}
\end{figure}

\textbf{Regularization Strategies.} We compare three regularization approaches for the inverse problem: no regularization, total variation (TV), and Tikhonov (L2 penalty on parameter magnitude). Supplementary Figure~\ref{fig:E3} shows recovered metrics for each strategy. Without regularization, the optimizer produces visible spatial oscillations as it overfits to observation noise. TV regularization promotes piecewise constant solutions that match the true metric structure. Tikhonov regularization fails because it penalizes deviation from the initial guess rather than spatial roughness, preventing the optimizer from finding the correct parameter values. This confirms TV as the appropriate regularization for piecewise-smooth parameter fields.

\begin{figure}[htbp]
\centering
\includegraphics[width=\textwidth]{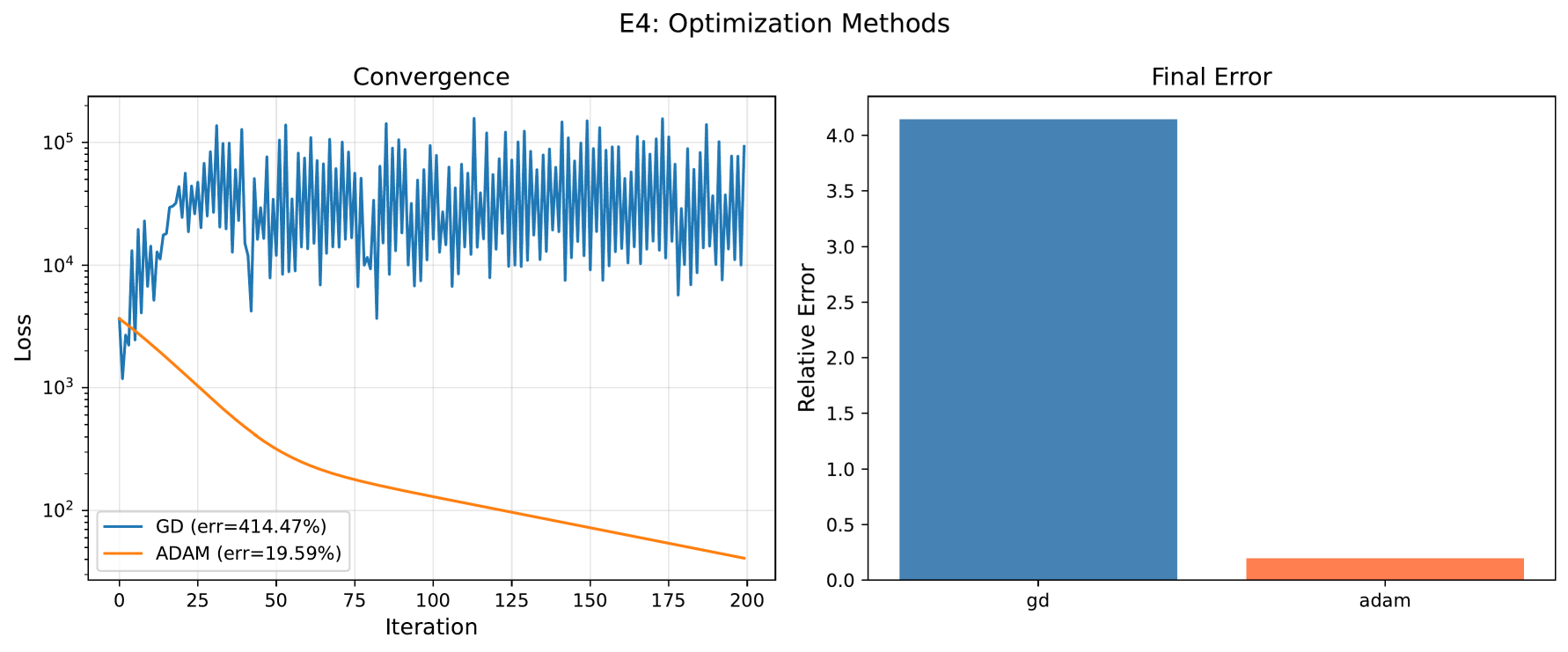}
\caption{\textbf{Optimization method comparison.} Left: loss curves showing Adam converges while gradient descent oscillates. Right: final error---Adam achieves 20\% error versus 414\% for gradient descent.}
\label{fig:E4}
\end{figure}

\textbf{Optimization Methods.} We compare gradient descent (GD) with fixed learning rate against Adam with adaptive learning rates and momentum. Supplementary Figure~\ref{fig:E4} shows dramatic differences in convergence behavior. Gradient descent fails to converge meaningfully, oscillating at high loss values. Adam converges smoothly over two orders of magnitude in loss. The poor performance of gradient descent reflects the ill-conditioned nature of the inverse problem: different spatial regions have vastly different gradient magnitudes, requiring adaptive per-parameter learning rates that Adam provides. This justifies our use of Adam throughout the experiments.

\textbf{Scalability.} We measure forward and backward pass times across grid sizes from $50 \times 50$ to $800 \times 800$ using the Numba JIT-compiled implementation.

\begin{figure}[htbp]
\centering
\includegraphics[width=\textwidth]{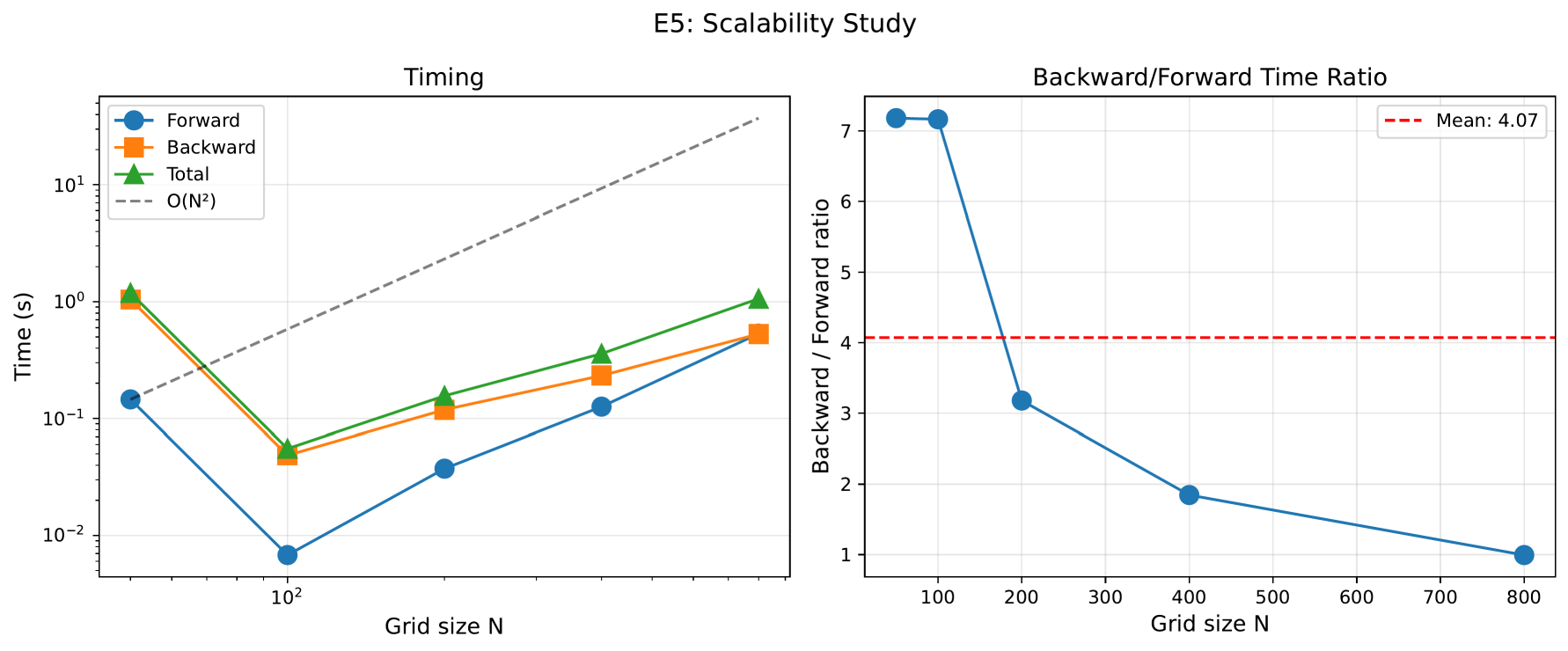}
\caption{\textbf{Scalability study (Numba-optimized).} Left: forward, backward, and total time versus grid size, all following $O(n)$ scaling for $n = N^2$ grid points. Right: backward/forward time ratio averages 4$\times$, demonstrating efficient gradient computation via implicit differentiation with JIT compilation.}
\label{fig:E5}
\end{figure}

Supplementary Figure~\ref{fig:E5} confirms that both forward and backward passes scale as $O(N^2)$, consistent with the fast sweeping algorithm visiting each grid point a constant number of times. With Numba optimization, the backward pass is approximately 4$\times$ slower than the forward pass on average. At the largest grid size ($800 \times 800$), the full forward-backward pass completes in approximately 1 second, enabling rapid iteration during optimization. This efficiency makes gradient-based learning feasible even for high-resolution wildfire domains.

\end{document}